\documentclass[twocolumn]{aastex63}

\usepackage{amsmath}

\newcommand\pcc{\,{\rm cm}^{-3}}
\newcommand\Msun{\, M_{\odot}}
\newcommand\kms{\, {\rm km}\,{\rm s}^{-1}}

\newcommand\cm{\,{\rm cm}}

\newcommand\second{\,{\rm s}}

\newcommand\Myr{\,{\rm Myr}}

\newcommand\pc{\,{\rm pc}}
\newcommand\kpc{\,{\rm kpc}}

\newcommand\Kel{\,{\rm K}}
\newcommand\eV{\,{\rm eV}}

\newcommand\simgt{\lower.5ex\hbox{$\; \buildrel > \over \sim \;$}}
\newcommand\simlt{\lower.5ex\hbox{$\; \buildrel < \over \sim \;$}}

\newcommand{\xHI}{x_{\rm H^0}}
\newcommand{\xHII}{x_{\rm H^+}}

\newcommand{\nH}{n_{\rm H}}

\newcommand{\HII}{\ion{H}{2}}

\newcommand{\CNM}{{\tt CNM}}

\newcommand{\WNM}{{\tt WNM}}

\newcommand{\WIM}{{\tt WIM}}

\newcommand{\CpU}{{\tt C+U}}

\newcommand{\hot}{{\tt hot}}

\newcommand{\epsPP}{\epsilon_{\rm PP}}
\newcommand{\dxymax}{d_{xy,{\rm max}}}
\newcommand{\zpp}{z_{\rm p\textnormal{-}p}}

\newcommand{\Pth}{P_\mathrm{th}}

\usepackage{amsmath}
\usepackage{bbm,bm}
\usepackage{multirow}
\usepackage{colortbl}
\usepackage{hyperref}
\usepackage{outlines}
\usepackage{comment,color}
\usepackage{enumitem}

\defcitealias{OML10}{OML10}
\defcitealias{OK22}{OK22}

\defcitealias{kim2020first}{Kim20}
\defcitealias{NCRI}{Kim23}

\graphicspath{{./}{figures/}}

\shorttitle{}
\shortauthors{}

\begin{document}

\title{Ultraviolet Radiation Fields in Star-Forming Disk Galaxies: Numerical Simulations with TIGRESS-NCR}

\author[0000-0001-8840-2538]{Nora B. Linzer}
\affiliation{Department of Astrophysical Sciences, Princeton University, 4 Ivy Lane, Princeton, NJ 08544, USA} \email{nlinzer@princeton.edu}

\author[0000-0001-6228-8634]{Jeong-Gyu Kim}
\affiliation{Division of Science, National Astronomical Observatory of Japan, Mitaka, Tokyo 181-0015, Japan}
\affiliation{Korea Astronomy and Space Science Institute, Daejeon 34055, Republic of Korea}
\affiliation{Department of Astrophysical Sciences, Princeton University, 4 Ivy Lane, Princeton, NJ 08544, USA}

\author[0000-0003-2896-3725]{Chang-Goo Kim}
\affiliation{Department of Astrophysical Sciences, Princeton University, 4 Ivy Lane, Princeton, NJ 08544, USA}

\author[0000-0002-0509-9113]{Eve C. Ostriker}
\affiliation{Department of Astrophysical Sciences, Princeton University, 4 Ivy Lane, Princeton, NJ 08544, USA}
\affiliation{Institute for Advanced Study, 1 Einstein Drive, Princeton, NJ 08540, USA}

\begin{abstract}
With numerical simulations that employ adaptive ray-tracing (ART) for radiative transfer at the same time as evolving gas magnetohydrodynamics, thermodynamics, and photochemistry, it is possible to obtain a high resolution view of ultraviolet (UV) fields and their effects in realistic models of the multiphase interstellar medium. Here, we analyze results from TIGRESS-NCR simulations, which follow both far-UV (FUV) wavelengths, important for  photoelectric heating and PAH excitation, and the Lyman continuum (LyC), which photoionizes hydrogen. Considering two models, representing solar neighborhood and inner galaxy conditions, we characterize the spatial distribution and time variation of UV radiation fields, and quantify their correlations with gas. We compare four approximate models for the FUV to simulated values to evaluate alternatives when full ART is infeasible. By convolving FUV radiation with density, we produce mock maps of dust emission. We introduce a method to calibrate mid-IR observations, for example from JWST, to obtain high resolution gas surface density maps. We then consider the LyC radiation field, finding most of the gas exposed to this radiation to be in ionization-recombination equilibrium and to have a low neutral fraction. Additionally, we characterize the ionization parameter as a function of environment. Using a simplified model of the LyC radiation field, we produce synthetic maps of emission measure (EM). We show that the simplified model can be used to extract an estimate of the neutral fraction of the photoionized gas and mean free path of ionizing radiation from observed EM maps in galaxies.
\end{abstract}

\section{Introduction}\label{sec:intro}

Ultraviolet (UV) radiation from young stars represents the largest energy input to the interstellar medium (ISM), exceeding that from supernovae (SNe) by a factor of $\sim 100$ \citep[e.g.][]{leitherer1999, kim2023photchem}. This radiation directly affects the physical state of the ISM, notably by heating and ionizing the gas, and produces observational diagnostics in both gas and dust. Because radiation fields vary in time and space \citep{parravano2003time}, it is important to include a time-dependent treatment of radiation in numerical ISM simulations.

Far-UV (FUV) radiation is critical to gas thermodynamics, with the equilibrium thermal pressure $\Pth$ in diffuse, neutral gas proportional to the FUV radiation field through photoelectric (PE) heating \citep[e.g.][]{1994ApJ...427..822B,1995ApJ...443..152W,2003ApJ...587..278W,2017ApJ...835..201H}; this relationship is key to the feedback loop that self-regulates star formation on large scales \citep{OML10}. Lyman continuum (LyC) radiation contributes to the dynamical state of the ISM since ionized gas pressure forces from this ``early feedback'' are believed to destroy most of the self-gravitating clouds where star formation occurs, with radiation pressure on dust becoming more important in the highest density clouds \citep[e.g.][]{1979MNRAS.186...59W,2009ApJ...703.1352K,sales2014stellar,2016ApJ...819..137K,2018ApJ...859...68K,
2019ARA&A..57..227K,
2023ASPC..534....1C,2024arXiv240500813M}. 

The inclusion of radiation modeling in numerical simulations enables more realistic thermodynamics in a variety of environments where heating and cooling depend on both the radiation itself and the properties of the gas that are sensitive to radiation, such as the chemical and ionic state.  In addition to enabling realistic thermodynamic and dynamic responses to stellar feedback, inclusion of radiation makes it possible to create synthetic observables that depend on radiation heating and photochemical reactions, including dust emission.  

Only very recently has time-dependent modeling begun to include radiative transfer to follow spatial variations in the FUV and LyC radiation intensities.  Many simulations employ approximate radiative transfer treatments such as two-moment/M1 or tree-based methods \citep[e.g.][]{2013ApJS..206...21S, 2013MNRAS.436.2188R, 2015MNRAS.449.4380R, 2019MNRAS.485..117K, 2019ApJS..242...20M, 2021MNRAS.506.5512F, wunsch2021tree,2022MNRAS.512.1430W, grudic2022dynamics,klepitko2023tree,wadsley2024trevr2}. Approximate methods are often sufficient for treating FUV radiation \cite[see comparisons of the two-moment implementation of \citealt{2013ApJS..206...21S} with ray-tracing in][]{kim2017modeling}.  However, given the short mean free path $\ell \sim 0.1 \pc (\nH/\pcc)^{-1}$ for LyC radiation, and corresponding exponential attenuation at optical depth $\tau >1$, methods with high spatial and angular resolution are crucial to an accurate solution for ionizing radiation. Methods that allow both high spatial and high angular resolution include Monte Carlo \citep[e.g.][]{2019A&C....27...63H}, variable Eddington tensor \citep[e.g.][]{2012ApJS..199...14J, 2022MNRAS.512..401M}, and adaptive ray tracing \citep[e.g.][]{2002MNRAS.330L..53A, rosen2017hybrid, kim2017modeling}.

The implementation by \cite{kim2017modeling} of adaptive ray tracing (ART) in {\it Athena} has made it possible, for the first time,  to follow radiative transfer with very high angular resolution at the same spatial resolution as MHD, at manageable computational cost. The ART solver was first applied to study giant molecular cloud (GMC) evolution under feedback \citep{2018ApJ...859...68K, kimjg2019, kim2021star}, and subsequently adapted to study the large-scale multiphase ISM with disk geometry and sheared rotation using the TIGRESS framework \citep{kim2017three,kim2023ncr}.  This new implementation is now being applied in an extensive simulation survey to study the dependence of star formation regulation under varying metallicity and environmental conditions \citep{2024arXiv240519227K}.  

The multiphase disk simulations presented in \citet{kim2023ncr} are similar in their environmental conditions to the Milky Way and other nearby star-forming galaxies. With these simulation results in hand, it is of interest to explore, for the first time, the detailed properties of the FUV and LyC radiation fields. This analysis of radiation fields, for two simulation runs,  is the subject of the present paper.  

We first summarize the ray-tracing methods used to evolve the UV radiation fields, and the model parameters for the two simulations we will explore (\autoref{sec:methods}). In \autoref{sec:sim} we provide an overview of the gas and radiation properties in both simulations. We quantify the statistical properties of the FUV radiation field in \autoref{sec:FUV}. In addition, \autoref{sec:FUV} compares the simulated FUV radiation field to different approximate models, evaluating them for potential usage as inexpensive alternatives to full radiative transfer. These approximate treatments include (1) the analytic slab model from \citet{OML10} that was adopted for setting the heating rate in \citet[][]{kim2020first}, (2) a plane-parallel model, (3) an analytic model from  \citet{bialy2020far}, and (4) an estimate of the radiation field taken from a sum over FUV source particles. The FUV radiation field produces IR emission through dust reprocessing \citep[including spectral features due to small grains -- ][]{2001ApJ...554..778L,Tielens2008}, so we also use our simulated results to characterize properties of the dust heating rate.  We explore potential usage of approximate radiation field models in calibrating PAH emission maps  \citep[such as those from {\it JWST}, e.g.][]{2023ApJ...944L...9L,pathak2023two}, in order to improve measurement of the gas surface density.
Finally, in \autoref{sec:LyC}, we examine the LyC radiation field and how it directly affects the ionization state of the gas. We present statistics of the radiation field itself and of the ionization parameter. We also consider an approximate model for the LyC radiation on large scales inspired by the model explored in \cite{belfiore2022tale}.  Our main conclusions are summarized in \autoref{sec:summary}.

\section{Numerical Methods and Models}\label{sec:methods}

This paper focuses on the UV radiation properties from the TIGRESS-NCR simulations for two galactic conditions presented in \citet{kim2023ncr}. TIGRESS-NCR is an extension of the ``TIGRESS-classic'' numerical framework \citep{kim2017three} for modeling local patches of vertically stratified, multiphase ISM disks with resolved star formation and feedback. The methodological enhancements are in the treatment of ISM photochemistry and thermodynamics based on \citet{Gong2017}, coupled to explicit UV radiation transfer via the ART method of \citet{kim2017modeling}. This enables us to explicitly model all major phases of the ISM and the UV radiation fields permeating them. \citet{kim2023photchem} describe photochemistry in determining time-dependent hydrogen and equilibrium C and O bearing species abundances from which we calculate cooling and heating rates. \citet{kim2023ncr} explain physics modules and their numerical algorithms as employed in the TIGRESS-NCR simulations. We refer interested readers to these papers. Here, we summarize the technical details regarding UV radiation that are most relevant to this paper.

\subsection{Radiation Transfer Methods}

In the TIGRESS-NCR simulations, we evaluate the UV radiation field from luminous point sources (representing star clusters) via ray-tracing. We track three frequency bins: (1) Lyman continuum (LyC) band ($\lambda < 912$ \AA) for hydrogen photoionization; (2) Lyman-Werner (LW) band (912 \AA $\;\le \lambda < 1108$ \AA) for H$_2$ dissociation, C ionization, and grain photoelectric heating; and (3) photoelectric (PE) band (1108 \AA $\;\le \lambda < 2066$ \AA) for grain photoelectric heating. When applied to our simulations, we use ``FUV'' to refer to the combined PE and LW bands. The input luminosities for these bands from each star cluster particle as a function of age ($t_{\rm age}$) is determined using Starburst99 (\citealt{leitherer2014}; see also Fig 21 in \citealt{kim2023photchem}). Although star cluster particles are active SN hosts for $t_{\rm age}=40\,{\rm Myr}$, $\sim$90\% of the FUV (PE+LW) radiation and essentially all of the LyC radiation are from star clusters younger than 20 Myr. We thus only consider particles with $t_{\rm age} < 20\,{\rm Myr}$ as UV radiation sources to save the computational expense of ray tracing.

Our radiative transfer method tracks frequency-dependent dust and gas absorption \citep[see Appendix B of][for crossections]{kim2023photchem}, neglecting scattering. In particular, the LyC band is subject to dust absorption and photoionization of atomic and molecular hydrogen; the LW band is subject to dust absorption and photodissociation of H$_2$, accounting for self-shielding along rays following \citet{DraineBertoldi1996}; the PE band is subject to dust absorption.

We adopt $\sigma_{\rm pi,H} = 3\times 10^{-18}\cm^2$ for the SED-averaged H photoionization cross section in the LyC band. The dust absorption cross sections per H are $\sigma_{\rm d}/(10^{-21}\cm^2)$ = (1, 2, 1) for the (LyC, LW, PE) bands. In the LW and PE bands, these values are appropriate for dust grains characteristic of the local diffuse ISM \citep[see Table 3 in][]{kim2023photchem}. The value in the LyC band is somewhat lower than the estimate based on a theoretical extinction model \citep{WeingartnerDraine2001extinction}. It should be noted, however, that the dust opacity (and other grain properties) in photoionized gas remains uncertain as grains may suffer collisional- and photo-destruction \citep{2010A&A...510A..37M,chastenet2019, egorov2023} and rotational disruption \citep{Hoang2019} in regions with intense UV radiation fields.

\subsubsection{Adaptive Ray Tracing}\label{sec:ART}

We solve for UV radiation transfer using the ART module implemented in {\it Athena} \citep{kim2017modeling}. Given information about the location and luminosity of point sources, photon packets are injected along rays at the location of each source and transported radially outward.  The direction of ray propagation is determined by the HEALPix scheme \citep{gorski2005healpix}, which divides a unit sphere into $12\times 4^{\ell}$ equal-area pixels at HEALPix level $\ell$. We adopt an initial HEALPix level of $\ell_0 = 4$. A minimum angular resolution of four rays per grid cell per source is maintained by dividing a ray into four sub-rays if the distance from the source becomes greater than $d_{\rm max}(\ell) = [3/(4\pi)]^{1/2}2^{\ell}\Delta x$, where $\Delta x$ is the grid resolution.  As photon packets propagate through the grid, they interact with each intersecting cell based on physical properties of the dusty gas (density; species abundance of H and H$_2$) and the path length of the ray through the cell. 
Accordingly, the optical depth through each grid cell and the volume-averaged radiation energy density ($\mathcal{E}$) and flux ($\pmb{F}$) are computed. These in turn are used to evaluate photochemical and photoheating rates and radiation pressure forces.

We apply shearing-periodic boundary conditions to rays crossing horizontal boundaries. For rays crossing the $\hat{x}$ (radial) boundaries, their $y$-position and source position are shifted by appropriate shear displacement, while the periodic boundary condition is applied to rays crossing the $\hat{y}$ (azimuthal) boundaries.

To alleviate the high computational cost of radiative transfer, we do not perform ray tracing every MHD time step, which is determined by the Courant--Friedrichs--Lewy (CFL) condition for the hot phase. The fastest wave speed in the hot phase can exceed $>1000\kms$ easily.  
Rather, we perform ray tracing at intervals determined by the CFL condition for the cold and warm phases with $T< 2\times 10^4\Kel$, whose maximum speed remains $\sim 100\kms$. This choice is justified by the fact that UV radiation has little interaction with the hot gas.

As described in \citet{kim2023ncr}, we employ the following ray termination conditions:
\begin{itemize}[itemsep=3pt,parsep=3pt,topsep=10pt]
  \item[(1a)] A ray hits the vertical boundaries
  \item[(1b)] The horizontal distance travelled from the source becomes larger than $\dxymax$
  \item[(2a)] (FUV only) The ratio between the photon packet luminosity and the total source luminosity in the computational domain falls below a small number $\epsPP$
  \item[(2b)] (FUV only) A ray in the FUV band is terminated if $|z| > \zpp$, where $\zpp$ is the height beyond which the plane-parallel approximation is used (described in \autoref{sec:planepar})
  \item[(3)] The optical depth from the source becomes larger than 20.
\end{itemize}
The adopted values of the termination parameters are summarized in \autoref{tab:model}. We stop following a ray completely if the condition (1a) or (1b) is met or if photons in all bands are used up, satisfying (2a) or (2b) for FUV and (3) for LyC\footnote{In practice, the condition (3) is hardly met for FUV photons, since the dust optical depth through the computational domain is order unity along most directions.}.

By adopting the conditions (1b) and (2a), we ignore contributions to UV radiation fields from distant and/or faint sources. This helps to reduce the cost of ART without significantly reducing its accuracy. A significant fraction of rays are completely terminated before reaching $z= \pm \zpp$, because both LyC and FUV photons are exhausted by the conditions (3) and (2b), respectively. This further helps to reduce the computational cost of ART.

\subsubsection{Plane Parallel Approximation}\label{sec:planepar}

We adopt the plane-parallel radiative transfer solution for the FUV (PE and LW) band in high-altitude regions with $|z| > \zpp$ \citep[see Appendix A.1 in][]{kim2023ncr}. Whenever a ray reaches the height $\zpp > 0 $ from below, we collect its luminosity as a function of the cosine angle $\mu = \hat{\pmb{k}} \cdot \hat{\pmb{z}} > 0$, where $\hat{\pmb{k}}$ and $\hat{\pmb{z}}$ are unit vectors along the ray propagation direction and $z$-axis, respectively. These are used to calculate the area-averaged mean intensity $\langle I \rangle (\mu; \zpp)$ at $\zpp$ as a function of $\mu$ (assuming symmetry of the intensity about the $z$-axis). The area-averaged radiation energy density and the vertical component of radiation flux at $z > \zpp$ can then be obtained as
\begin{equation}\label{e:Erad_pp} \langle \mathcal{E}_{\rm p\textnormal{-}p} \rangle(z) = (2\pi/c) \int_{-1}^{1} \langle I \rangle (\mu; \zpp) e^{-\Delta \tau/\mu}d\mu
\end{equation}
\begin{equation}\label{e:Frad_z_pp}
\langle F_{z,{\rm p\textnormal{-}p}} \rangle(z) = 2\pi \int_{-1}^{1} \langle I \rangle (\mu; \zpp) e^{-\Delta \tau/\mu} \mu d\mu.
\end{equation}
Here $\Delta \tau(z ; \zpp)$ is the area-averaged vertical dust optical depth from $\zpp$ to $z$.   
In evaluating $\langle I \rangle (\mu; \zpp)$ and Equations~\eqref{e:Erad_pp} and \eqref{e:Frad_z_pp}, we discretize $\mu$ uniformly into 64 bins. A similar calculation is done for the high-altitude region below the midplane ($z < -\zpp$).

We note that a small fraction of rays with non-zero FUV luminosity are terminated before reaching $\zpp$ by conditions (1b) and (2a). Due to these ``lost'' photons, our approach somewhat underestimates the true plane-parallel solution by $\sim 10$--$15 \%$. We also note that the LyC radiation field is still followed with the ART, to the extent that the restrictions imposed by termination conditions (1b) and (3) allow.

\subsection{Simulation Physical and Numerical Parameters}

We analyze the two simulations presented in \citet{kim2023ncr}: {\tt R8-4pc} and {\tt LGR4-2pc}. The key simulation parameters for these models are listed in \autoref{tab:model} \citep[for full information see Table 1 in][]{kim2023ncr}. We analyze $\sim 200$ snapshots in the time range $250$--$450 \Myr$ ({\tt R8-4pc}) and $250$--$350 \Myr$ ({\tt LGR4-2pc}), during which star formation/feedback cycles and the ISM multiphase structure are generated self-consistently. These time ranges are used for all time-averaged quantities. The average gas surface densities over these time intervals are $11 \Msun \pc^{-2}$ and $38 \Msun \pc^{-2}$. These values roughly correspond to the area-averaged and molecular mass-weighted conditions representative of nearby star-forming galaxies studies by the PHANGS survey \citep{sun2022}, which also roughly correspond to the solar neighborhood ({\tt R8-4pc}) and an inner-galaxy environment ({\tt LGR4-2pc}). 
The grid spacing is uniform and is $\Delta x = 4\pc$ and $2\pc$ for {\tt R8-4pc} and {\tt LGR-2pc}, respectively.

Columns (6)--(10) in \autoref{tab:model} give box sizes and parameters for ray termination. The full region extends to $z = \pm 3072 \pc$ for the {\tt R8-4pc} simulation, and $z = \pm 1536 \pc$ for {\tt LGR4-2pc}. \citet{kim2023ncr} performed a convergence test of radiation fields with respect to $\dxymax$ and $\epsPP$ for models ({\tt R8-8pc} and {\tt LGR4-4pc}). From that test, the midplane mean radiation energy density in both LyC and FUV bands exhibit converging trend when $\dxymax \gtrsim L_{x}$ and $\epsPP \lesssim 10^{-8}$ \citep[Fig 14 in][]{kim2023ncr}.

\begin{deluxetable*}{lCCCCCCCCC}
\tablecaption{Simulation Parameters\label{tab:model}}
\tablehead{
\colhead{Model} &
\dcolhead{R_0} &
\dcolhead{\Sigma_{\rm gas,0}} &
\dcolhead{\Sigma_{\rm *}} &
\dcolhead{z_{\rm *}} &
\dcolhead{L_{x}, L_{y}} &
\dcolhead{L_z} &
\dcolhead{\dxymax} &
\dcolhead{\epsPP} &
\dcolhead{\zpp} \\
\colhead{} &
\dcolhead{({\rm kpc})} &
\dcolhead{(\Msun\pc^{-2})} &
\dcolhead{(\Msun\pc^{-2})} &
\dcolhead{({\rm pc})} &
\dcolhead{({\rm pc})} &
\dcolhead{({\rm pc})} &
\dcolhead{({\rm pc})} &
\dcolhead{} &
\dcolhead{({\rm pc})}
}
\colnumbers
\startdata
{\tt R8-4pc}    & 8 & 12 & 42 & 245 & 1024 & 6144 & 2048 & 10^{-8} & 300 \\
{\tt LGR4-2pc}  & 4 & 50 & 50 & 500 & 512  & 3072 & 1024 & 10^{-8} & 300
\enddata
\tablecomments{
Column (1): model name with spatial resolution.
Column (2): nominal galactocentric radius.
Column (3): initial gas surface density.
Column (4): stellar surface density.
Column (5): scale height of the stellar disk.
Column (6): horizontal box size.
Column (7): vertical box size.
Column (8): maximum horizontal distance from the source for ray termination.
Column (9): fractional photon packet luminosity relative to the source luminosity for ray termination (FUV only).
Column (10): height at which FUV photons are collected and above which the plane-parallel radiative transfer solution is applied.
}
\end{deluxetable*}

\section{Simulation Overview}\label{sec:sim}

In this section, we provide an overview of the properties of the gas and radiation field. We include snapshots illustrating the morphology and physical relationship of the gas components and UV radiation fields; horizontally-averaged vertical profiles of both gas and radiation; and temporal histories and statistics of key physical parameters. The results we present here highlight the similarities and differences when compared to the radiation field and ionization state of gas from the post-processing ray-tracing results of the TIGRESS-classic solar neighborhood model \citep{kado2020diffuse}.  

\subsection{Morphology of Gas Components and UV Radiation Fields}\label{sec:morphology}

\begin{figure*}
    \centering
	\includegraphics[scale = 0.45]{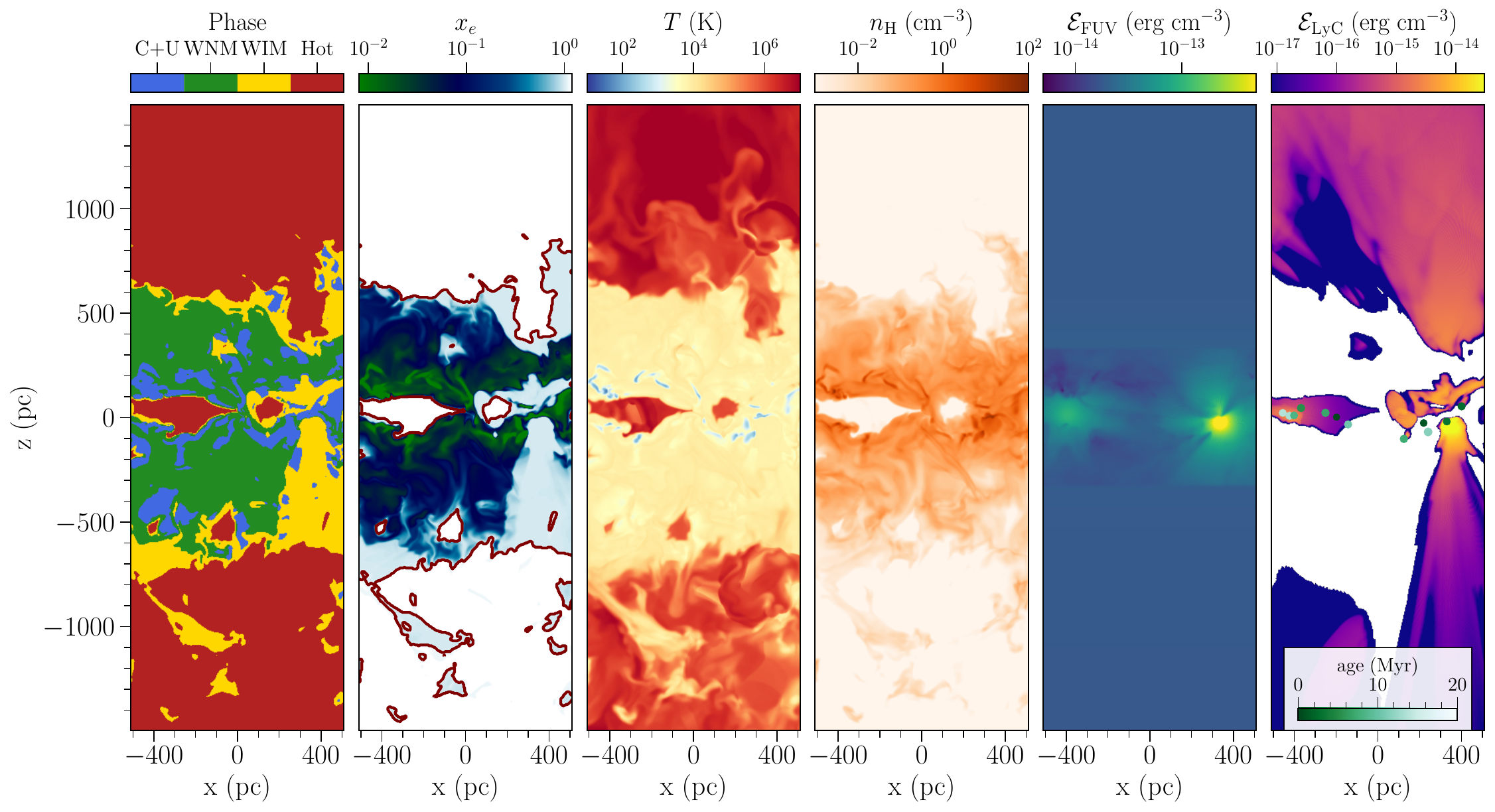}
    \caption{
    Vertical slices of the {\tt R8-4pc} simulation at $t = 430 \Myr$ through $y = 200$ pc. From left to right, the panels show ISM phase (see \autoref{tab:phase} for phase definitions), electron abundance ($x_e$), temperature ($T$), number density of hydrogen ($\nH$), FUV energy density ($\mathcal{E}_{\rm FUV}$), and LyC energy density ($\mathcal{E}_{\rm LyC}$). The points in the rightmost panel represent radiation source particles projected onto the $y=200$ pc plane. The age of each source is given by its color, and the points are limited to sources with ages less than 20 Myr. In the panel showing $x_e$, the red contour marks the boundary between the \hot\ and \WIM\ phases.}
    \label{fig:snap_R8}
\end{figure*}

\begin{figure*}
    \centering
	\includegraphics[scale = 0.45]{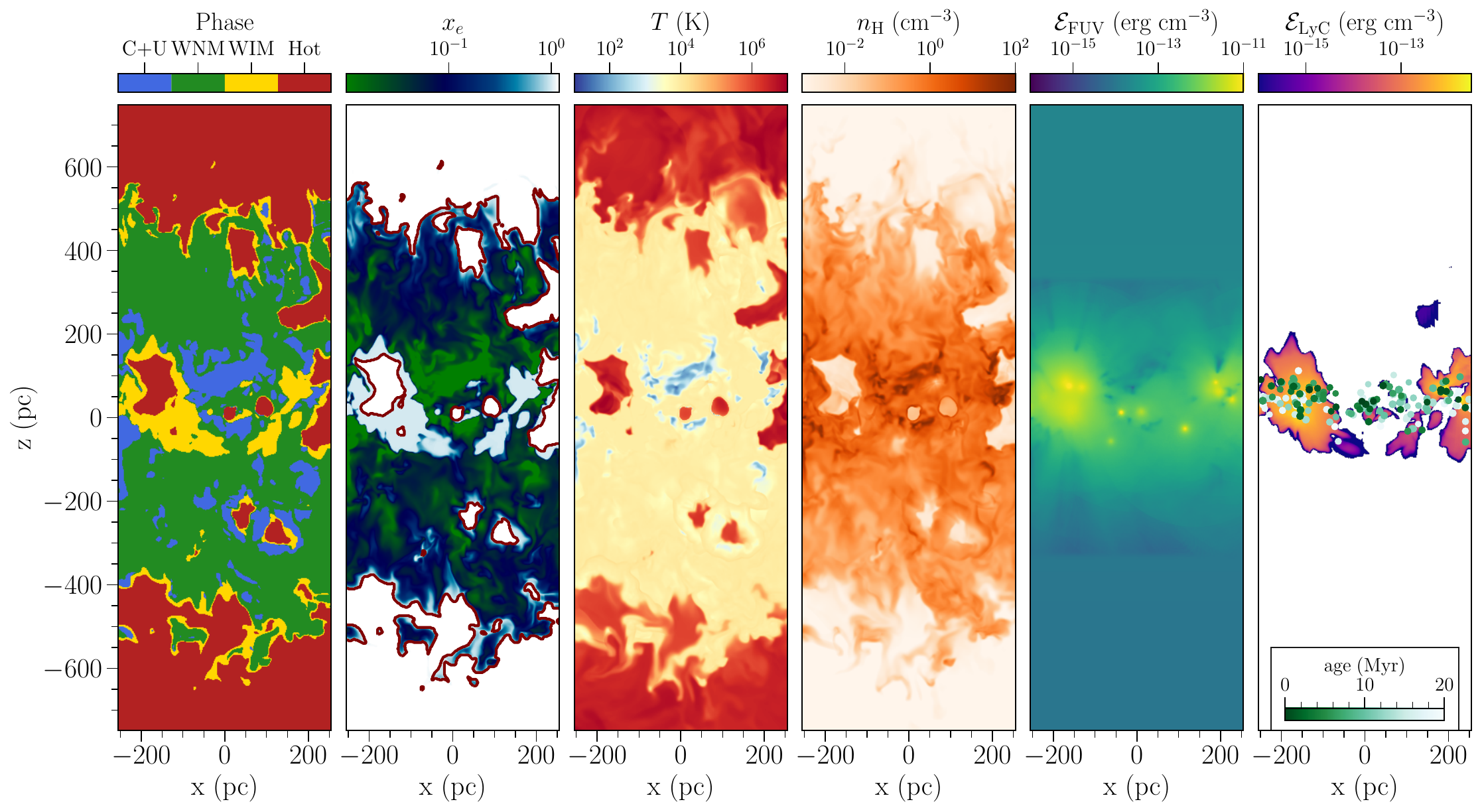}
    \caption{Same as \autoref{fig:snap_R8} but for the {\tt LGR4-2pc} model at $t = 298 \Myr$ through $y=0$.}
    \label{fig:snap_LGR4}
\end{figure*}

\begin{figure*}
    \centering
        \includegraphics[width=\linewidth]{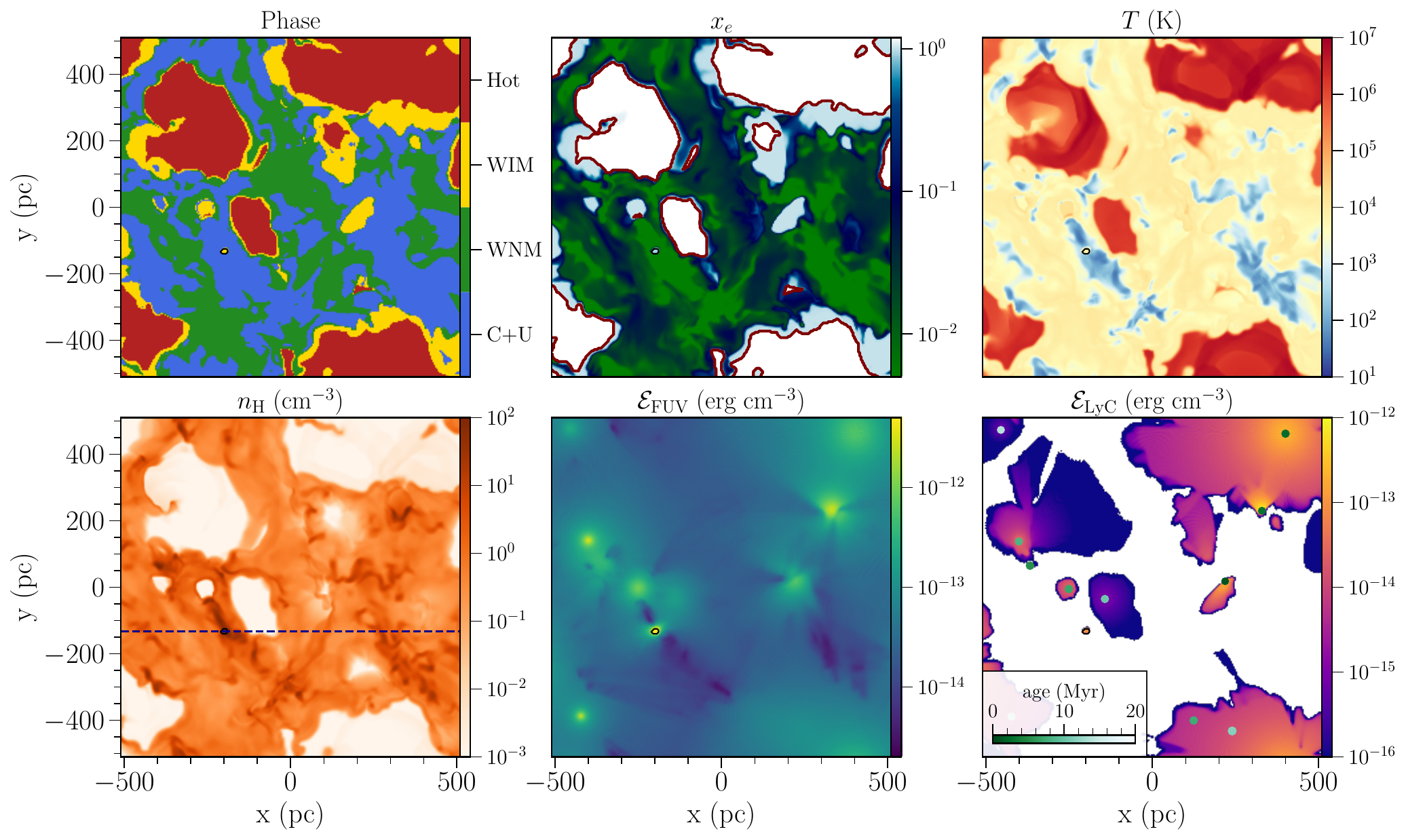}
    \caption{Horizontal slices of the {\tt R8-4pc} model at $z = 0 \pc$ and $t = 430 \Myr$, with the same quantities shown in \autoref{fig:snap_R8}. Clockwise from top left the panels represent ISM phase, $x_e$, $T$, $\mathcal{E}_{\rm LyC}$, $\mathcal{E}_{\rm FUV}$, and $\nH$. Source particles are shown in the bottom right panel. The red contour in the $x_e$ panel represents the boundary between \hot{} and \WIM{}. Additionally, the black contour in each panel represents a classical \ion{H}{2} region, which contains a star cluster source of mass $1.3\times 10^3\; M_\odot$ (this cluster particle is omitted from the  $\mathcal{E}_{\rm LyC}$ panel for clarity). The dashed horizontal line shown in the bottom left panel represents the 1D, horizontal ray used in \autoref{fig:slice_1D}.}
    \label{fig:snap_h}
\end{figure*}

\begin{figure*}
    \centering
        \includegraphics[width=\linewidth]{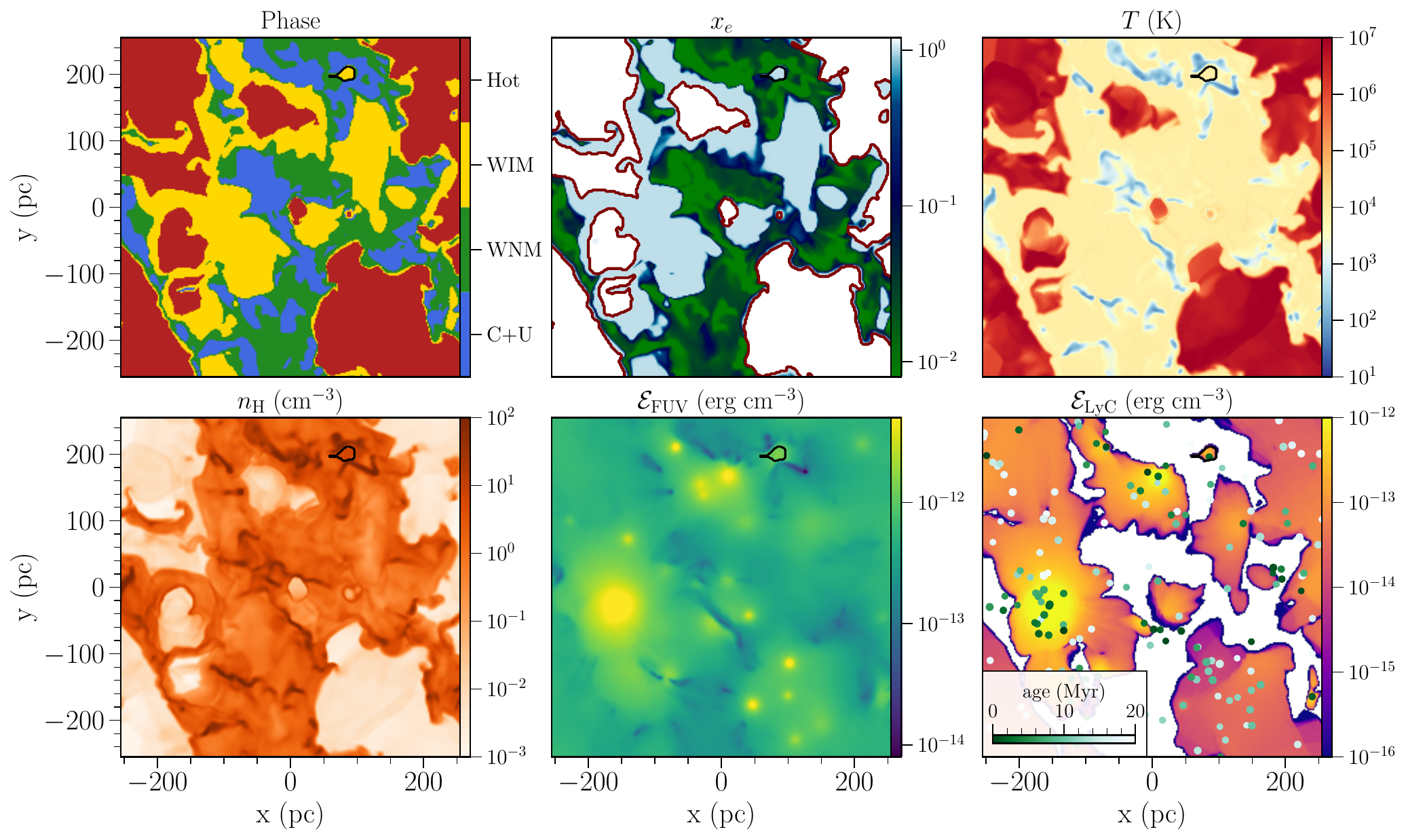}
    \caption{Same as \autoref{fig:snap_h} but for the {\tt LGR4-2pc} model at $t = 298 \Myr$. The black contour in each panel represents a classical \ion{H}{2} region, which contains a star cluster source of mass $1.2\times 10^3\; M_\odot$.}
    \label{fig:snap_h_LGR4}
\end{figure*}

We present individual snapshots of the {\tt R8-4pc} and {\tt LGR4-2pc} TIGRESS-NCR simulations in \autoref{fig:snap_R8}--\autoref{fig:snap_h_LGR4} at times of $430$ and $298 \Myr$, respectively. \autoref{fig:snap_R8} and \autoref{fig:snap_LGR4} show vertical slices at either $y = 200$ or $0 \pc$ limited to a region within 1500 and 750 pc of the midplane ($z = 0$) respectively. \autoref{fig:snap_h} and \autoref{fig:snap_h_LGR4} present horizontal snapshots in the x-y plane at $z = 0$. 

From left to right (and from top to bottom for \autoref{fig:snap_h} and \autoref{fig:snap_h_LGR4}), the panels present the ISM phases, fractional abundance of free electrons ($x_e$)\footnote{We define the fractional abundance of species s relative to hydrogen nuclei as $x_{\rm s} = n_{\rm s}/n_{\rm H}$ where $n_{\rm s}$ denotes the number density of species s.}, gas temperature ($T$), number density of hydrogen ($\nH$), FUV energy density ($\mathcal{E}_{\rm FUV}$), and LyC energy density ($\mathcal{E}_{\rm LyC}$). Additionally, locations of radiation source particles (clusters) are shown in the LyC panel as circles.

We focus on the vertical slices first then take a closer look at the midplane structure using the horizontal slices.

\paragraph{Gas Phases}

The ISM can be divided into distinct gas phases as determined by the temperature and ionization state of hydrogen. For this analysis, we define four phases as given in \autoref{tab:phase}. These include the cold medium and unstable medium (we group this as \CpU\ here), warm neutral medium (\WNM{}), warm ionized medium (\WIM{}), and hot ionized gas (\hot{}). Each cell in the simulation can be divided into one of these four phases, as shown in the first panel of each snapshot. We note that this is a simplified categorization compared to our analysis of the same simulations as presented in \citet{kim2023ncr}.

\begin{deluxetable*}{lCCc}
\tablecaption{Phase Definition \label{tab:phase}}
\tablehead{
\colhead{Name} &
\dcolhead{{\rm Temperature}} &
\dcolhead{{\rm Abundance}} &
\colhead{{\rm Shorthand}}
}
\startdata
Cold and Unstable Medium & T<6\cdot10^3\Kel & \cdots &  \CpU\\
Warm Neutral Medium & 6\cdot10^3\Kel<T<3.5\cdot10^4\Kel & \xHI>0.5 & \WNM{} \\
Warm Ionized Medium & 6\cdot10^3\Kel<T<3.5\cdot10^4\Kel & \xHII>0.5 & \WIM{} \\
Warm-Hot and Hot Ionized Medium & 3.5\cdot10^4\Kel<T & \cdots & \hot{} \\
\enddata
\end{deluxetable*}

The {\tt R8-4pc} and {\tt LGR4-2pc} models present similar phase structure. The coldest gas (\CpU) is in high density structures found near the midplane, with the majority of the volume within 500~pc of the midplane filled by \WNM. Within the \WNM\ are pockets of both \hot\ and \WIM\ gas. The \WNM\ and hot gas are separated by a layer of \WIM, of varying thickness.
Most of the extraplanar region is filled with hot gas.

\paragraph{Electron Abundance}

The second panel of each snapshot shows the electron abundance ($x_e = n_e/\nH$) of the gas. A red contour representing the separation between the \hot{} and \WIM{} phases is also included in this panel. In the \WIM{} phase, $x_e\sim1$  due to electrons from nearly completely ionized hydrogen, while $x_e\sim1.2$ in the \hot{} phase due to additional contributions from collisionally ionized helium and metals.\footnote{Our model assumes collisional ionization equilibrium for helium and metals as a function of temperature in the hot gas. Most helium becomes doubly ionized above $T \sim 10^5 \Kel$ by collisional ionization alone.} The \WNM\ and \CpU\ phases are mostly neutral, where the majority of the gas has $x_e\sim0.1$ and $<0.01$, respectively. 

Comparing the first and second panels in \autoref{fig:snap_R8} and \autoref{fig:snap_LGR4}, we see that there is a sharp drop in ionization between the \WIM\ and \WNM\ near the midplane, but for higher-altitude gas the ionization transition can be more gradual. The reason for this difference is that the mean free path for LyC photons to photoionize neutral gas becomes very short, $\ell = (\nH \sigma_{\rm pi})^{-1} = 0.1\pc ~(\nH/1\pcc)^{-1}$, at densities comparable to those found in midplane regions, but is longer for lower density, extraplanar gas. Additionally, the recombination timescale $t_\mathrm{rec}=(\alpha_{\rm B} n_e)^{-1} \sim 1\Myr (n_e/0.1\pcc)^{-1}$ becomes longer at the lower densities that are present at high altitude (see the fourth panel).  When this timescale is longer than the dynamical time, gas that is intermittently exposed to radiation (as is often the case -- see below) may remain in a partially-ionized state (bright blue).  The time-dependent treatment 
of ionization/recombination and ray-tracing allows us to recover this partially-ionized, low-density gas.  We note that this is a new capability of the current simulations compared to our previous post-processing simulations \citep{kado2020diffuse}, where the ionization state in each location was evolved until equilibrium was attained. With fully time-dependent simulations, more of the gas remains partially ionized (J.-G. Kim et al., in prep.).

\paragraph{Radiation Fields}

In the final two panels of \autoref{fig:snap_R8} and \autoref{fig:snap_LGR4} (see also \autoref{fig:snap_h} and \autoref{fig:snap_h_LGR4}), we show the energy density of FUV and LyC radiation. Individual young cluster particles colored by age are overlaid on the LyC panel. These clusters act as the sources of UV photons. In regions with sources surrounded by hot gas, radiation is essentially unattenuated until it begins to interact with denser gas and dust at larger distance. FUV and LyC photons are absorbed by dust in both neutral and ionized gas; neutral (or partially ionized) gas absorbs LyC only. In addition to a sharp transition between \WIM\ and neutral phases, the photoionization front is also visible as a sharp drop in the $\mathcal{E}_{\rm LyC}$ panel. LyC is strongly attenuated in neutral gas because of the high photoionization crossection. Therefore, a significant volume is not instantaneously exposed to any ionizing radiation. Nonetheless, as noted above, the recombination timescale in low-density gas can be long enough for intermittently-irradiated gas to remain partially ionized. Thus, there are locations where \WIM\ gas is present without LyC radiation.

Since the dust absorption crossection is three orders of magnitude lower than the photoionization crossection, FUV radiation is not attenuated as strongly as the LyC radiation, and therefore fills the whole simulation volume. For TIGRESS-NCR simulations, however, the FUV radiation is followed by ray tracing only within 300 pc of the midplane. Above this height, the plane parallel approximation is used to reduce the computational cost of the simulations (see \autoref{sec:planepar}). This can be seen in the vertical slices as a transition from a radiation field that is fully resolved spatially to a horizontally averaged value. 

\paragraph{Midplane Structure}

In the horizontal slices shown in \autoref{fig:snap_h} and \autoref{fig:snap_h_LGR4}, there are classical \ion{H}{2} regions at approximately $x = -200\pc$, $y = -130\pc$ for the {\tt R8-4pc} simulation, and $x = 85\pc$, $y = 200\pc$ for the {\tt LGR4-2pc} simulation. Both regions are outlined with black contours as defined by the separation between the \WIM{} and \CpU\ phases. Each \ion{H}{2} region is marked by a young source in the center which produces the high ionizing radiation density seen in $\mathcal{E}_{\rm LyC}$. These \ion{H}{2} regions are distinct from diffuse ionized gas because they appear as bubbles of \WIM{} embedded in colder  gas, unlike the majority of the \WIM{} in our simulations which is generally bounded by \WNM\ and by \hot\ gas.  \ion{H}{2} regions have similar density to the surrounding gas (this is also true for diffuse \WIM), as can be seen in the bottom left panel, but they have much higher $x_e$ and radiation density. 

The projection of cluster positions shows that radiation sources are primarily found within the hot and warm gas that fills most of the volume.  As noted above, LyC is able to propagate unimpeded through the hot gas.  In denser warm gas, provided the LyC radiation intensity is sufficient to maintain a highly ionized state, photon packets continue to propagate along rays while being attenuated subject to the rate at which photons are used up by balancing recombinations. If in a given cell the LyC intensity is too low to maintain a fully ionized state, rays will be strongly attenuated. 
In the midplane slices, we see sharp gradients in $x_e$ separating the \WIM{} and \WNM{}.  These sharp gradients in $x_e$ arise due to strong local attenuation of the LyC.  The \hot\ regions that contain LyC sources will always be surrounded by a diffuse \WIM\ layer separating the \hot\ gas from the \WNM.

The FUV radiation can propagate throughout most of the midplane region.  Nevertheless, there are darker regions evident in the FUV energy density slice.  Some of these coincide with regions of very high gas density. A few of these low-${\cal E}_\mathrm{FUV}$ regions correspond to shadows which are created when a dense clump intercepts radiation from a source.

\begin{figure*}
    \centering
	\includegraphics[scale = 0.45]{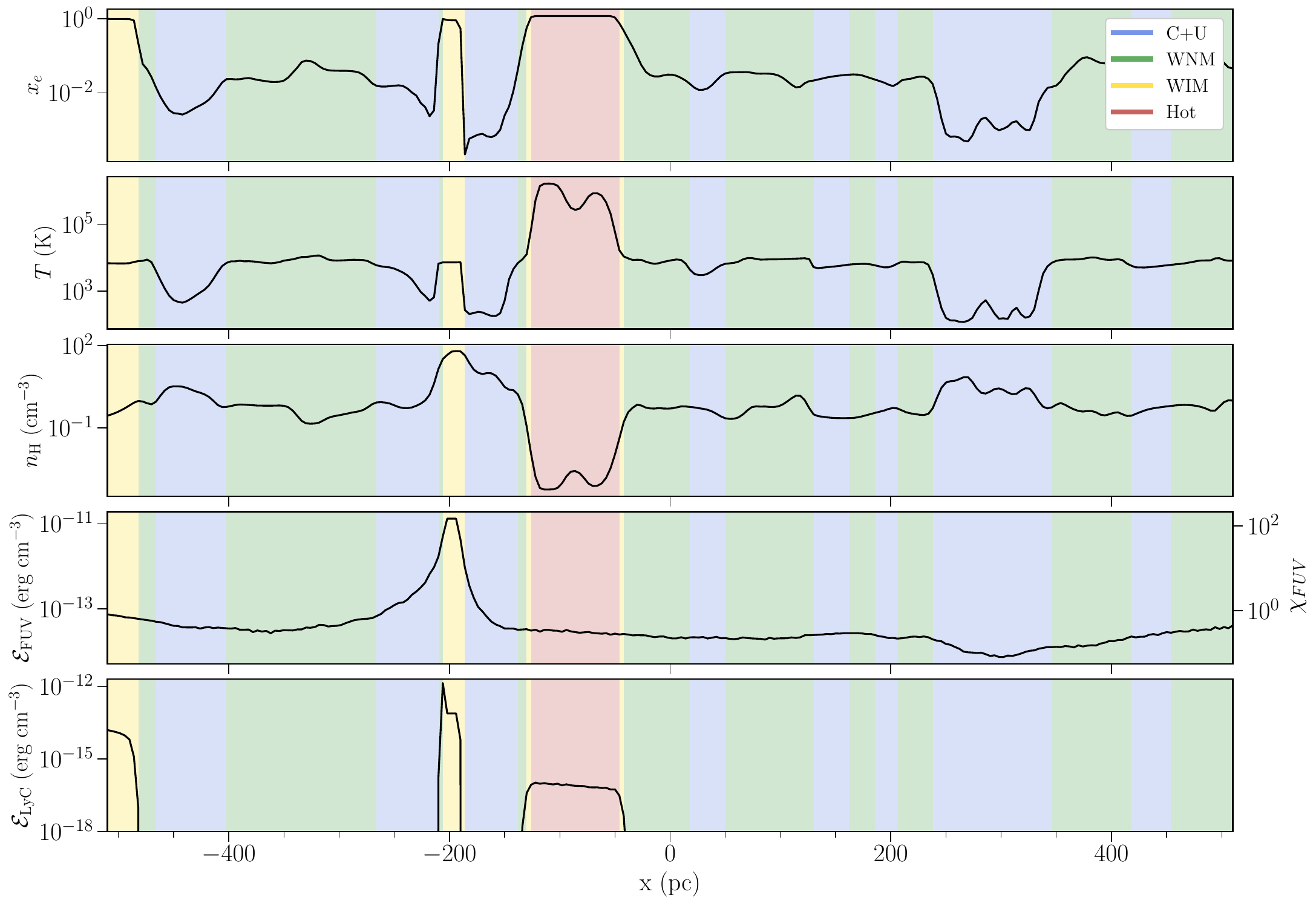}
    \caption{Physical quantities along a 1D, horizontal ray through $z = 0$, $y = -134$ pc of the {\tt R8-4pc} model at $t = 430 \Myr$. From top to bottom, the panels represent $x_e$, $T$, $\nH$, $\mathcal{E}_{\rm FUV}$, and $\mathcal{E}_{\rm LyC}$. The gas phase is represented by the background color. Sharp photoionization fronts are clearly seen in the $\mathcal{E}_{\rm LyC}$ and  $x_e$ and $T$ panels, which separate \WIM\ from \CNM\ gas. These represent the boundary of the \ion{H}{2} region visible in \autoref{fig:snap_h} at $x = -200 \pc$.}  
    \label{fig:slice_1D}
\end{figure*}

In addition to the midplane slices, we also present one-dimensional cuts of the {\tt R8-4pc} simulation variables in \autoref{fig:slice_1D}. We plot the simulation variables in the snapshot shown in \autoref{fig:snap_h} at $z = 0\pc$, $y = -134\pc$ (horizontal dashed line in the bottom-left panel). This cut was chosen such that it passes through the \ion{H}{2} region described previously. We include the values of $x_e$, $T$, $\nH$, $\mathcal{E}_{\rm FUV}$, and $\mathcal{E}_{\rm LyC}$. Along with $\mathcal{E}_{\rm FUV}$, we also show the value of $\chi_{\rm FUV}$, the FUV radiation density divided by the reference value defined in \citet{draine1978}, $8.94\times 10^{-14}$ erg cm$^{-3}$. Background shading in each panel indicates the defined gas phases.

The values of $\mathcal{E}_{\rm LyC}$ and $x_e$ are highly correlated, as would be expected. The attenuation of LyC in regions of high density can be seen by the absence of $\mathcal{E}_{\rm LyC}$ corresponding to most regions with high $\nH$. One exception to this behavior is at $x \sim -200 \pc$ where the slice passes through the classical \ion{H}{2} region. Both the LyC and FUV radiation are much higher in this region due to the presence of a source particle. The ionization fraction is similarly increased. The density, however, is similar to the surrounding gas ($\sim 10^2$ cm$^{-3}$) and the temperature is $\sim 10^4$ K, as would be expected in a typical \ion{H}{2} region. 

\subsection{Vertical Profiles}

\begin{figure*}
    \centering
	\includegraphics[scale = 0.39]{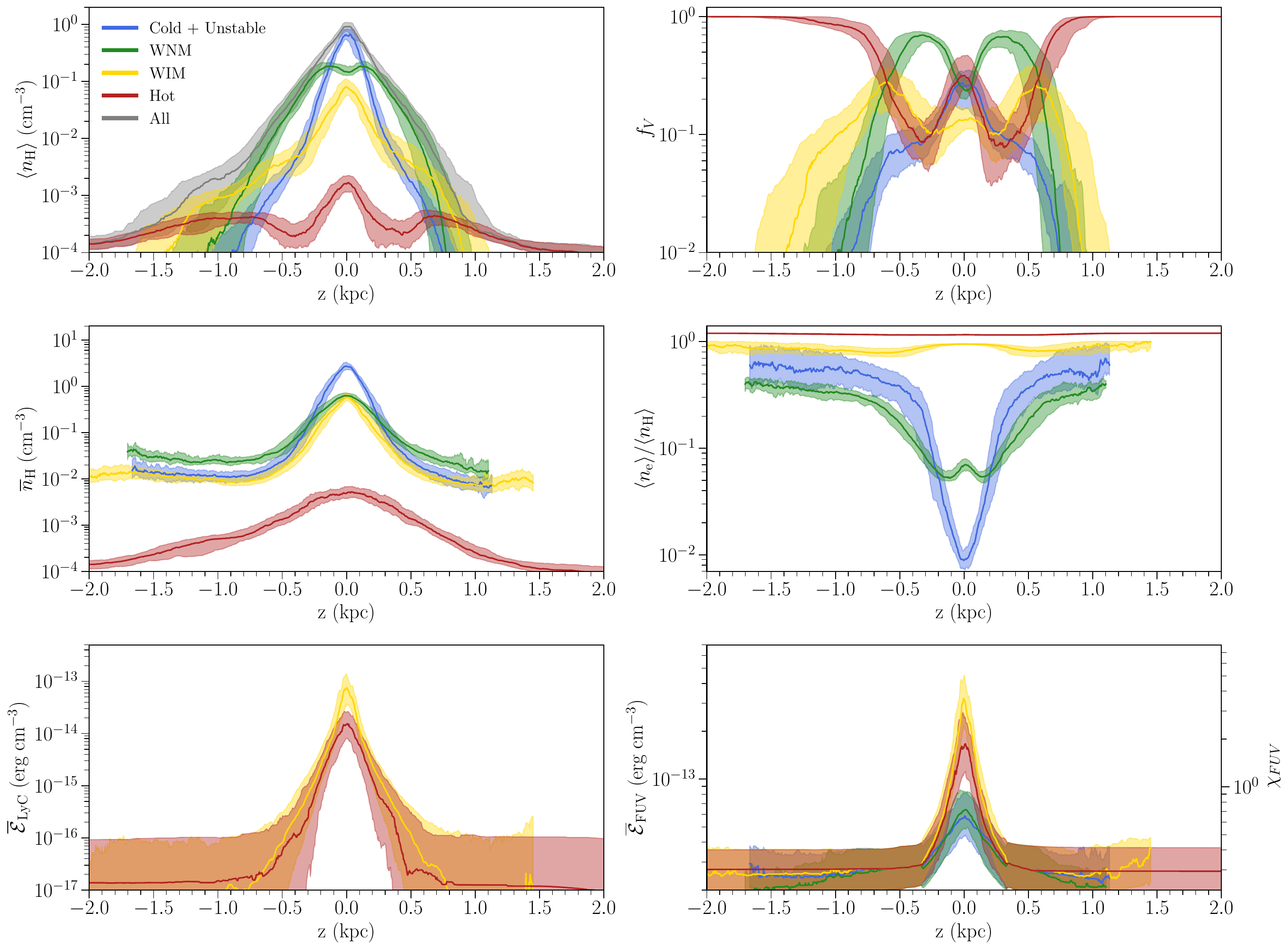}
    \caption{Vertical profiles of contribution to the mean density $\langle \nH \rangle$ and volume filling factor $f_V$ for each of four ISM phases, as well as density $\bar{n}_{\rm H}$, mass weighted electron abundance $x_e = \langle n_e \rangle / \langle \nH \rangle$, and LyC and FUV radiation density $\bar{\cal E}_\mathrm{LyC}$ and $\bar{\cal E}_\mathrm{FUV}$ within defined gas phases. Profiles are based on $\sim 200$ simulation snapshots between $250$--$450 \Myr$ for the {\tt R8-4pc} model. The lines represent the median value and the shaded regions indicate 25th--75th percentiles. See text for details.}
    \label{fig:z_prof}
\end{figure*}

\begin{figure*}
    \centering
	\includegraphics[scale = 0.39]{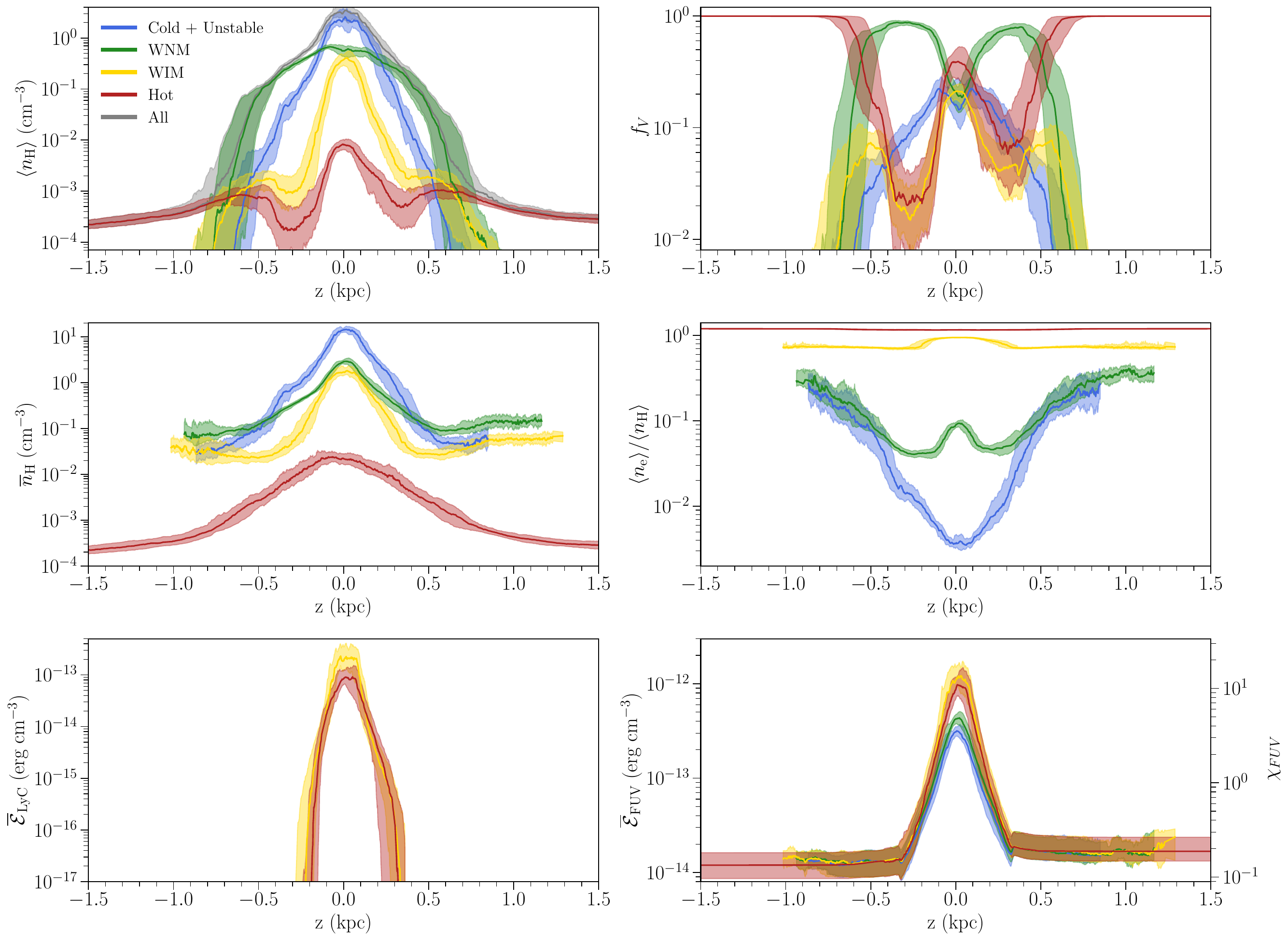}
    \caption{Same as \autoref{fig:z_prof} but for the {\tt LGR4-2pc} model between times 250--350 Myr.}
    \label{fig:z_prof_LGR4}
\end{figure*}

In this section, we consider the distribution of simulation variables over time as a function of the vertical ($z$)  coordinate. For each snapshot at every height $z$ we take the horizontal ($x$--$y$) average of various quantities as in \citet{kim2023ncr} defined by
\begin{equation}
    \langle q \rangle_{\textrm{ph}} = \frac{\iint q \Theta(\textrm{ph}) dxdy}{L_x L_y}
    \label{eq:q}
\end{equation}
where $\Theta(\textrm{ph}) = 1$ if the gas is in the given phase (as defined by \autoref{tab:phase}) and 0 otherwise. The volume filling factor ($f_V$) is defined similarly as $f_{\rm V,ph} = \iint \Theta(\textrm{ph}) dxdy/L_x L_y$. We define the characteristic values of $\nH$, $\mathcal{E}_{\rm LyC}$, and $\mathcal{E}_{\rm FUV}$ as the average within a phase, $\bar{q}_{\rm ph} = \langle q \rangle_{\rm ph} / f_{\rm V,ph}$. We also include the mass weighted value of $x_e = \langle n_e \rangle / \langle n_H \rangle$. These vertical profiles are shown for the {\tt R8-4pc} and {\tt LGR4-2pc} simulations in  \autoref{fig:z_prof} and \autoref{fig:z_prof_LGR4},  respectively. In each panel, the central, dark line represents the median value while the shaded area shows the 25--75th percentile region across all snapshot times. For each phase, results for a given $z$ coordinate are only included if at least 25\% of the snapshot times have gas of that phase at that altitude. Therefore, there are regions at high $z$ where, for example, no \CpU\ gas is shown.

The top left panel represents $\langle\nH\rangle$ for each gas phase. In all phases, the highest density is found close to the midplane, then generally decreasing with increasing $z$. The \CpU\ gas shows a significant drop in $\langle \nH \rangle$ moving away from the midplane because the averaging is done over the full horizontal area, and this gas has a low volume filling fraction $f_V$ at large altitudes (see top right panel).

Above $\sim$1 kpc, the hot gas completely dominates the volume, with the other three phases having $f_V<10\%$. At even higher $z$, the filling fraction of the colder phases is further reduced, with each occupying less than $\sim$1\% of the volume. Near the midplane, all four phases have more similar values of $f_V$, between approximately 10--50\% for each of the phases defined in \autoref{tab:phase}.  This can also be seen in the horizontal snapshot (\autoref{fig:snap_h}).  At $z = 0$, the hot gas filling fraction increases while the warm gas occupies less volume because hot gas is generated there via shock heating \citep{kado2020diffuse}.
For values of the mass fraction and the volume filling factors of each phase within $|z|<300\pc$, see Table 4 in \cite{kim2023ncr}. 

The remaining four panels of  \autoref{fig:z_prof} and \autoref{fig:z_prof_LGR4} show characteristic values of simulation variables within each phase. Like the volume-averaged density, the characteristic density ($\bar{n}_\mathrm{H}$) within each phase peaks at the midplane. The value of $\bar{n}_\mathrm{H}$ in the hot gas drops exponentially over all $z$. The characteristic densities within the cold and warm gas decline at larger $z$, until reaching the extraplanar fountain region at $|z|\gtrsim 500\pc$ where densities plateau at $\bar{n}\sim$0.01 - 0.1 cm$^{-3}$.  In \autoref{sec:ionmass}, we explain why the majority of the ionized gas mass in galaxies is expected to be in the diffuse component outside of classical \ion{H}{2} regions.    

The middle-right panel represents the mass weighted electron abundance $x_e = \langle n_e \rangle / \langle \nH \rangle$ within each phase. The hot gas is completely ionized with $x_{\rm e} \approx 1.2$ (also seen in  \autoref{fig:snap_h}). The \WIM{} is almost fully ionized with $x_{\rm e} \approx 1$ at all $z$ values, despite being defined only as gas with $\xHII>0.5$. The \CpU\ gas, as well as the \WNM, is partially ionized at all heights, but the ionization fraction is low ($<0.1$) at the midplane.

The LyC and FUV radiation energy density profiles are presented in the bottom row. These show similar behavior, both strongly peaked at the midplane, but the dynamic range in LyC is much larger than FUV. For the LyC radiation, we include only the hot gas and \WIM{} as the other (neutral) phases have much lower radiation density in this wavelength regime due to the high attenuation in regions with higher density. All four phases are shown for the FUV radiation, and each has centrally peaked profiles with lower radiation energy density in the neutral phases compared to the ionized. The FUV profiles flatten at $|z| \gtrsim 300\pc$ as gas density is sufficiently low and FUV photons become optically thin. For the FUV profiles, we include a scale normalized to the Draine value ($\chi_{\rm FUV}={\cal E}_{\rm FUV}/8.94\times 10^{-14}$ erg cm$^{-3}$) on the right axis.

We fit the median radiation energy density in both the LyC and FUV bands with an exponential distribution given by 
\begin{equation}
    {\cal E}(z) = {\cal E}_0 e^{-\lvert z \rvert / b}.
\end{equation}
For the FUV radiation, we fit the energy density profile in the \WNM\, since this vertical dependence is most relevant for heating through the photoelectric effect; the profile in the \CpU\ is quite similar.  For the LyC radiation, we look at the energy density profile in the \WIM\ as this phase is most relevant for photoionization. In both cases, the fit is limited to $|z| < 300$ pc as above this height we are not performing the full ray-tracing for the FUV radiation and the disk enters an optically thin regime.

The fit to the LyC radiation in the {\tt R8-4pc} simulation gives an exponential scale height of $b = 60$ pc with  normalization ${\cal E}_0 = 6.8\times10^{-14} \;{\rm erg\;cm^{-3}}$. The FUV radiation has $b = 350$ pc with normalization ${\cal E}_0 = 6.6\times10^{-14} \;{\rm erg\;cm^{-3}}$, corresponding to $\chi_{\rm FUV}=0.74$. That is, the midplane FUV radiation energy density is very close to the Draine value (see \autoref{fig:pp} for additional comparisons). We note that the peak value of $\chi_{\rm FUV}$ is $\sim 5$ times higher in the \WIM\ than in the \WNM, while decreasing to match the value in the \WNM\ at larger $z$.  This is not surprising, since the \WIM\ near the midplane includes \ion{H}{2} regions immediately surrounding radiation sources, where the FUV is preferentially enhanced.  

The FUV radiation in the {\tt LGR4-2pc} simulation is best fit by an exponential scale height $b = 90$ pc with normalization ${\cal E}_0  = 5.2\times10^{-13} \;{\rm erg\;cm^{-3}}$, corresponding to $\chi_{\rm FUV}=5.8$.
We find the LyC component of the {\tt LGR4-2pc} simulation to drop rapidly away from $z = 0 \pc$. Therefore, we fit only $|z| < 200$ pc and find the best fit to have an exponential scale height $b = 30$ pc with normalization ${\cal E}_0  = 5.7\times10^{-13} \;{\rm erg\;cm^{-3}}$.

\subsection{History}

\begin{figure*}
    \centering
	\includegraphics[scale = 0.4]{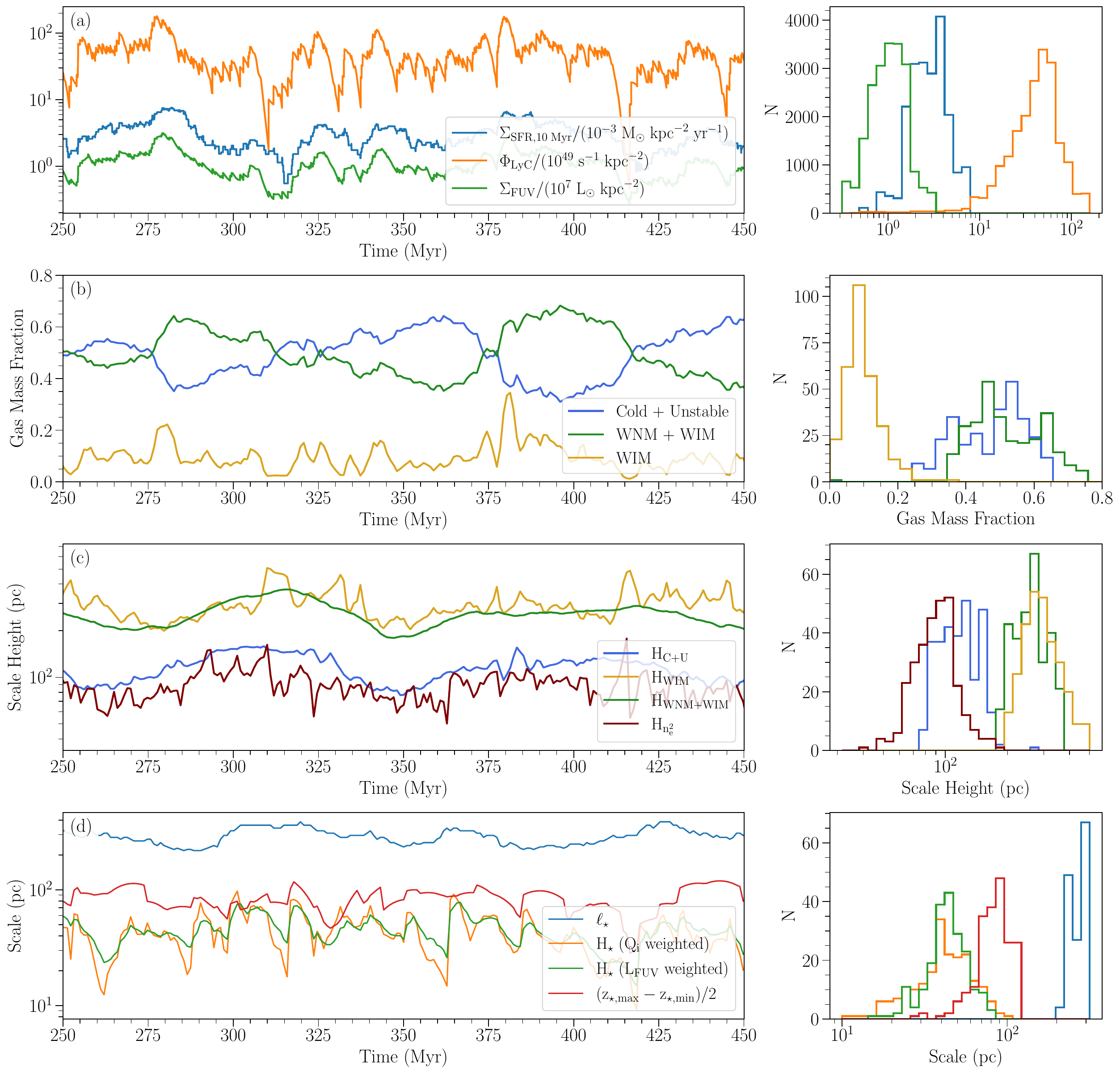}
    \caption{Temporal evolution and statistics of key global simulation quantities, for the {\tt R8-4pc} model. From top to bottom, the left column shows (a) the LyC (ionizing) photon rate, star formation rate, and FUV energy input rate per unit area averaged over the whole simulation domain, (b) total mass fractions of gas in different thermal phases, (c) scale heights of different phases of gas, as well as $n_e^2$ ($\propto$ emission measure) (d) mean in-plane separations $\ell_{\star}$ of cluster particle sources (with $t_{\rm age} < 20\Myr$)
    as well as their scale heights and maximum and minimum $z$ locations. The right panels show statistical distributions of these same quantities as histograms including all times.}
    \label{fig:history}
\end{figure*}

\begin{figure*}
    \centering
	\includegraphics[scale = 0.4]{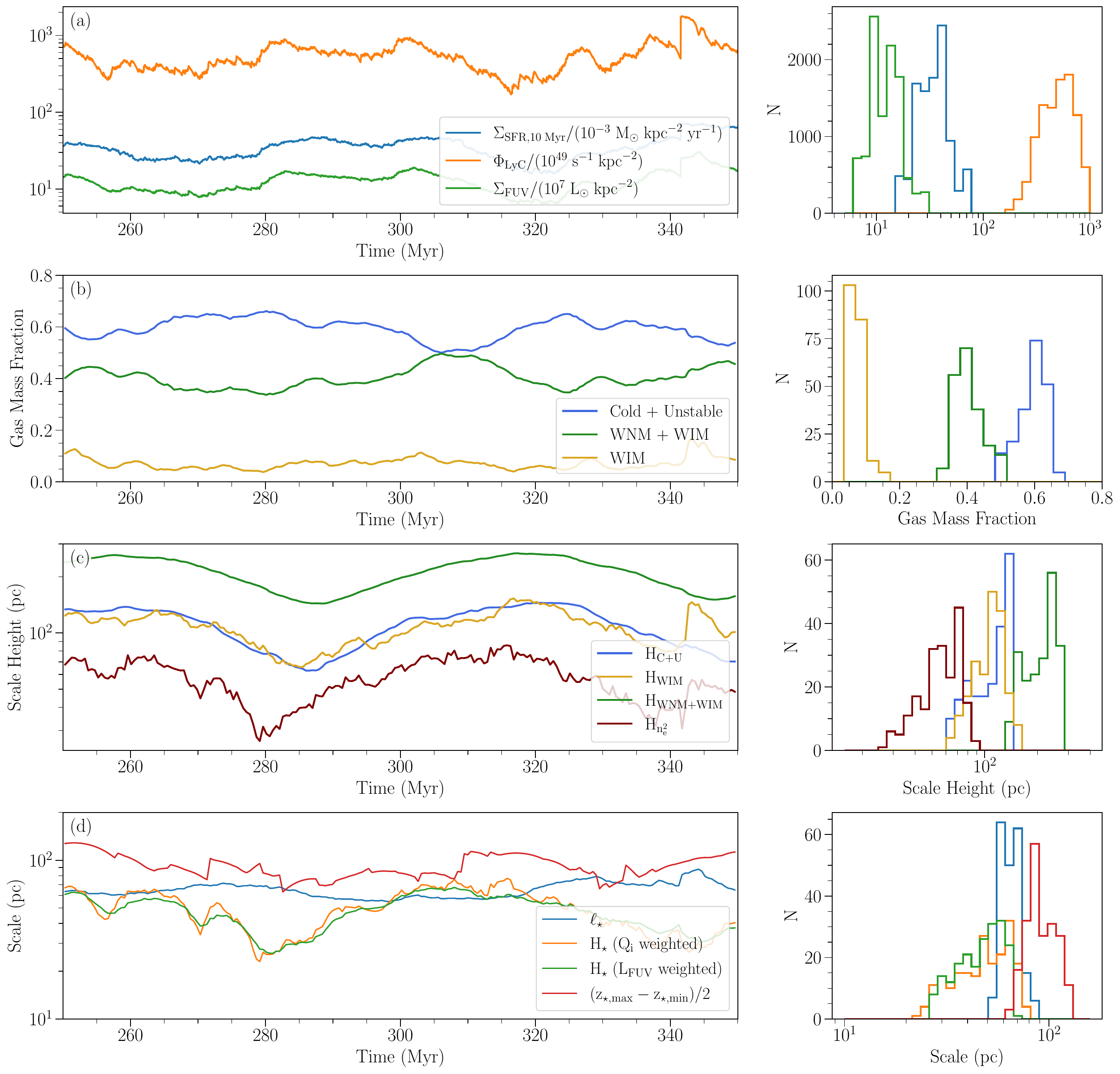}
    \caption{Same as \autoref{fig:history}, but for the {\tt LGR4-2pc} model.}
    \label{fig:history_L}
\end{figure*}

\begin{deluxetable*}{lCCl}
\tablecaption{Physical Quantities Measured in Simulations\label{tab:quantities}}
\tablehead{
\colhead{Value} &
\dcolhead{${\tt R8-4pc}$} &
\dcolhead{${\tt LGR4-2pc}$} &
\colhead{Description}
}
\startdata
$\Sigma_{\rm gas}$   & 10.6_{-0.2}^{+0.2} & 37.9_{-0.9}^{+1.3}   & Gas surface density in $M_{\odot}\,{\rm pc}^{-2}$\\
$\Sigma_{\rm SFR} $  & 2.8_{-1.0}^{+1.5}  & 34.8_{-10.7}^{+10.4} & 10 Myr averaged SFR surface density in $10^{-3} M_{\odot}\,{\rm kpc}^{-2}\,{\rm yr}^{-1}$\\
$\Sigma_{\rm FUV} $  & 1.0_{-0.4}^{+0.5}  & 12.3_{-3.4}^{+4.5}   & FUV luminosity per unit area in $10^7\,L_{\odot}\,{\rm kpc}^{-2}$\\
$\Phi_{\rm LyC}$     & 4.3_{-2.2}^{+2.6}  & 53_{-20}^{+23}       & Ionizing photon rate per unit area in $10^{50}\,{\rm photons}\,{\rm s}^{-1}\,{\rm kpc}^{-2}$\\
$H_{w}$              & 260_{-50}^{+40}    & 220_{-60}^{+40}      & Scale height in pc for total warm (\WNM+\WIM) gas\\
$H_{c+u}$            & 110_{-30}^{+30}    & 120_{-40}^{+20}      & Scale height in pc for \CpU\ phase\\
$l_*$                & 280_{-40}^{+60}    & 65_{-7}^{+8}         & Mean horizontal distance between FUV sources ($t_{\rm age} < 20\Myr$) in pc \\
$H_*$                & 40_{-10}^{+10}     & 50_{-10}^{+10}       & Scale height of FUV sources in pc weighted by FUV luminosity \\
$z_*$                & 80_{-20}^{+20}     & 90_{-11}^{+20}       & Approximate vertical extent of the particles, defined as $(z_\mathrm{max} - z_\mathrm{min})/2$\\
$J_{\rm FUV, mid}$   & & & Midplane FUV intensity in $10^{-4} \;\rm{erg}\;\rm{cm}^{-2}\;\rm{sr}^{-1}\;\rm{s}^{-1}$ limited to gas with $T < 3.5 \times 10^4 \Kel$ \\
\quad mass weighted        & 1.1_{-0.7}^{+3.0} & 6.5_{-4.5}^{+12.0} \\
\quad volume weighted      & 1.2_{-0.7}^{+2.2} & 9.3_{-5.2}^{+11.3} \\
$\cal{J}_{\rm mid}$  & & & Midplane normalized FUV intensity $\mathcal{J}$ (\autoref{eq:calJdef}) for gas with $T < 3.5 \times 10^4 \Kel$\\
\quad mass weighted        & 0.3_{-0.2}^{+0.8} & 0.2_{-0.1}^{+0.3} \\
\quad volume weighted      & 0.4_{-0.2}^{+0.6} & 0.2_{-0.1}^{+0.3} \\
$\mathcal{J}_{|z| < \rm 300 \; pc}$
& 0.5_{-0.1}^{+0.1} & 0.15_{-0.03}^{+0.04} & Volume-weighted average of $\mathcal{J}$ within $300 \pc$ of the midplane including all gas \\
\enddata
\tablecomments{
The reported values are median and 16th and 84th percentiles over $t = 250$--$450\Myr$ for {\tt R8-4pc} and $t= 250$--$350\Myr$ for {\tt LGR4-2pc}.}
\end{deluxetable*}

We present the time evolution and statistics of different physical parameters from the {\tt R8-4pc} and {\tt LGR4-2pc} simulations in  \autoref{fig:history} and \autoref{fig:history_L}, respectively. For each quantity, we show temporal histories in the left panel. On the right, we show the same variables but as a histogram representing the distribution of each value over time. Since all quantities represent horizontal averages, and at any given time our box does not fully sample from the conditions appearing over the ISM/star formation lifecycle, the level of temporal variation is subject to the horizontal size of the simulation domain; a larger (smaller) horizontal domain would naturally have less (more) variation. \autoref{tab:quantities} summarizes the median values and 16th--84th percentiles of these quantities over time.

The first row represents the surface densities of ionizing photon rate $\Phi_{\rm LyC} = Q_{\rm i,tot}/(L_x L_y)$, the star formation rate (SFR) $\Sigma_{\rm SFR, 10 Myr}$, and the FUV luminosity $\Sigma_{\rm FUV} = L_{\rm FUV,tot}/(L_x L_y)$. Here, $Q_{\rm i,tot}$ and $L_{\rm FUV,tot}$ are the total ionizing photon rate and FUV luminosity summed over source particles, respectively. The SFR is calculated using only star particles with ages less than $10 \Myr$. These three values appear to be well correlated in time as would be expected because young stars are the major source of the FUV and LyC radiation. The LyC radiation has greater variation in amplitude because it is coming from even younger stars ($t < 5 \Myr$).  

The median value of the SFR in {\tt R8-4pc} is $\Sigma_{\rm SFR, 10 Myr}=0.0028 \Msun\pc^{-2}\Myr^{-1}$ (\autoref{fig:history}, top right panel), which is very similar to observed estimates for the solar neighborhood in the Milky Way \citep[e.g.][]{fuchs2009, mor2019, zari2023}. The median value of the FUV luminosity per unit area is $\Sigma_{\rm FUV}=1.0 \times 10^7 \; L_\odot \textrm{ kpc}^{-2}$; this is consistent with expectations given the above median $\Sigma_{\rm SFR, 10 Myr}$ and the expected FUV luminosity to SFR ratio, which for constant SFR is $L_{\rm FUV}/{\rm SFR} = 4.3\times 10^3 L_{\odot}/(\Msun~ \Myr^{-1})$ based on our adopted initial mass function and spectral population synthesis choices (see Appendix C in \citealt{kim2023photchem}). The median ionizing photon rate (and the 16th and 84th percentiles) is $\Phi_{\rm LyC} = 4.3_{-2.2}^{+2.6} \times 10^{50}$ s$^{-1}$ kpc$^{-2}$ with a mean (and standard deviation) of $\Phi_{\rm LyC} = 4.8 \pm 3.0 \times 10^{50}$ s$^{-1}$ kpc$^{-2}$. These values are similar to the post-processing results presented in \cite{kado2020diffuse} and are consistent with the observational estimate for the solar neighborhood of $\Phi_{\rm LyC} = 5.0 \times 10^{50}$ s$^{-1}$ kpc$^{-2}$ \citep{mckee1997luminosity}.

The second row presents the gas mass fraction for the cold gas, total warm gas, and for the \WIM{} alone. For {\tt R8-4pc}, the majority of the gas mass is contained in the warm phase, with an average of 50$\pm$9\%. A fraction of this warm gas is \WIM{}, which makes up 9$\pm$5\% of the total mass. 
Most of the remaining gas mass is in the cold and unstable medium, making up an average 49$\pm$9\%. A small fraction of the total gas mass is in the hot medium. In the inner-disk {\tt LGR4-2pc} model, the fraction of cold gas increases compared to {\tt R8-4pc}. For a more detailed breakdown of the mass distribution by phase refer to Table 4 of \citet{kim2023ncr}.

The scale heights of the \CpU, total warm, and \WIM\ phases are shown in the third row. We also show the scale height of $n_e^2$ (observable as emission measure EM per unit length along the line of sight), which is proportional to the emissivity of the \WIM{}. The scale height is defined as
\begin{equation}
    H = \sqrt{\frac{\int \langle q \rangle_{\textrm{ph}} z^2 dz}{\int \langle q \rangle_{\textrm{ph}} dz}}
    \label{eq:H}
\end{equation}
Here $\langle q \rangle_{\textrm{ph}}$ (\autoref{eq:q}) defines the horizontal average over a given phase at a given $z$ value. To find the scale height of each phase, we set $q = \nH$. We also find the scale height of $n_e^2$ by setting $q = n_e^2$. For {\tt R8-4pc}, the median scale height of the \CpU\ gas is $110 \pc$. The combined warm gas has a median scale height of $260 \pc$ while \WIM{} alone has a slightly larger median scale height of $280 \pc$. The warm gas scale height is somewhat smaller (by $\sim 15\%$) in ${\tt LGR4-2pc}$ owing to the higher surface density and therefore higher gravity. The higher gravity is partly offset by slightly larger velocity dispersion (see Table 2 in \citealt{kim2023ncr}). The median value of the EM scale height is much smaller, approximately $90 \pc$. As in \cite{kado2020diffuse}, the scale height we measure for the \WIM{} can be much smaller than observational estimates because our definition of the phase does not distinguish between diffuse and dense (\HII\ regions) ionized gas and the emission is dominated by the latter. All of the scale heights vary in time by a factor $\sim 2$--$3$ from minimum to maximum.

In the final row, we show different source particle scales. The first is the average distance between sources, $\ell_\star \equiv N_\star^{-1/2}$ where $N_\star$ is the surface number density of FUV sources younger than $20 \Myr$ that account for $90\%$ of the total FUV luminosity cumulating downward from the most luminous sources. The median values are $\ell_\star= 280\pc$ and $65\pc$ for {\tt R8-4pc} and {\tt LGR4-2pc}, respectively. If we consider only ionizing sources, defined as the sources which account for more than 90\% of the total LyC luminosity, the separation is much larger, averaging approximately $460 \pc$ for {\tt R8-4pc} and $110 \pc$ for {\tt LGR4-2pc}. The characteristic source luminosity is then $\Sigma_\mathrm{FUV}/N_\star \sim 10^6 L_\odot$ and $\Phi_\mathrm{LyC}/N_\star \sim 10^{50} \second^{-1}$ for FUV and LyC sources, respectively. The quantity $\ell_\star$ will be used in \autoref{sec:Bialy} in testing an estimator of the FUV radiation field. 

The figure also includes the scale height of sources ($H_\star$). Similar to gas scale heights, $H_\star$ is calculated as the luminosity-weighted rms distance from the midplane, using \autoref{eq:H}. The result for both weightings (FUV and LyC) is comparable, with median value $H_*=40^{+10}_{-10}$~pc and $50^{+10}_{-10}$~pc for {\tt R8-4pc} and {\tt LGR4-2pc}, respectively. The last measure of the source distribution is the approximate vertical extent of the particles, defined as $(z_{\rm max} - z_{\rm min})/2$. Here, $z_{\rm max}$ and $z_{\rm min}$ represent the maximum and minimum $z$ values among the sources at each time.

The vertical scale height of the cluster particles (radiation sources) is comparable in the {\tt R8-4pc} and {\tt LGR4-2pc} simulations, while the higher SFR in the latter also corresponds to a smaller horizontal separation between sources, $\ell_\star$.  With $\Sigma_{\rm SFR} \propto N_\star = \ell_\star^{-2}$, we would expect a ratio of $\ell_\star$[{\tt LGR4-2pc}]/$\ell_\star$[{\tt R8-4pc}] $\sim 0.29$, which is close to the actual ratio of 0.23.

\section{FUV Radiation Field}
\label{sec:FUV}

We now focus on the radiation fields in the TIGRESS-NCR simulations, beginning with FUV radiation. First, we present the general characteristics of the FUV radiation field. Then, we evaluate a number of simplified models for the FUV radiation which may be used when full ray tracing might not be feasible.

\subsection{Radiation Field Distribution}

\begin{figure*}
    \centering
	\includegraphics[scale = 0.4]{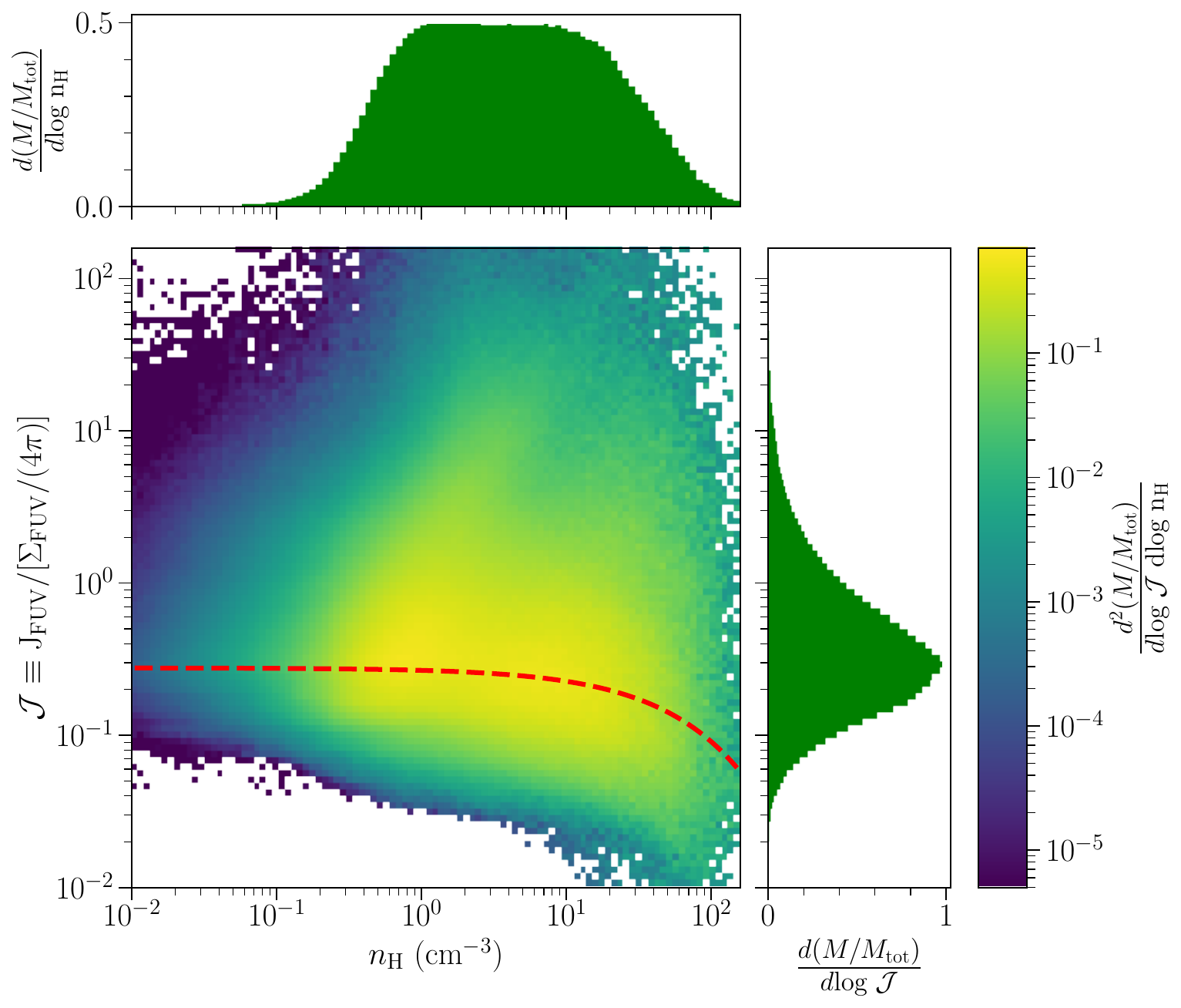}
         \includegraphics[scale = 0.4]{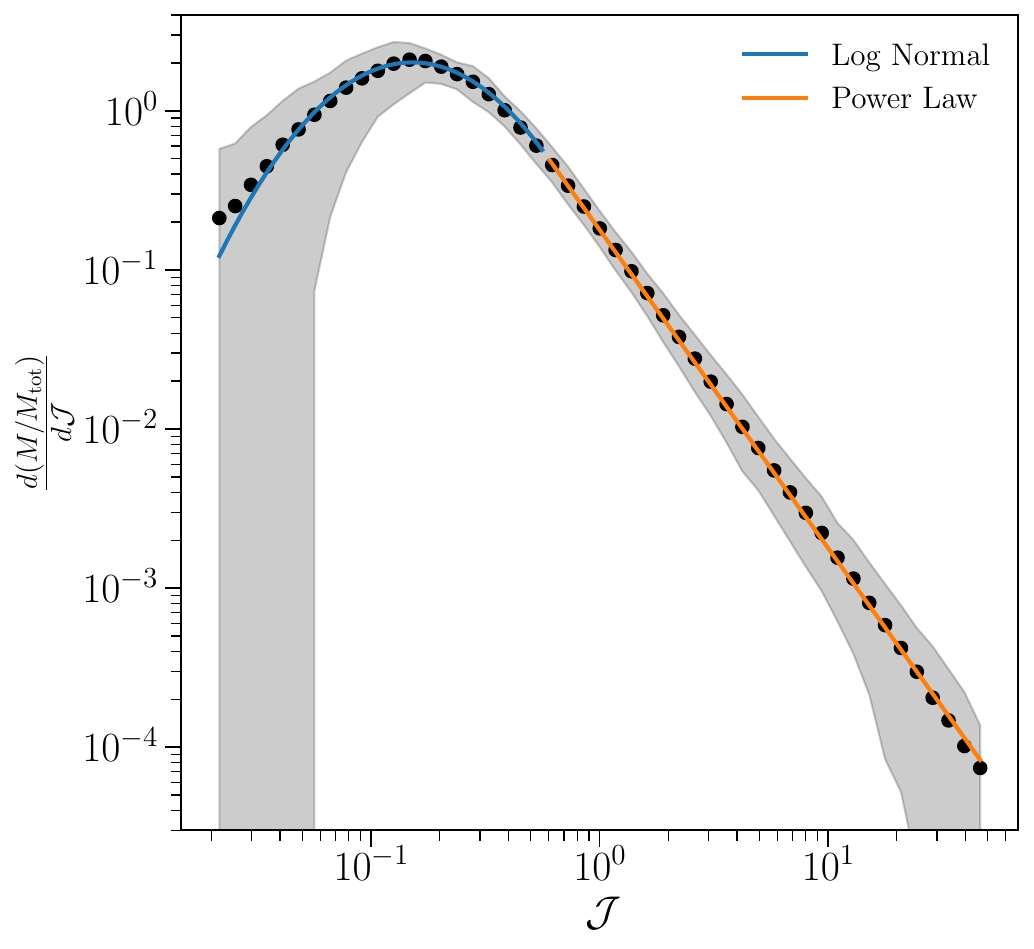}
    \caption{Left: Distribution of mass with normalized mean intensity in the FUV band $\mathcal{J}$ and hydrogen number density $\nH$ for the midplane gas in model {\tt R8-4pc} including all snapshots between 250-450 Myr. We limit to gas with $T < 3.5\times 10^4$ K. Insets show mass-weighted PDFs of log$\, \mathcal{J}$ and log$\, \nH$. The red dashed line represents the best fit of \autoref{eq:calJfit} to the mode of the distribution. Right: The black points represent the mass-weighted PDF of normalized mean intensity $\mathcal{J}$ (rather than log$\, \mathcal{J}$ as in the left panel). The overplotted lines represent the best fit of the model given in \autoref{eq:PDF_fit} with the log-normal and power law components represented in blue and orange respectively. The shaded region represents the 16th-84th percentile of the two PDFs of each individual snapshot between 250-450 Myr.}
    \label{fig:FUV_nH}
\end{figure*}

\begin{figure*}
    \centering
	\includegraphics[scale = 0.4]{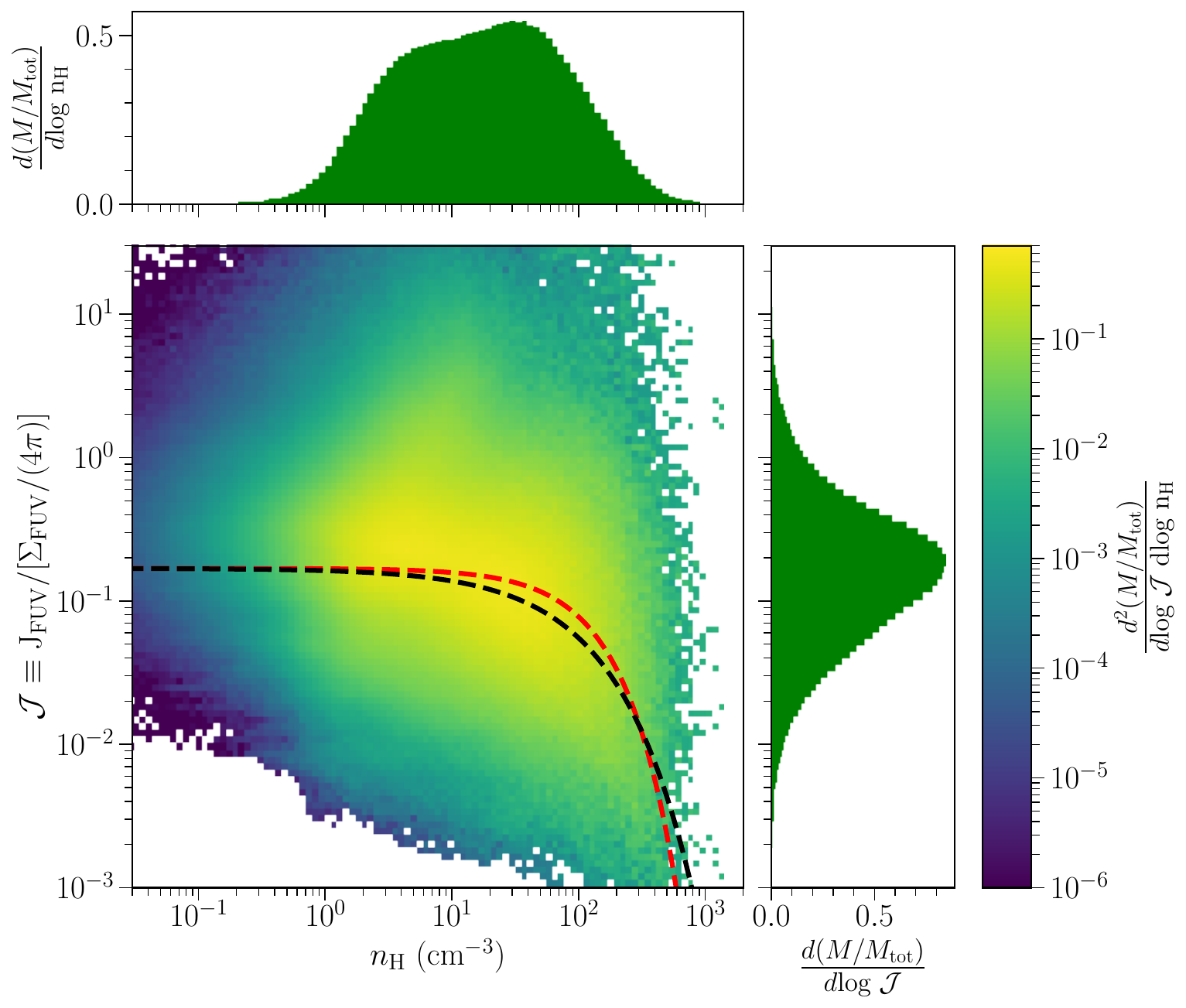}
         \includegraphics[scale = 0.4]{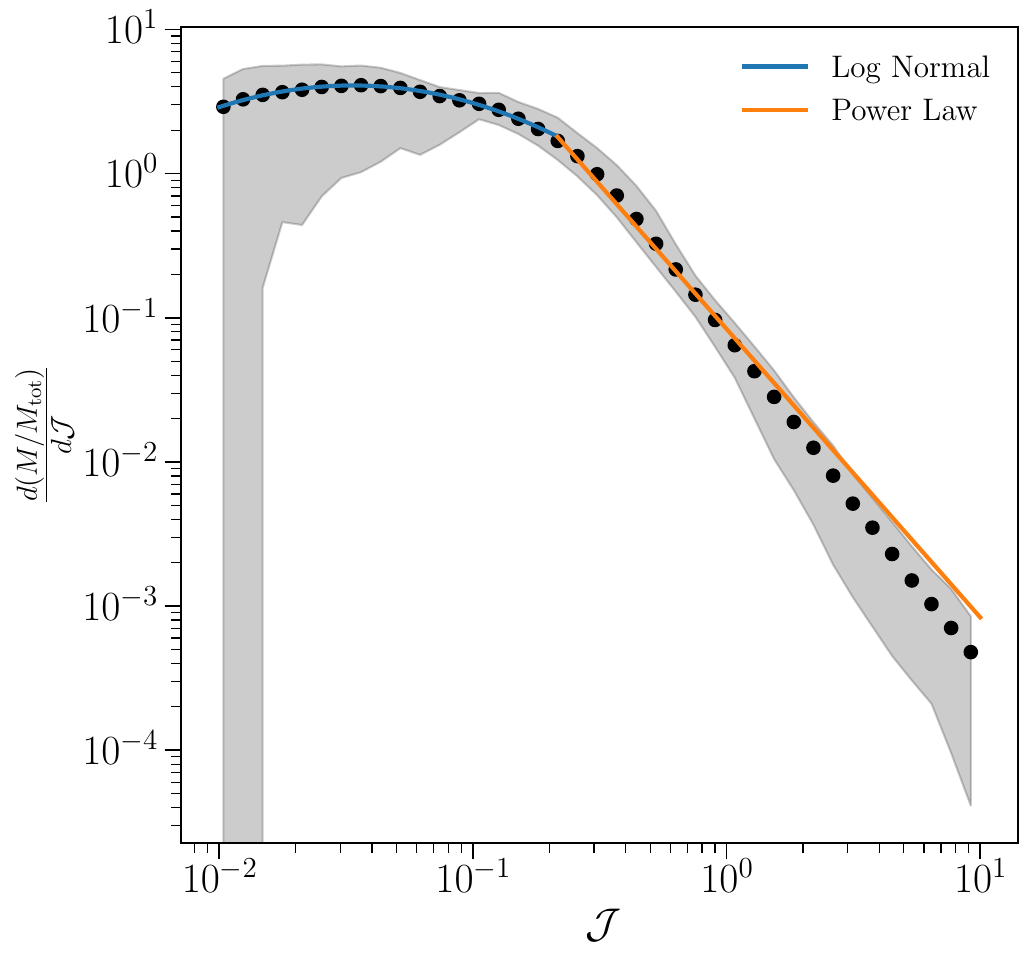}
    \caption{Same as in \autoref{fig:FUV_nH} but for model {\tt LGR4-2pc} over time range 250-350 Myr. The red dashed line represents the best fit of \autoref{eq:calJfit} to the mode of the distribution. The black dashed line represents the value of \autoref{eq:calJfit} using the value of $\tau_{\rm FUV, eff}$ fit from the {\tt R8-4pc} model (\autoref{eq:tau_R8}).}
    \label{fig:FUV_nH_L}
\end{figure*}

To lowest order, we expect the radiation intensity to vary proportional to the total (time-varying) FUV luminosity per unit area,  
$\Sigma_\mathrm{FUV} = L_{\rm FUV, tot}/(L_x L_y)$, where $L_{\rm FUV,tot}$ is the total FUV luminosity from the source particles and $L_xL_y$ is the horizontal area.\footnote{In observations, $\Sigma_\mathrm{FUV}$ on $\kpc$ scales may be obtained directly from (extinction-corrected) UV mapping surveys  \citep[e.g.][]{2007ApJS..173..267S,2018ApJ...859...11S,2019ApJS..244...24L,2024ApJS..271....2H}, or else 
estimated from traditional star formation tracers such as extinction-corrected H$\alpha$  emission, radio continuum from bremsstrahlung, or other tracers of photoionized gas \citep[e.g.][]{2012ARA&A..50..531K}. It should be borne in mind, however, that the duration of LyC emission is much shorter than that of FUV for a coeval stellar population, so $\Phi_\mathrm{LyC}$ (and resulting tracers of photoionized gas) will be more variable than $\Sigma_\mathrm{FUV}$ \citep[see e.g. Table 5 of][and \autoref{fig:history}(a) here]{kim2023photchem}. Catalogues of star clusters and associations \citep[e.g.][]{2017ApJ...841..131A,2022ApJS..258...10L} with age and mass measurements can also be combined with theoretical population synthesis to obtain individual source FUV luminosities, averaged over area to obtain $\Sigma_\mathrm{FUV}$.} We therefore define a normalized mean intensity in the FUV band by
\begin{equation}\label{eq:calJdef}
    \mathcal{J} \equiv \frac{J_\mathrm{FUV}}{\Sigma_\mathrm{FUV}/(4\pi)}.
\end{equation}
where $J_{\rm FUV} = (c/4\pi)\mathcal{E}_{\rm FUV}$ is the mean intensity in the FUV band.
In \autoref{fig:FUV_nH}, we show the mass-weighted distribution of $\mathcal{J}$ at the midplane for model {\tt R8-4pc}. The figure presents the two-dimensional distribution in $\mathcal{J}$ and $\nH$, as well as the individual, mass-weighted PDF of each. We include all gas in the midplane ($|z| < 4 \pc$) with $T < 3.5\times 10^4$ K across all snapshot times. \autoref{fig:FUV_nH_L} shows the same for model {\tt LGR4-2pc} at the midplane ($|z| < 2 \pc$). For both models, there is a unimodal distribution of  $\mathcal{J}$ with a well-defined central peak, but tails extending over more than two orders of magnitude. 
The modes of each radiation distribution, ${\cal J} \approx 0.3$ for {\tt R8-4pc} and  ${\cal J} \approx 0.2$ for {\tt LGR4-2pc}, are quite similar to the mass-weighted means at the midplane reported in \autoref{tab:quantities}. 
The peak of the $\mathcal{J}$ distribution shifts to lower values at higher $n_\mathrm{H}$, with this shift more pronounced in {\tt LGR4-2pc} as would be expected given the higher densities and optical depth of this model.  

We fit this peak in $\cal{J}$ as a function of $n_{\rm H}$ with a local shielding approximation similar to that suggested by \cite{kim2021star}
\begin{equation}
\label{eq:calJfit}
\mathcal{J} = \mathcal{J}_0 e^{-\tau_{\rm FUV,eff}} \,,
\end{equation}
where the value of $\mathcal{J}_0$ is given by the mode of the (one-dimensional) mass-weighted PDF of $\mathcal{J}$. At each value of $\nH$ in the two-dimensional distribution, we find the local peak in $\mathcal{J}$. We fit these points to find the optical depth, $\tau_{\rm FUV, eff}$, as a power-law function of gas density for the {\tt R8-4pc} model:
\begin{equation}
\label{eq:tau_R8}
\tau_{\rm FUV, eff} = 1.1\bigg( \frac{\nH}{10^2\textrm{ cm}^{-3}} \bigg)^{0.7}\,.
\end{equation}
For the {\tt LGR4-2pc} model we find
\begin{equation}
\tau_{\rm FUV, eff} = 0.8\bigg( \frac{\nH}{10^2\textrm{ cm}^{-3}} \bigg)^{1.1}\,.
\end{equation}

\autoref{eq:calJfit} using these two best fit values is displayed in \autoref{fig:FUV_nH} and \autoref{fig:FUV_nH_L} as red lines. In \autoref{fig:FUV_nH_L}, we also show \autoref{eq:calJfit} using the best fit value of $\tau_{\rm FUV, eff}$ from the {\tt R8-4pc} model (\autoref{eq:tau_R8}) with $\mathcal{J}_0$ given by the {\tt LGR4-2pc} distribution as a black dashed line.

The distribution of $J_\mathrm{FUV}$ (or $\cal J$) is of interest because FUV radiation is responsible for heating the gas through the photoelectric effect, and variations in $J_\mathrm{FUV}$ correspond to variations in the heating rate.  Since FUV also heats the dust, and the dust reprocesses this emission into the IR, the distribution of FUV radiation is reflected in the spectral energy distributions (SEDs) of thermal IR emission and PAH emission rate. For example, \cite{draine2007b} apply a simplified model of the global FUV intensity distribution with two distinct components to fit the total dust emission in the SINGS galaxy sample. The diffuse radiation has a single uniform value, represented as a delta function distribution,  while regions near stars are represented separately with a power law distribution for the radiation intensity.

We fit the distribution of the FUV radiation field in our simulations, considering the mass fraction at the midplane as a function of ${\mathcal J}$. For both the {\tt R8-4pc} and {\tt LGR4-2pc} models, the functional form we adopt for this fit is:
\begin{equation}
\frac{1}{M_{\rm tot}}\frac{dM}{d\mathcal{J}} = \left\{
\begin{array}{ll}
      A_1 \frac{1}{\mathcal{J} \sigma \sqrt{2 \pi}} e^{-\frac{(\textrm{ln}(\mathcal{J}) - \mu)^2}{2 \sigma^2}}& \mathcal{J} < \mathcal{J}_{\rm trans} \\
      A_2 \mathcal{J}^{-\alpha} & \mathcal{J} \geq \mathcal{J}_{\rm trans} \\
\end{array} 
\right.
\label{eq:PDF_fit}
\end{equation}
Here, $M_{\rm tot}$ is the total mass over the whole distribution. For high values of $\mathcal{J}$, we use a power law with exponent fixed to be $\alpha = 2$ as in \cite{draine2007b}, since this value was found to fit most observed galaxies well. We also find $\alpha = 2$ to describe our data well. To represent the peak of the distribution, we use a log-normal function. In our model, the amplitudes, $A_1$ and $A_2$, are determined by the requirement for the piecewise function to be continuous and for the total integral to be equal to one. This leaves three free parameters including the mean and width of the log-normal distribution ($\mu$ and $\sigma$) as well as the location of the transition from the log-normal to power law distribution ($\mathcal{J}_{\rm trans}$). In the right panels of \autoref{fig:FUV_nH} and \autoref{fig:FUV_nH_L}, we show the fits together with the PDFs of the normalized intensity $\mathcal{J}$ as measured from the simulations. Parameters for the fits are listed in \autoref{tab:density_J_fit}.

\begin{deluxetable}{lCCCCCC}
\tablecaption{FUV PDF Fit Parameters \label{tab:density_J_fit}}
\tablehead{
\colhead{Model} &
\dcolhead{\mu} &
\dcolhead{\sigma} &
\dcolhead{\mathcal{J}_{\rm trans}} &
\dcolhead{A_1} &
\dcolhead{A_2} &
\dcolhead{{\rm PL \; Fraction}}
}
\colnumbers
\startdata
{\tt R8-4pc} &  -1.2 & 0.82 & 0.6 & 0.88 & 0.18 & 0.3 \\
{\tt LGR4-2pc} & -1.3 & 1.5 & 0.2 & 1.4 & 0.08 & 0.4  \\
\enddata
\tablecomments{Columns (2)--(4): best fit values of the function described in \autoref{eq:PDF_fit} for both the {\tt R8-4pc} and {\tt LGR4-2pc} models. Columns (5)--(6): normalizations of the log-normal and power law components of the model required for the function to be continuous and integrate to one. Column (7): mass fraction described by the power law component rather than the log-normal.}
\end{deluxetable}

\subsection{Model Comparisons}

Ray tracing accurately solves the equations of radiative transfer and resolves the radiation field, but it is computationally intensive. Therefore, it is useful to compare the results of the full TIGRESS-NCR simulations to simplified models of the FUV radiation field, which potentially may be used when full radiative transfer is not practical.
We test four simple models using the {\tt R8-4pc} model, but we find similar results for {\tt LGR4-2pc}.

\begin{figure}
    \centering
	\includegraphics[scale = 0.55]{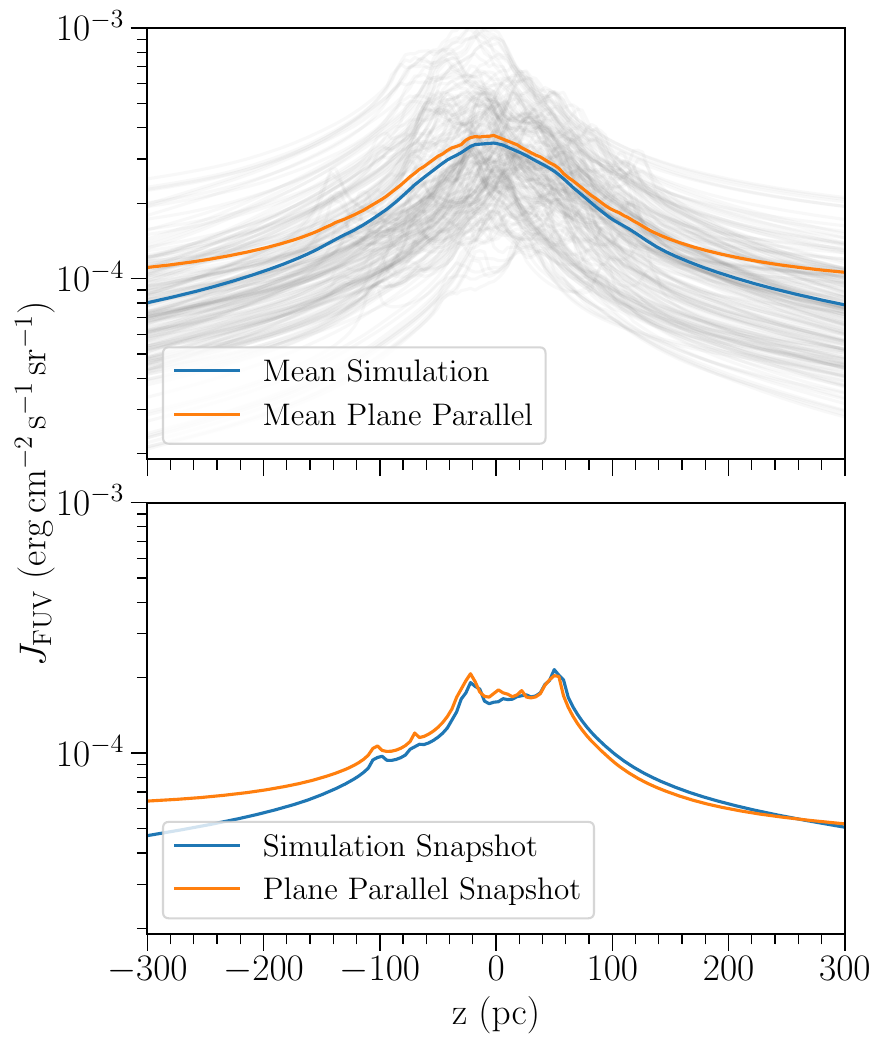}
    \caption{Comparison of the vertical profile of the mean FUV intensity to the plane parallel approximation (see \autoref{sec:planeparmod}) for model {\tt R8-4pc}. The upper panel shows the average mean intensity as a function of $z$ with individual snapshots shown in gray. The lower panel shows a comparison for a single snapshot at $t = 430 \Myr$.}
    \label{fig:pp_z}
\end{figure}

\begin{figure*}
    \centering
	\includegraphics[scale = 0.5]{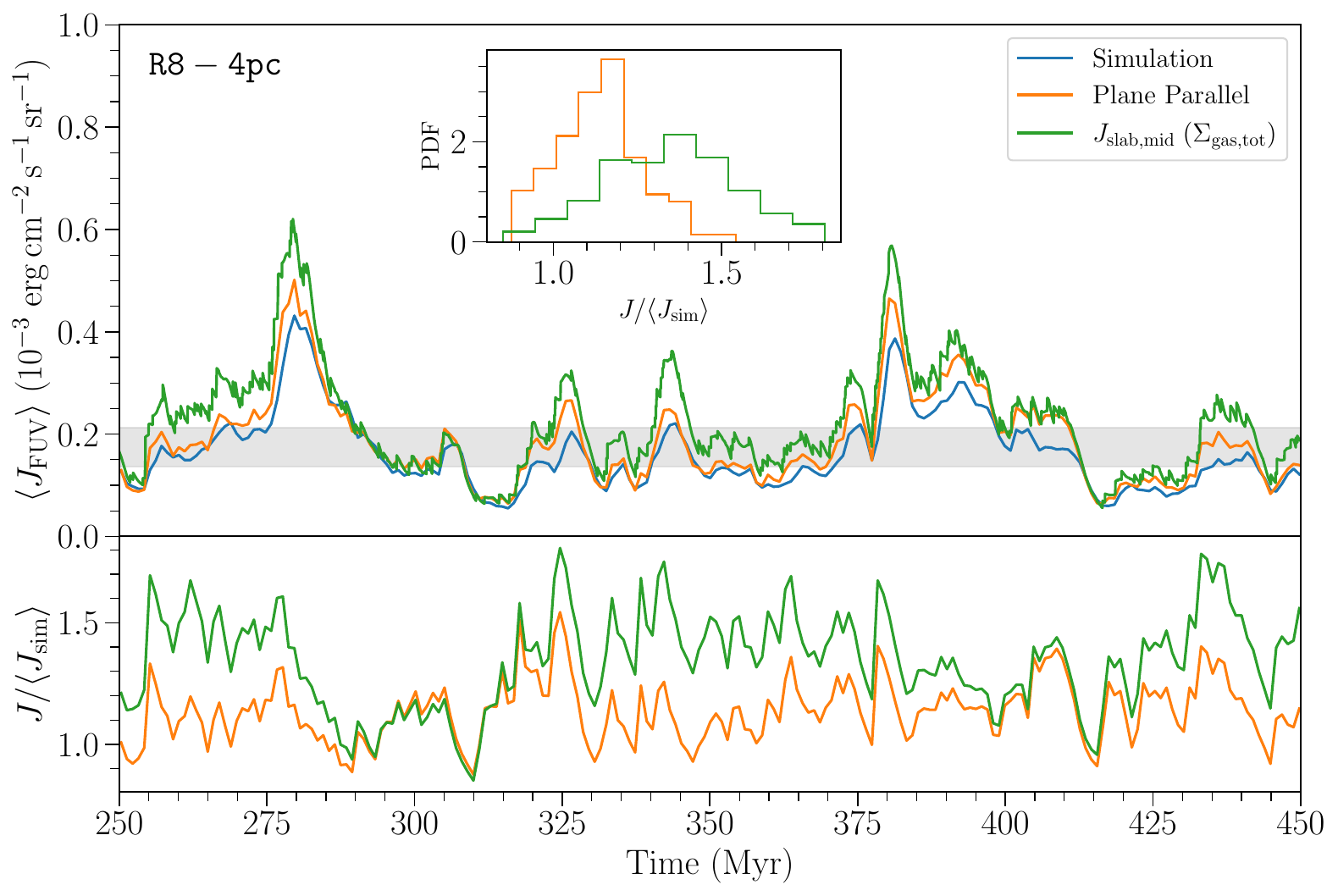}
	\includegraphics[scale = 0.5]{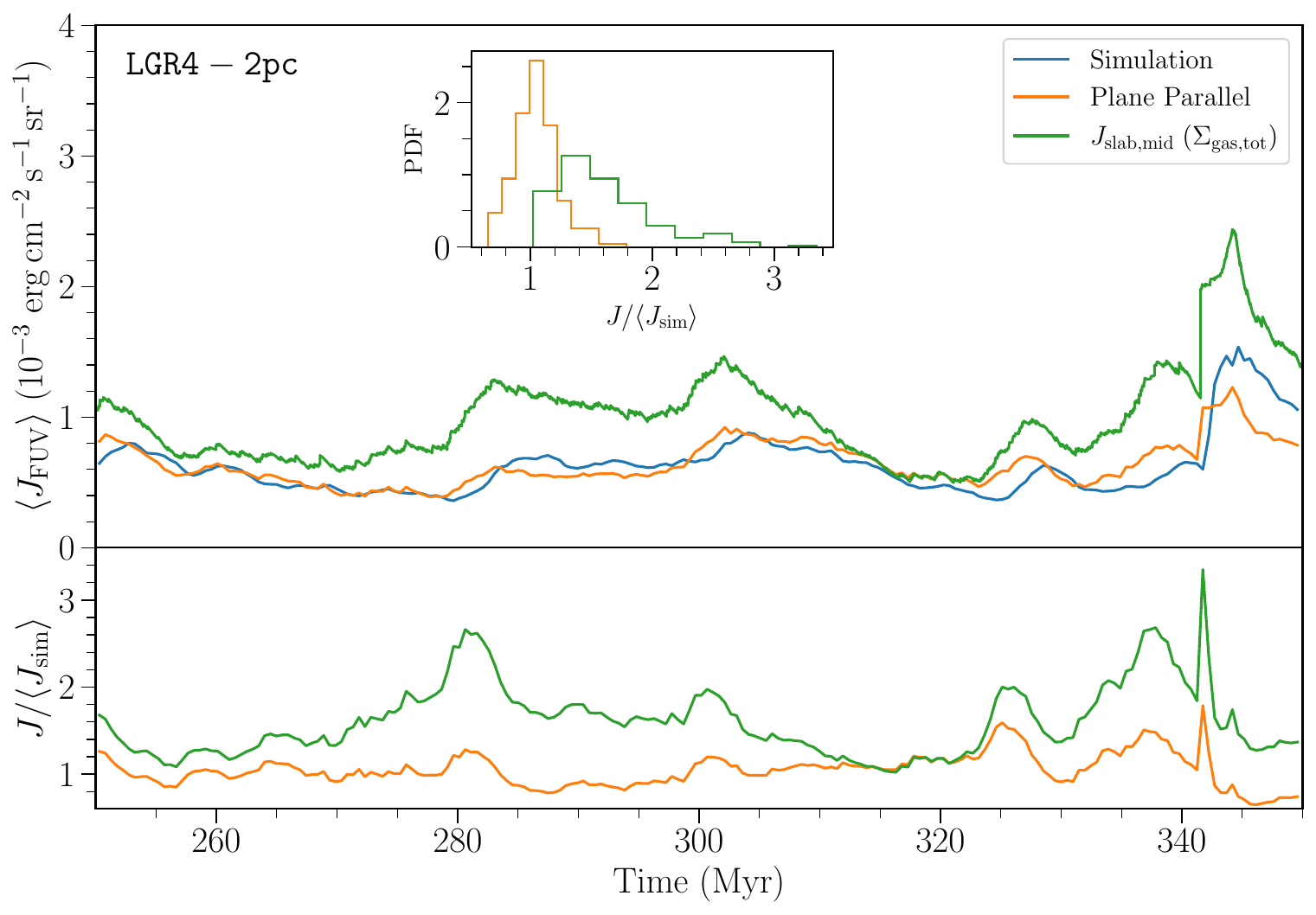}
    \caption{Comparison of the temporal history of the mean FUV intensity from ray tracing to the plane parallel approximation and to the slab approximation (\autoref{eq:J_norm_OML}) adopted for TIGRESS-classic. The comparison is for the volume-weighted averages within $|z| = 300 \pc$ of the midplane and is shown for both the {\tt R8-4pc} model (top) and {\tt LGR4-2pc} model (bottom). In the upper panel, the gray band represents the range of observed estimates for the solar neighborhood including \cite{1978ApJS...36..595D}, \cite{1982ApJ...258..533V}, and \cite{1983A&A...128..212M}. The insets show the distributions of the ratio between the model and simulation results.}
    \label{fig:pp}
\end{figure*}

\subsubsection{Plane Parallel Model}\label{sec:planeparmod}

In the TIGRESS-NCR simulations, a plane parallel approximation is used to calculate the FUV radiation field beyond $|z|=300\pc$ (see \autoref{sec:planepar}) as it is too computationally intensive to resolve the full box with ray tracing.  In the simulation, this plane-parallel solution has horizontally-averaged intensity as its inner boundary condition (\autoref{e:Erad_pp} and \autoref{e:Frad_z_pp}).

A related plane-parallel model can be applied to compute the angle-averaged radiation intensity throughout the domain. For each radiation source particle with luminosity $L_{\rm sp}$ and vertical position $z_{\rm sp}$, we consider a thin plane-parallel layer that has the source function $S = S_{\rm sp}\delta (\tau_{\perp})$, where $S_{\rm sp} = L_{\rm sp}/(4\pi L_{x}L_{y})$ and $\tau_{\perp}(z; z_{\rm zp}) = |\int_{z_{\rm sp}}^{z} \sigma_{\rm d} \langle \nH \rangle dz|$ is the dust optical depth measured from $z_{\rm sp}$ in the vertical direction. Solving the plane-parallel radiative transfer equation $\mu dI/d\tau_{\perp} = S - I$ gives the intensity $I(z; \mu) = S_{\rm sp} e^{-\tau_{\perp}/|\mu|}/|\mu|$ 
($0$ for $\mu \gtrless 0$ if $z \lessgtr z_{\rm zp}$)
and the mean intensity $J_{\rm p\textnormal{-}p}(z) = S_{\rm sp} E_1(\tau_{\perp})/2$, where $E_1$ is the exponential integral. The total FUV mean intensity can be obtained as a superposition of thin layers for individual sources and the summation over frequency bins.

We compare the results of the {\tt R8-4pc} simulation to the plane parallel approximation in \autoref{fig:pp_z}. The upper panel shows the horizontally and time averaged FUV mean intensity, $\langle J_{\rm FUV} \rangle$, as a function of $z$. The faint gray lines represent individual simulation snapshots and the blue line represents the mean value. We show the mean of the plane parallel model approximation for comparison in orange. The peak of the mean plane-parallel profile agrees well with the full ray-tracing solution. However, by $z=\pm$300 pc the plane-parallel model exceeds the ray-tracing based result by $\sim 50$\% averaged in time. This is most likely because $10\%$ of FUV photons are lost due to the ray terminations based on $\epsilon_{\rm PP}$ and $d_{xy,{\rm max}}$ (see \autoref{sec:ART}), meaning that the FUV radiation field is slightly underestimated far from the sources in the ray-tracing based result.

Similar behavior is also seen for individual snapshots rather than the average. This is shown in the lower panel of \autoref{fig:pp_z} for $t=430 \Myr$ (a radial-vertical slice at the same time is portrayed in \autoref{fig:snap_R8}). 

\subsubsection{Slab Model}\label{sec:slab}

The plane parallel model gives a $z$-dependent approximation of the FUV radiation field. From \citet{OML10}, an even simpler model is that of a slab with uniform density and opacity such that the total perpendicular optical depth is $\tau_\mathrm{\perp,FUV}$, and uniform emissivity such that the source function is $S_\mathrm{FUV}=\Sigma_\mathrm{FUV}/(4\pi \tau_\mathrm{\perp,FUV})$.  
For this system, the intensity at the midplane of the slab is given by $J_\mathrm{FUV}=S_\mathrm{FUV}[1-E_2(\tau_\mathrm{\perp,FUV}/2)]$ for $E_2$ the second exponential integral (see \autoref{sec:slab_deriv}).  
The resulting normalized intensity at the midplane is thus 
\begin{equation}\label{eq:J_norm_OML}
\mathcal{J}_\mathrm{slab, mid} \equiv \frac{1 - E_2(\tau_\mathrm{\perp,FUV}/2)}{\tau_\mathrm{\perp,FUV}}.
\end{equation}
The FUV optical depth perpendicular to the slab, $\tau_\mathrm{\perp,FUV}$ is defined in terms of the gas surface density in the slab, $\Sigma_{\rm{slab}}$, and the FUV opacity, $\kappa_\mathrm{FUV}$, as
\begin{equation}
\tau_\mathrm{\perp,FUV} = \kappa_\mathrm{FUV}\Sigma_{\rm{slab}} 
\end{equation}
where $\kappa_\mathrm{FUV} \equiv \sigma_\mathrm{FUV}/(\mu_{\rm H} m_{\rm H})$ for  $\sigma_\mathrm{FUV}$ the FUV cross-section and $\mu_{\rm H}=1.4$ the mean molecular weight per hydrogen nucleus.
The TIGRESS-classic simulations \citep{kim2020first,OK22} adopt \autoref{eq:J_norm_OML} as the radiation model, with $\Sigma_\mathrm{slab}=\Sigma_\mathrm{gas,tot}$, the total (instantaneous) surface density of all gas.\footnote{In \citet{kim2020first} and \citet{OK22}, we use the notation $f_\tau$ for ${\cal J}_{\rm slab,mid}$, and refer to this as an attenuation factor.}

Using the same uniform-slab model, one can also compute the mean intensity within the slab (as opposed to the midplane).  It is straightforward to show that this is:
\begin{equation}\label{eq:J_norm_slab_ave}
\mathcal{J}_\mathrm{slab, avg} \equiv \frac{1}{\tau_\mathrm{\perp,FUV}}\left[1 - \frac{1}{2 \tau_\mathrm{\perp,FUV}} + \frac{E_3(\tau_\mathrm{\perp,FUV})}{{\tau_\mathrm{\perp,FUV}}}\right],
\end{equation}
for $E_3$ the third exponential integral.  The functions $\mathcal{J}_\mathrm{slab,mid}$ and  $\mathcal{J}_\mathrm{slab,avg}$ are shown in  \autoref{fig:rad_model_comp}. Both monotonically decrease with increasing optical depth. It is worth noting that both exceed unity at low optical depth.

To get some idea of what the slab model predicts, we can consider typical numbers for these simulations. With a time-averaged gas surface density in the {\tt R8-4pc} model of $\Sigma_\mathrm{gas, tot}= 10.5\ \Msun\ {\rm pc}^{-2}$, and for $\sigma_\mathrm{FUV} = 10^{-21}$ cm$^2$, we find $\tau_\mathrm{\perp,FUV}=0.94$, $\mathcal{J}_\mathrm{slab,mid}=0.70$, and $\mathcal{J}_\mathrm{slab,avg}=0.63$.  For model {\tt LGR4-2pc}, $\Sigma_\mathrm{gas, tot}= 38.0\ \Msun\ {\rm pc}^{-2}$, $\tau_\mathrm{\perp,FUV}= 3.4$, 
$\mathcal{J}_\mathrm{slab,mid}=0.28$, and
$\mathcal{J}_\mathrm{slab,avg}=0.25$.

\autoref{fig:pp} compares the result from full ray tracing to the $z$-dependent plane parallel approximation (\ref{sec:planeparmod})  and the simple slab model (\autoref{eq:J_norm_OML}). Although \autoref{eq:J_norm_slab_ave} is nominally more appropriate to compare with the vertically averaged radiation field, we compare with \autoref{eq:J_norm_OML} as this was the model adopted for all warm and cold gas in the TIGRESS-classic simulations analyzed in \cite{kim2020first} and \cite{OK22}. The difference between $\mathcal{J}_\mathrm{slab,avg}$ and $\mathcal{J}_\mathrm{slab,mid}$ is at most 10\% (at $\tau_\perp =2$). The simulation and plane parallel values shown in the upper main panel are volume-weighted averages over all gas within $|z| = 300 \pc$ of the midplane (this choice is motivated by the measured scale heights in the simulation -- see \autoref{tab:quantities}). The ratio of the model values, either $z$-dependent plane parallel or slab, to the simulated values is shown in the lower panel. We include the distribution of these ratios in the inset panel.

Both approximate models track the temporal variations in the mean radiation field well, although the slab model generally exceeds the plane-parallel model, and both exceed the ray-tracing solution. As noted above, the ray-tracing solution will underestimate the true solution at larger $|z|$ due to ray truncation. The plane parallel approximation differs from the simulated value by an average factor of $\sim1.14$ ({\tt R8-4pc}) or $\sim1.05$ ({\tt LGR4-2pc}). The slab model overestimates the simulated value by a mean factor of $\sim 1.35$ ({\tt R8-4pc}), or $\sim 1.61$ ({\tt LGR4-2pc}).  Despite these differences, both approximations do well for being simplified models. Indeed, the slab approximation is remarkably good, considering that it relies only on the average optical depth through the disk and total luminosity per unit area.  Since these quantities are straightforward to calculate, the slab approximation could be readily employed for an inexpensive estimate of radiation field in simulations that do not follow radiative transfer and is useful for observational estimates.  Our  results also show that the radiation prescription adopted for the TIGRESS-classic simulations only modestly overestimated the values that would have been obtained with ray tracing. However, there is substantial local variation in the intensity that cannot be captured by the mean field model, as evidenced by the scatter in the distribution of $\mathcal{J}$ in \autoref{fig:FUV_nH} and \autoref{fig:FUV_nH_L}.

\subsubsection{Bialy Model}
\label{sec:Bialy}

\cite{bialy2020far} presents another model which estimates the FUV radiation field based on the surface density of FUV  luminosity, the mean distance between FUV radiation sources, and the gas density. In the analytic model of \cite{bialy2020far}, the average distance between sources in a region is defined as
\begin{equation}
    l_* \equiv N_*^{-1/2}
\end{equation}
for $N_*$ the surface density of source counts. 
For randomly distributed sources, the median distance between sources is $x_0 l_*$ for $x_0 $ = 0.47, which represents the distance to the nearest source from any location.
With $n_\mathrm{H}$ a mean gas number density, 
\begin{equation}\label{eq:taustar}
    \tau_* \equiv l_*  \nH \sigma_{\rm d,FUV} = l_* \rho \kappa_{\rm FUV}
\end{equation}
is a characteristic optical depth, and $x_0 \tau_*$ is taken as the optical depth to the nearest source.

In \citet{bialy2020far}, the FUV radiation flux at any point is the sum of two terms. The first represents the contribution from the closest point source, and the second represents the total contributions from all other sources treated as a spatially-uniform distribution located at the midplane with a minimum distance $x_0 l_*$.  Equation 18 of \cite{bialy2020far} gives the radiation flux $F$ assuming a distribution of $\Sigma_\mathrm{FUV}$ that declines exponentially with radius with a scale length $R_d=X l_*$. The flux $F$ can be divided by $4\pi$ to obtain $J_\mathrm{FUV}$, or by $\Sigma_\mathrm{FUV}$ ($\Sigma_*$ in the notation of  \cite{bialy2020far}) to obtain the normalized intensity,
\begin{equation}
\mathcal{J}_\mathrm{Bialy} = \frac{1}{4 \pi x_0^2}e^{-x_0 \tau_*} + \frac{1}{2} E_1(x_0 \tau_* + x_0/X).
\label{eq:bialy_F}
\end{equation}

Similar to the slab model, $\mathcal{J}_\mathrm{Bialy}$ monotonically decreases with increasing optical depth $\tau_*$ (we compare the two functions in \autoref{sec:slab_deriv} -- see \autoref{fig:rad_model_comp}). For application to our simulation, $X$ is effectively infinite because we are using a local simulation box. Therefore, we ignore the second term in the argument of the exponential integral. Physically, $1/(\tau_* X) = 1/(\nH \sigma_{\rm FUV} R_d)$ is equal to the ratio of photon mean free path to disk radial scale length, which is small for massive star-forming galaxies. In \autoref{eq:bialy_F}, the second term is larger than the first term for $\tau_*<1.6$, but becomes increasingly subdominant for larger $\tau_*$. We note that as $N_*$ increases, $\tau_*$ decreases, which leads to an overall increase in $\mathcal{J}_\mathrm{Bialy}$ and an increase in the second term relative to the first.

A caveat regarding this model is that for the purpose of making observational estimates, the gas volume density is not directly accessible.  Estimating a midplane or volume-weighted value of the density requires dividing a measured gas surface density by an estimate of the gas disk's thickness, which generally depends on an assumption of vertical dynamical equilibrium and therefore does not track temporal variations in $\nH$. That is, an empirical estimate would be $\tau_*\sim \tau_{\rm \perp,FUV} l_*/(2H)$. Here, we are able to use measured values of $\nH$ which are available from the simulation, and we also compare to the case where $\nH$ is estimated from the surface density.   

In our simulation, we determine $N_*$ by dividing the number of star cluster particles in the simulation box by the horizontal area of the region. We limit the number of sources to those which contribute 90\% of the total FUV luminosity. This results in a time-averaged source density $N_* = 12.7\,{\rm kpc}^{-2}$ for the {\tt R8-4pc} model, and an average value of $l_*= 290\pc$ (\autoref{fig:history}(d) shows temporal variations of $l_*$). For \autoref{eq:taustar} we either set the gas number density equal to the mean midplane value or to the mean value within $|z|<H$, where $H$ is the scale height at each time (with an average value of $220\pm40\pc$). As for the slab model we include all gas, and we average over the whole horizontal area.  Respectively, this yields $n_\mathrm{H}=0.93\,{\rm cm}^{-3}$ so that $\tau_*=0.84$, or $n_\mathrm{H}=0.57\,{\rm cm}^{-3}$ so that $\tau_*=0.51$, averaging over time. Using midplane (or $|z|<H$) numbers, the first term in \autoref{eq:bialy_F} is equal to 0.24 (0.28), the second term is equal to 0.36 (0.54), and their sum is $\mathcal{J}_\mathrm{Bialy}=0.60$ (0.82). This total is $\sim 14\%$ smaller ($\sim 30 \%$ larger) than $\mathcal{J}_\mathrm{slab,mid}$ ($\mathcal{J}_\mathrm{slab,avg}$) for the same model. 

\begin{figure*}
    \centering
        \includegraphics[scale = 0.6]{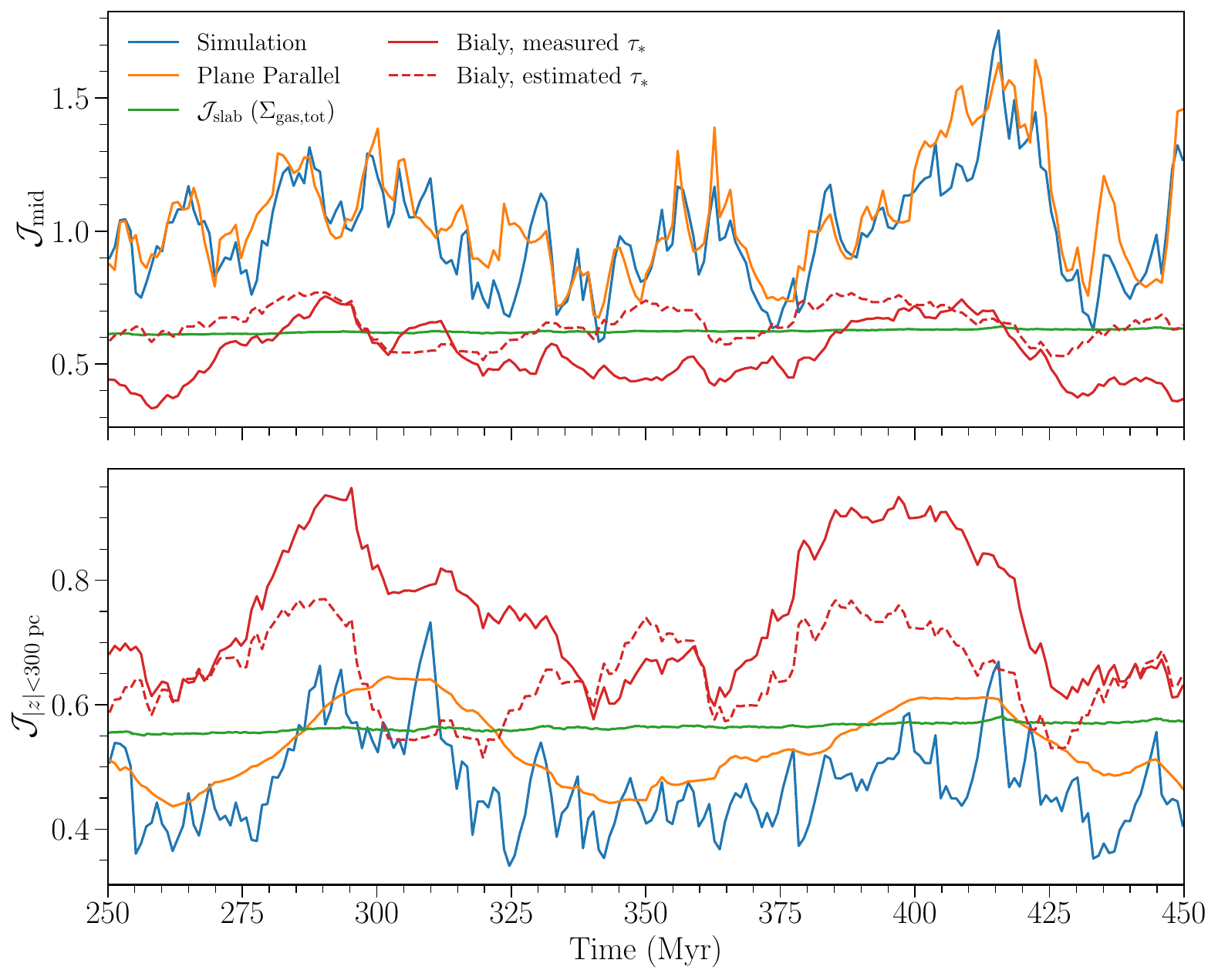}
    \caption{Comparison of the normalized mean FUV radiation intensity $\mathcal{J}$ from the simulation ({\tt R8-4pc}) with different approximate models, as described in \autoref{sec:planeparmod}- \autoref{sec:Bialy}. {\it Top}: Values are computed at the midplane. {\it Bottom}: Values represent a volume weighted average within $300\pc$ of the midplane. In the upper panel, $\cal{J}_{\rm slab}$ uses \autoref{eq:J_norm_OML}; in the lower panel, $\cal{J}_{\rm slab}$ uses \autoref{eq:J_norm_slab_ave}. The solid red curves use measured midplane or volume-averaged densities to evaluate $\tau_*=l_* \rho \kappa_\mathrm{FUV}$, while the dashed curves use the measured surface density and the mean value of $\langle H \rangle$ to estimate $\tau_*=l_* \Sigma_\mathrm{gas} \kappa_\mathrm{FUV}/(2 \langle H \rangle)$.}
    \label{fig:f_comp}
\end{figure*}

For a more detailed comparison, in \autoref{fig:f_comp} we show for model {\tt R8-4pc} both the midplane and volume-weighted mean normalized intensity, $\cal J$, in the simulation, in comparison to the three approximate models. These averages are done including all gas. As noted above, the temporal tracking of variations in the radiation field seen here in $\mathcal{J}_\mathrm{Bialy}$ is possible in large part because $\nH$ is directly measured. If we instead use $\tau_*= \tau_{\rm \perp,FUV} l_*/(2H)$ for $H$ an estimated mean disk thickness (adopting $220$ pc here), the majority of variation in the radiation would be due to the changing value of $l_*=N_*^{-1/2}$. This does not track the variations in the simulated $\cal J$ as closely, as shown in the dashed curves in \autoref{fig:f_comp}.

Considering ${\cal J}_{\rm |z|<300pc}$ with directly measured values of $\nH$, the \cite{bialy2020far} model overestimates the ray-tracing value by an average factor of approximately 1.6. Although this approximation still does well as a simplified estimate, the slab approximation (\autoref{eq:J_norm_slab_ave}) is somewhat closer in mean value (a factor 1.2 times the simulation value, on average). For the prediction of \autoref{eq:bialy_F} to more closely match the intensity found with full radiative transfer, $\tau_*$ would have to be larger by a factor of two. This would require a much lower source density, $N_*\sim 4\ {\rm kpc}^{-2}$, i.e. four sources in the simulation domain. If we instead consider only the midplane, the average value of $\nH$ is greater so the modeled value of $\mathcal{J_{\rm Bialy}}$ is smaller. At the same time, the simulated $\cal J_\mathrm{sim}$ is larger at the midplane. These effects compound so that $\mathcal{J}_\mathrm{Bialy} \sim 0.6 \mathcal{J}_\mathrm{sim}$. The mean value of $\mathcal{J}_\mathrm{Bialy}$ using the measured midplane density is similar to that from \autoref{eq:J_norm_OML}.

The plane parallel model follows the simulation very closely, both at the midplane and volume-averaged, and for this reason we recommend it as an inexpensive alternative when full radiative transfer is not practical but close tracking of the mean radiation field is desirable. 

\subsubsection{FUV Sum of Sources Model}
\label{sec:FUV_source}

The previous models of the FUV radiation field are useful in describing the average intensity across the simulation volume. To find an estimate for the FUV radiation at any point when the locations and luminosities of sources are known, we can instead model the contribution of each individual source particle to the total flux. In this case, the approximate FUV intensity is given by
\begin{equation}\label{eq:J_sum_source}
    J_{\rm FUV, src} = \frac{1}{4\pi}\sum_{\mathrm{sources}\ i, b} \frac{L_{i,b}}{4 \pi r_i^2}e^{- r_i \langle \nH \rangle \sigma_{\rm d,b}}\,,
\end{equation}
where $L_{i,b}$ is the luminosity of source $i$ in band $b$ (LW and PE), $r_i$ is the distance to the source, $\langle \nH \rangle$ is the mean ambient ISM density within one scale height of the midplane (\autoref{eq:H}), and $\sigma_\mathrm{d, b}$ is the dust absorption cross section in band $b$. For this comparison, we include only the sources with $t_{\rm age} <20 \Myr$ and horizontal distance less than $d_{xy,{\rm max}}$ from each cell, consistent with our practice for ray-tracing. We consider sources from a horizontally extended domain by applying the shearing-periodic boundary conditions. For $\langle \nH \rangle$, we take the average density within one scale height of the midplane. 

\autoref{eq:J_sum_source} models geometric attenuation correctly, but it is approximate because it uses a single  average density $\langle \nH \rangle$ rather than allowing the density to vary along rays. This may lead to an overestimation of $J_{\rm FUV}$ in dense regions, where the density exceeds $\langle \nH \rangle$ and the photon mean free path decreases. To address this issue, we also consider a model including local attenuation
\begin{equation}
    \label{eq:J_sum_source_att}
    J_{\rm src, att} = J_{\rm FUV, src} e^{-\tau_{\rm eff, local}}\,,
\end{equation}
where $\tau_{\rm eff, local}$ is given by  the best fit relationship between $\mathcal{J}$ and density in \autoref{eq:calJfit}, using the local value of $\nH$ in each cell in \autoref{eq:tau_R8}.  

\begin{figure*}
    \centering
        \includegraphics[scale = 0.53]{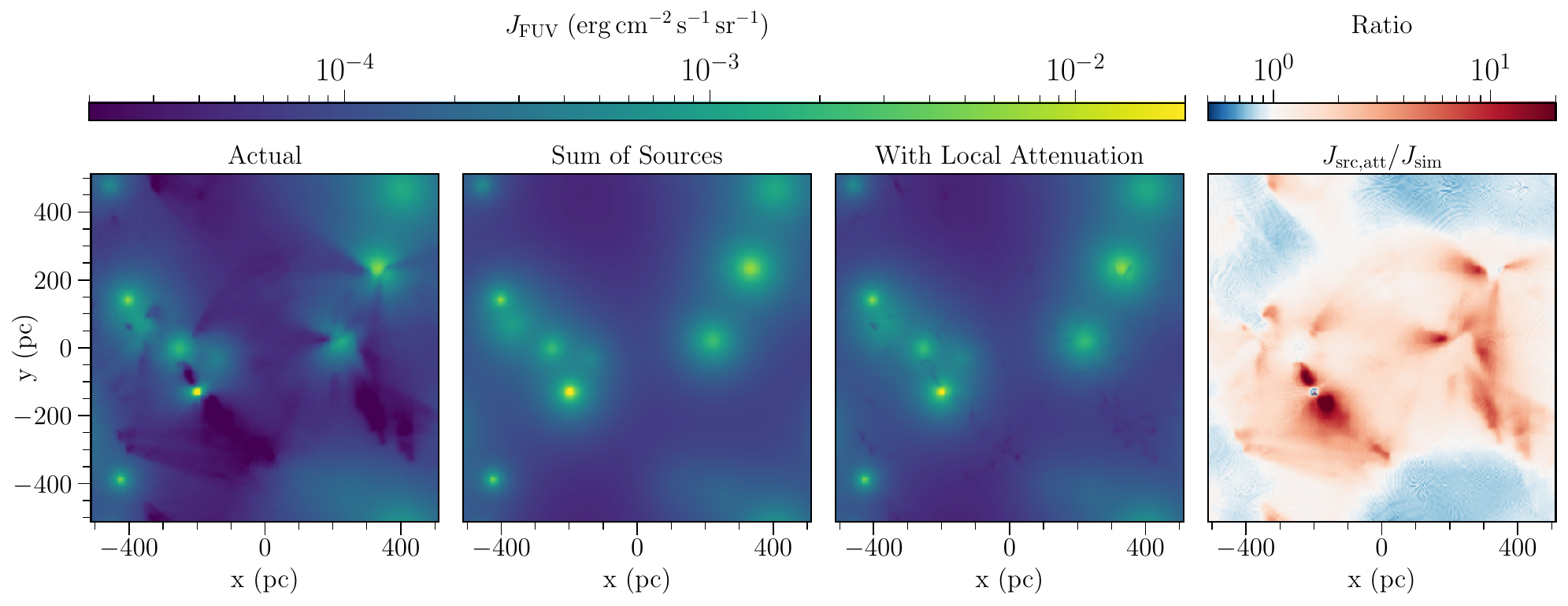}
    \caption{From left to right, the midplane FUV intensity $J_\mathrm{FUV}$ for model {\tt R8-4pc} at  $t = 430 \Myr$ in comparison to the model based on the contribution from a sum over attenuated FUV sources (\autoref{eq:J_sum_source}) and the same model with local attenuation (\autoref{eq:J_sum_source_att}). The far right panel represents the ratio of the attenuated sum of sources model to the true FUV radiation field.}
    \label{fig:Jsource}
\end{figure*}

\begin{figure*}
    \centering
	\includegraphics[scale = 0.35]{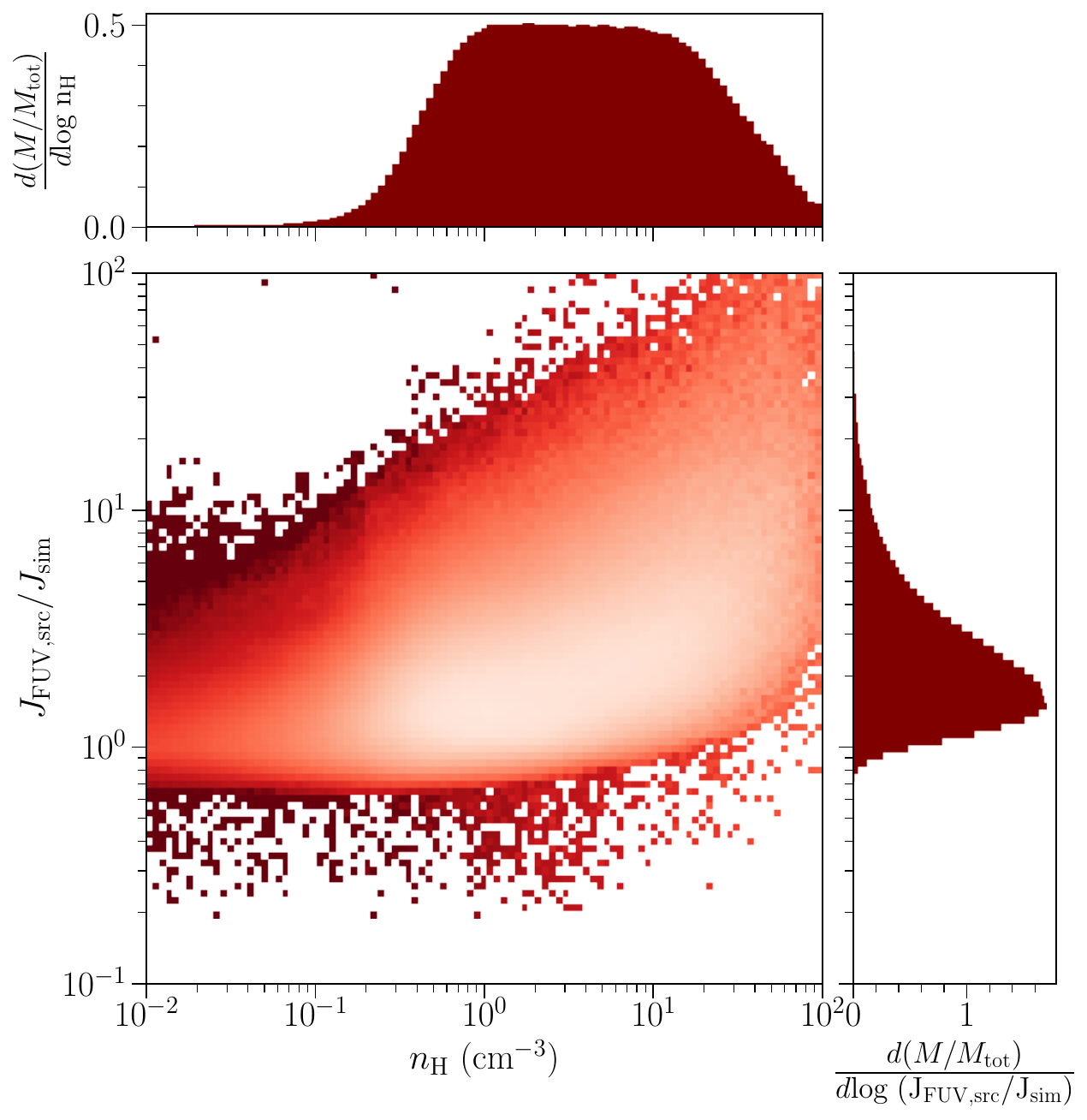}
 	\includegraphics[scale = 0.35]{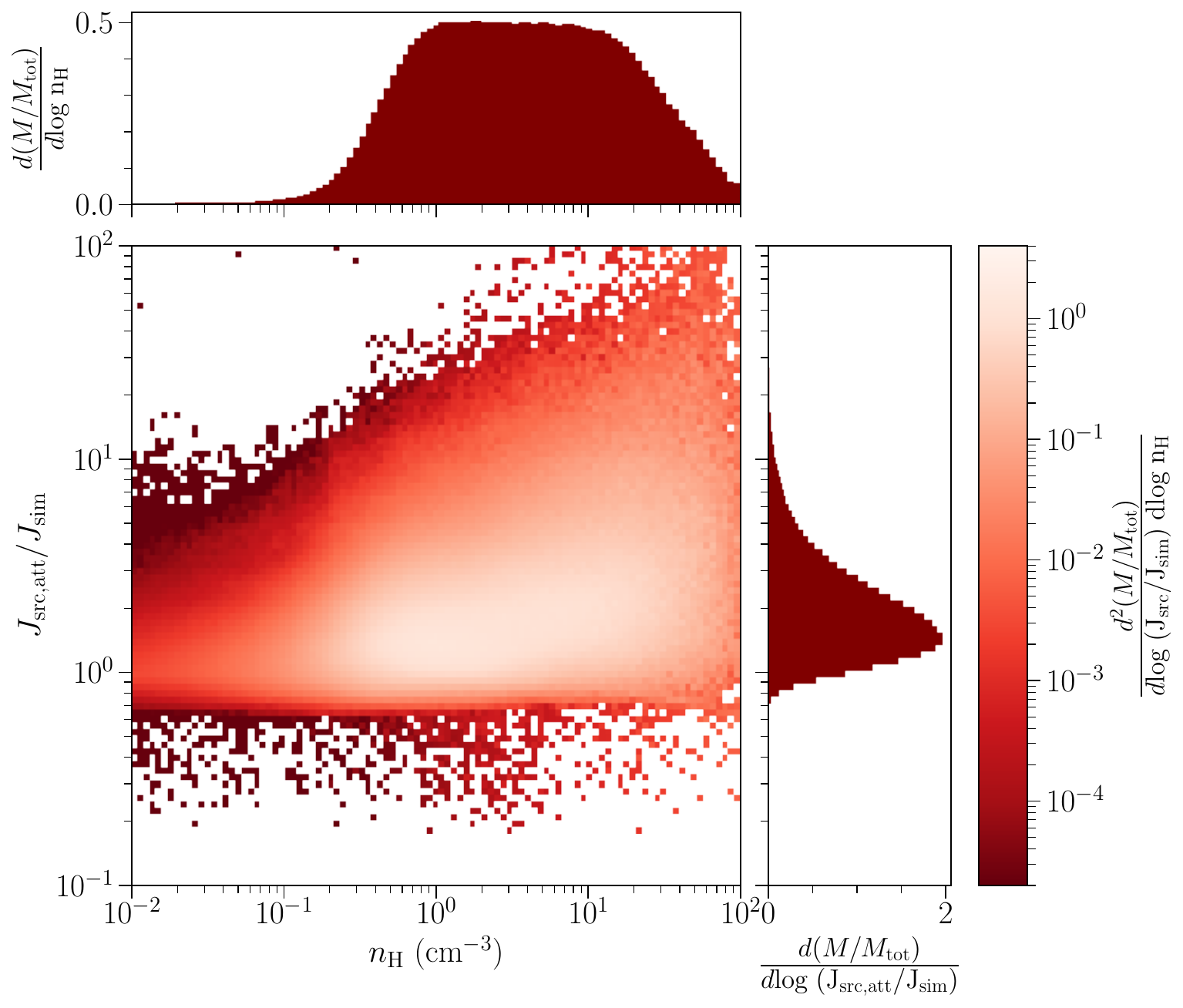}
    \caption{Distribution of mass as a function of the ratio of the true FUV radiation field to the estimate based on attenuated sum over sources models and the local density. On the left is the model given by \autoref{eq:J_sum_source} and on the right is the model with additional local attenuation given by \autoref{eq:J_sum_source_att}. Results are based on $\sim$200 snapshots from the {\tt R8-4pc} model between $t=250$--$450\Myr$ considering only the midplane ($z = 2\pc$).}
    \label{fig:Jsourceerr}
\end{figure*}

An example of the midplane intensity from ray-tracing compared to the sum of sources models for the {\tt R8-4pc} simulation is given in  \autoref{fig:Jsource} for $t = 430 \Myr$. Qualitatively, the model described in \autoref{eq:J_sum_source} matches the simulated values well, especially in the warm-phase diffuse gas. In particular, for the snapshot shown here, $\sim$82\% of cells are within a factor of two of the true FUV radiation value; this is true for $\sim83\pm9\%$ of cells including all snapshots. The distribution of the ratio between the approximate and simulated values of $J_\mathrm{FUV}$ is shown in the left panel of \autoref{fig:Jsourceerr}. 

With the addition of local attenuation in the model given by \autoref{eq:J_sum_source_att}, 
some high density regions show significantly more attenuation, and now  $\sim$84\% of cells are within a factor of two of the true FUV radiation value. 
However, there are still significant differences between the true FUV radiation and the model in regions with higher than average density. This can be seen for one snapshot in the far right panel of \autoref{fig:Jsource}, and across all snapshots in \autoref{fig:Jsourceerr}.  

Even at lower density, the true intensity is lower than \autoref{eq:J_sum_source_att} ($J_{\rm sim} \sim 0.79 J_{\rm src, att}$ averaged over the warm phase for a single snapshot, or a factor of $\sim 0.76$ more generally), because of the higher-than average attenuation in dense gas around sources.  Still, the local attenuation factor is helpful in reducing the bias toward overestimating the radiation field in dense regions, as comparison of the right with the left of \autoref{fig:Jsourceerr} shows. 

\subsection{FUV Heating Rate and Dust Emission}

Interstellar dust is heated by stellar FUV radiation\footnote{Stellar optical (and near-UV) radiation also contributes to dust heating, although in the present work we do not consider optical radiative transfer since the optical does not heat the gas. In the local interstellar medium, the dust heating of larger grains by optical photons is at a level of $\sim 50\%$ of the FUV \citep[e.g., see Fig. 26 in][]{kim2023photchem}}, and re-radiates the energy that is absorbed over a wide range of wavelengths.  Dust SEDs therefore reflect both the distribution of physical properties of grains (which vary in size over orders of magnitude, with composition including both hydrocarbons and silicates  -- see e.g. \citet{hensley_draine2023} and references therein), and the radiation intensity $J_\nu$ over a range of frequencies that the grains are exposed to.  Additionally, the spatial distribution of dust emission at any given wavelength depends not only on the distribution of the dust but also of the radiation field.  

Considering a given grain type with abundance $q$, the observed emission at a given wavelength per unit area per unit time will be proportional to an integral over path length and frequency of  $q n_\mathrm{H} J_\nu$.  If we take our simulations as representing a patch of the ISM in a face-on disk galaxy, the dust emission due to FUV absorption will therefore be proportional to a dust heating parameter
\begin{equation}\label{eq:dustheat}
{\cal N}_d(x,y)\equiv 
\int \mathcal{J} n_\mathrm{H} dz
\end{equation}
where we use the normalized FUV intensity $\cal J$ of \autoref{eq:calJdef}. That is, the dust emission at $(x,y)$ will convolve over $n_\mathrm{H}(x,y,z)$ and ${\cal J}(x,y,z)$. 
To obtain the FUV-induced dust emission at a given frequency and location, one would multiply ${\cal N}_d$ by the dust abundance, by the mean local UV production rate $\Sigma_{\rm FUV}/(4\pi)$, and by factors for FUV absorption crossection and emissivity at the frequency of interest.   While larger grains absorb essentially all radiation and emit primarily at far-IR wavelengths, PAHs are believed to be excited primarily by absorption of FUV and have strong mid-IR emission \citep{Tielens2008,hensley_draine2023}. Mid-IR emission maps, as are now available at high resolution with {\it JWST} \citep[e.g.][]{2023ApJ...944L...9L,2024AJ....167...39P}, are therefore sensitive to a similar radiation distribution (in space, time, and frequency) to the radiation field that heats the gas via PE.  

\autoref{fig:snap_R8}--\autoref{fig:snap_h_LGR4} show that there are large spatial variations in both density and the FUV radiation field.  The variation in the gas density, however, is larger overall than the variation in normalized radiation intensity (see \autoref{fig:FUV_nH} and \autoref{fig:FUV_nH_L}), so the spatial structure in dust emission may potentially be employed as a proxy of spatial structure in the gas \citep[e.g.][]{2024AJ....167...39P}. 

\begin{figure*}
    \centering
    \includegraphics[scale = 0.5]{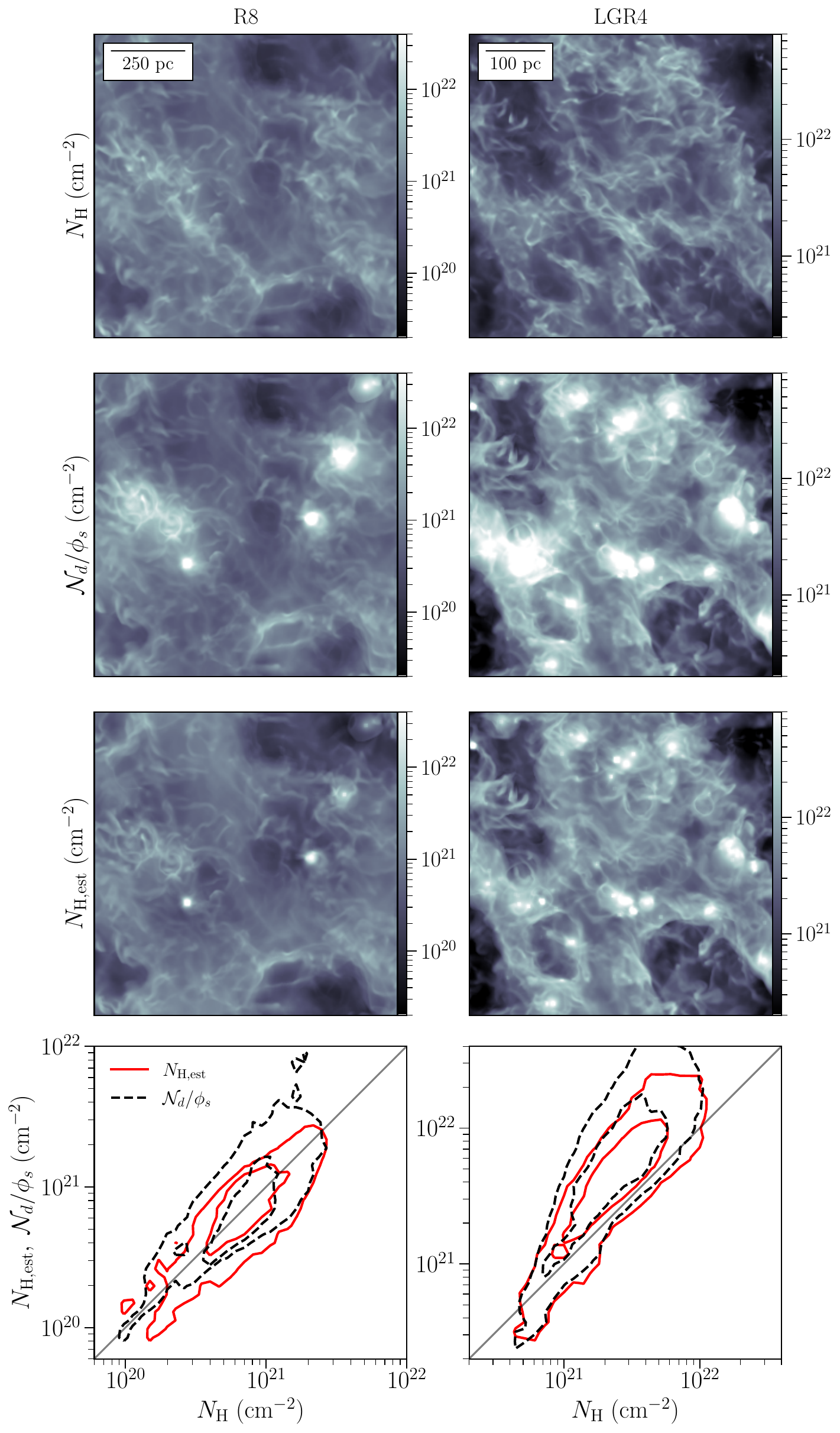}
    \caption{Snapshots of the column density, $N_H$ (first row) and the dust heating parameter ${\cal N}_d$ from \autoref{eq:dustheat} renormalized by a constant factor $\phi_s$ (second row), and the estimate $N_\mathrm{H,est}$ from \autoref{eq:NHest} that applies a correction based on the sum of sources radiation map (third row). The final row gives contours of both $N_{\rm H, est}$ and ${\cal N}_d/\phi_s$ compared to $ N_H$ representing the 50th and 90th percentiles by area. These are shown for the {\tt R8-4pc} simulation at time $t = 430 \Myr$ (left), and for the  {\tt LGR4-2pc} simulation at time $t = 298 \Myr$ (right). The gray lines in the final row show where ${\cal N}_d/\phi_s$ or $N_\mathrm{H,est}$ are equal to $N_\mathrm{H}$.}
    \label{fig:heatingrate_slice}
\end{figure*}

Based on our simulations, in the first two rows of \autoref{fig:heatingrate_slice} we compare the spatial structure of gas column density $N_\mathrm{H}(x,y) \equiv \int \nH dz$ with \autoref{eq:dustheat}, for the same snapshots shown in \autoref{fig:snap_h} and \autoref{fig:snap_h_LGR4}. We remove the hot gas by restricting to $T < 3.5\times 10^4$ K. In hot gas, the PAH lifetime would be short so we would not expect significant dust emission \citep[e.g.][]{2012A&A...545A.124B}. We renormalize the dust heating parameter (${\cal N}_d$) by a constant ($\phi_s$) such that it has the same median value as $N_{\rm H}$ for $N_{\rm H} < 10^{21} \textrm{ cm}^{-2}$; operationally this divides \autoref{eq:dustheat} by the typical ${\cal J}$ of atomic regions.\footnote{In observations, the equivalent would be to renormalize mid-IR maps such the median value matches the median column of $N_H$ from 21 cm emission in atomic-dominated regions; this empirically calibrates for dust abundance as well as UV absorption and IR emission coefficients and the mean $J_{\rm FUV}$. \cite{2017ApJ...846...38L}, for example, find a linear relation between $N_{\rm H}$ and dust reddening for $N_{\rm H} < 4\times10^{20} \textrm{ cm}^{-2}$.} Evidently, ${\cal N}_d/\phi_s$ shows small-scale structures in the gas column density quite well but is systematically elevated compared to $N_\mathrm{H}$ in the neighborhood of sources, as  
would be expected from the enhancement of the radiation field in these regions (see also \autoref{fig:Jsource}).

Considering the potential of using high-resolution dust emission maps to trace gas column, it is interesting to test whether approximate radiation solutions can be used to correct for varying heating.  To this end, the third row of \autoref{fig:heatingrate_slice} shows 
\begin{equation}\label{eq:NHest}
    N_\mathrm{H,est} \equiv \frac{{\cal N}_d}{{\cal J}_\mathrm{src,\perp} \phi_s}, 
\end{equation}
where
${\cal J}_\mathrm{src,\perp}\equiv J_{\rm src,\perp}/(\Sigma_{\rm FUV}/4\pi)$ for $J_{\rm src,\perp}$ as in \autoref{eq:J_sum_source}, but with $r_i$ the projection of distance onto the $(x,y)$ plane at a $z$ value representing the average height of a source particle above the midplane ($z=45$ pc). We again renormalize by a value of $\phi_s$ such that the median value of $N_\mathrm{H,est}$ is the same as $N_{\rm H}$ for $N_{\rm H} < 10^{21} \textrm{ cm}^{-2}$.

Compared to the map of $\mathcal{N}_d/\phi_s$, the map of $N_{\rm H, est}$ shows reduced elevation around the source particles particularly in the {\tt R8-4pc} model. The distributions of both $\mathcal{N}_d/\phi_s$ and $N_{\rm H,est}$ compared to $N_{\rm H}$ are shown in the final row \autoref{fig:heatingrate_slice} as contours. Both proxies of dust emission probe $N_{\rm H}$ well at low column densities. This would be expected as the values are normalized for $N_{\rm H} < 10^{21}\cm^{-2}$ and the radiation field is relatively uniform in low-density gas. At higher column, $\mathcal{N}_d/\phi_s$ is more biased towards overestimating $N_{\rm H}$ when compared to $N_{\rm H, est}$.

\begin{figure*}
    \centering
    \includegraphics[scale = 0.36]{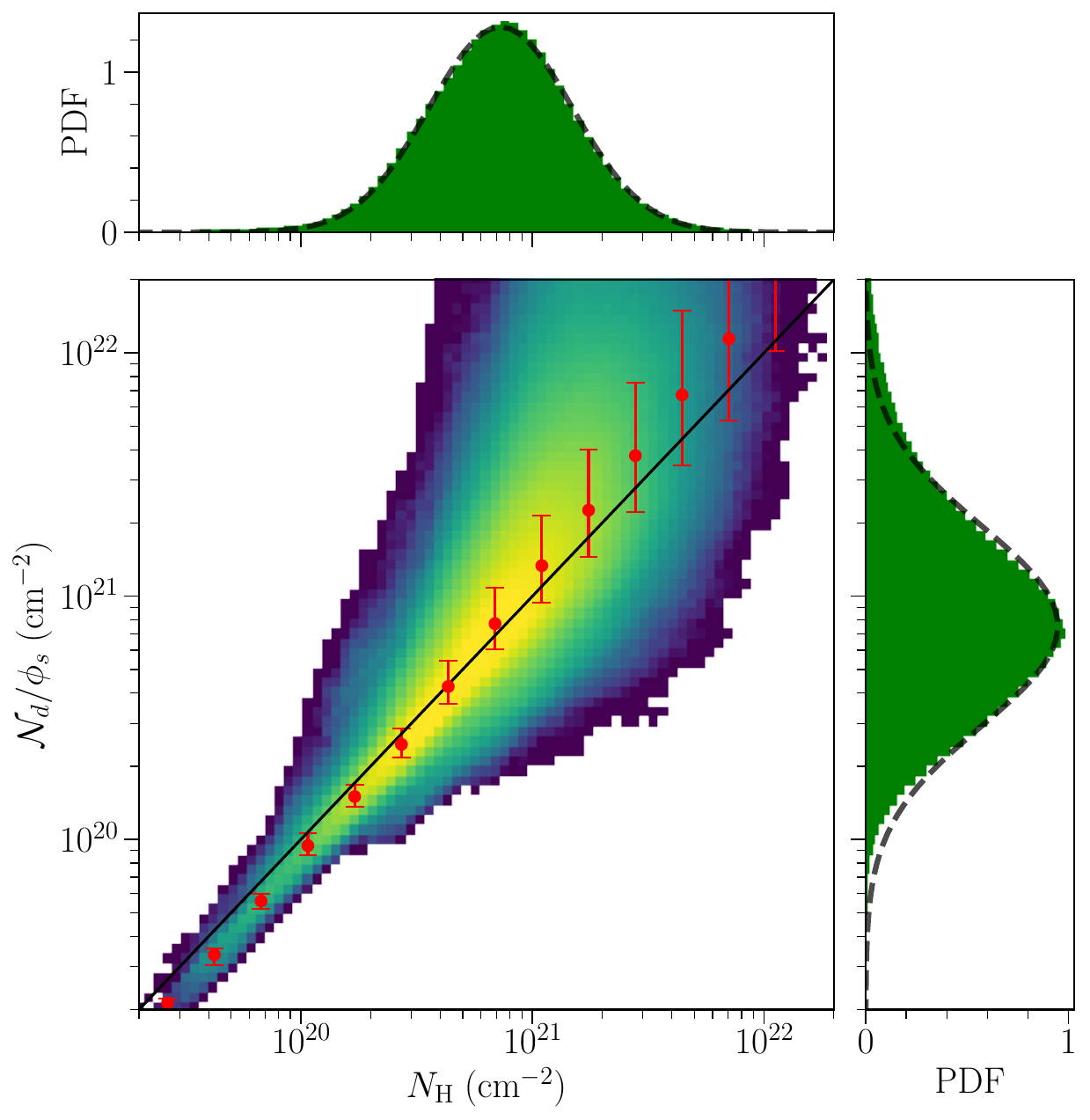}
    \includegraphics[scale = 0.36]{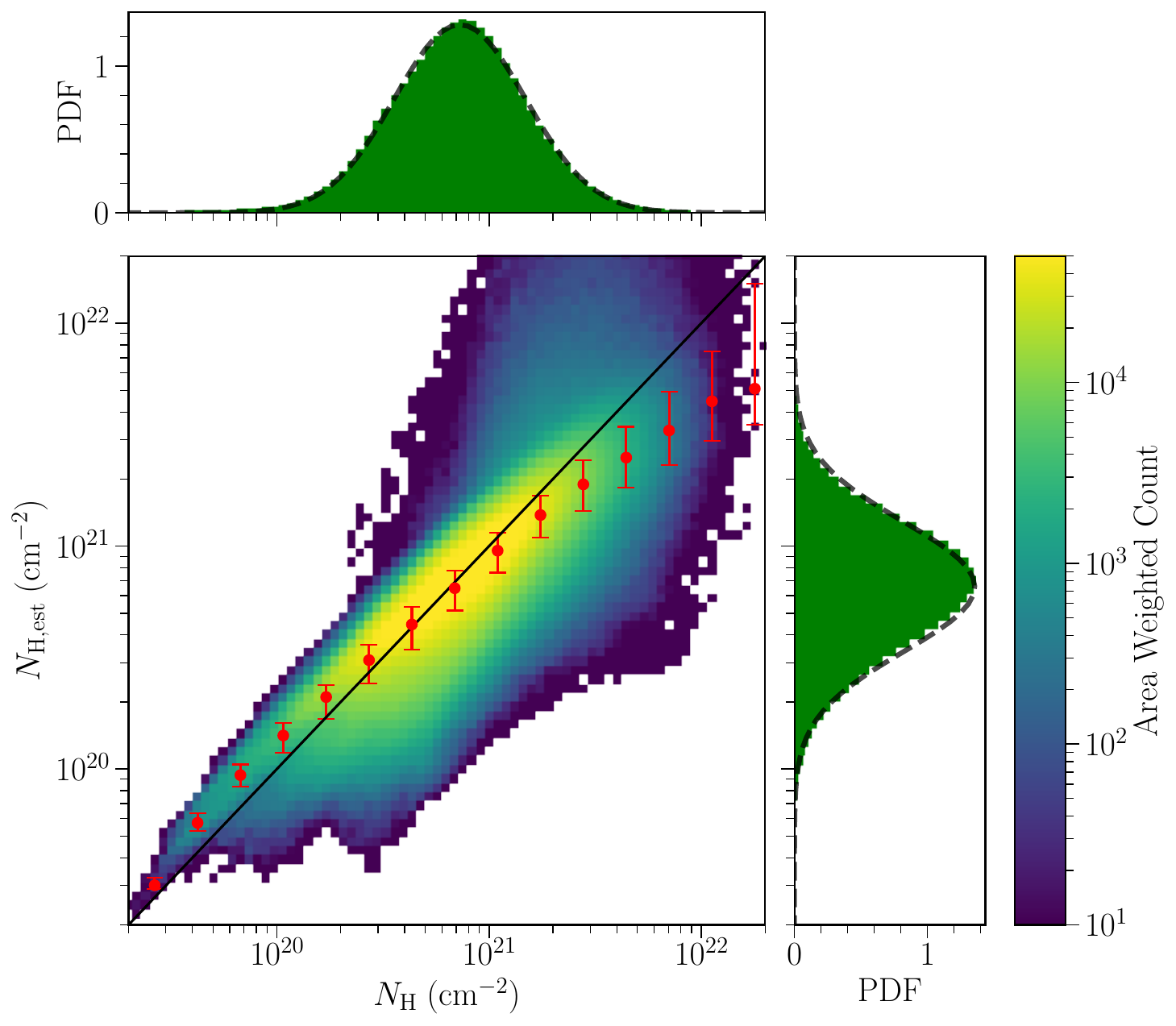}
    \caption{Distribution of the renormalized dust heating parameter ${\cal N}_d/\phi_s$ (left, see \autoref{eq:dustheat}) and $N_\mathrm{H,est}$ (right, see \autoref{eq:NHest})
    vs. gas column $N_\mathrm{H}$. Both distributions include all snapshots from the {\tt R8-4pc} simulation between $250$--$450 \Myr$. The inset axes represent the projections of the two dimensional histogram into one dimension. A log-normal fit is plotted over each of the one dimensional distributions. The red points represent the median and 25th-75th percentile values of ${\cal N}_d/\phi_s$ or $N_\mathrm{H,est}$ within individual $N_H$ bins. The black lines show where ${\cal N}_d/\phi_s$ or $N_\mathrm{H,est}$ are equal to $N_\mathrm{H}$.}
    \label{fig:heatingrate}
\end{figure*}

\begin{figure*}
    \centering
    \includegraphics[scale = 0.36]{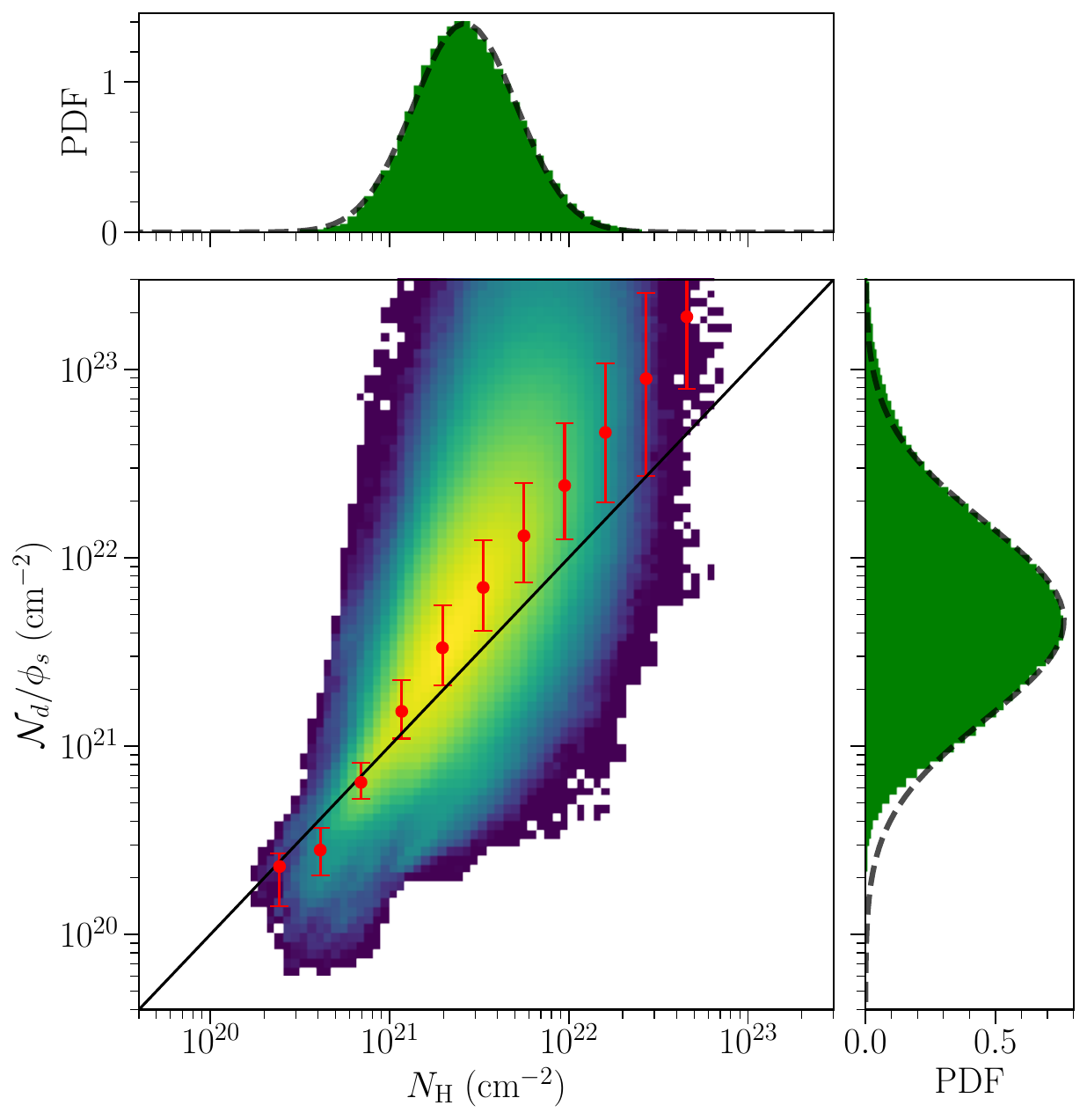}
    \includegraphics[scale = 0.36]{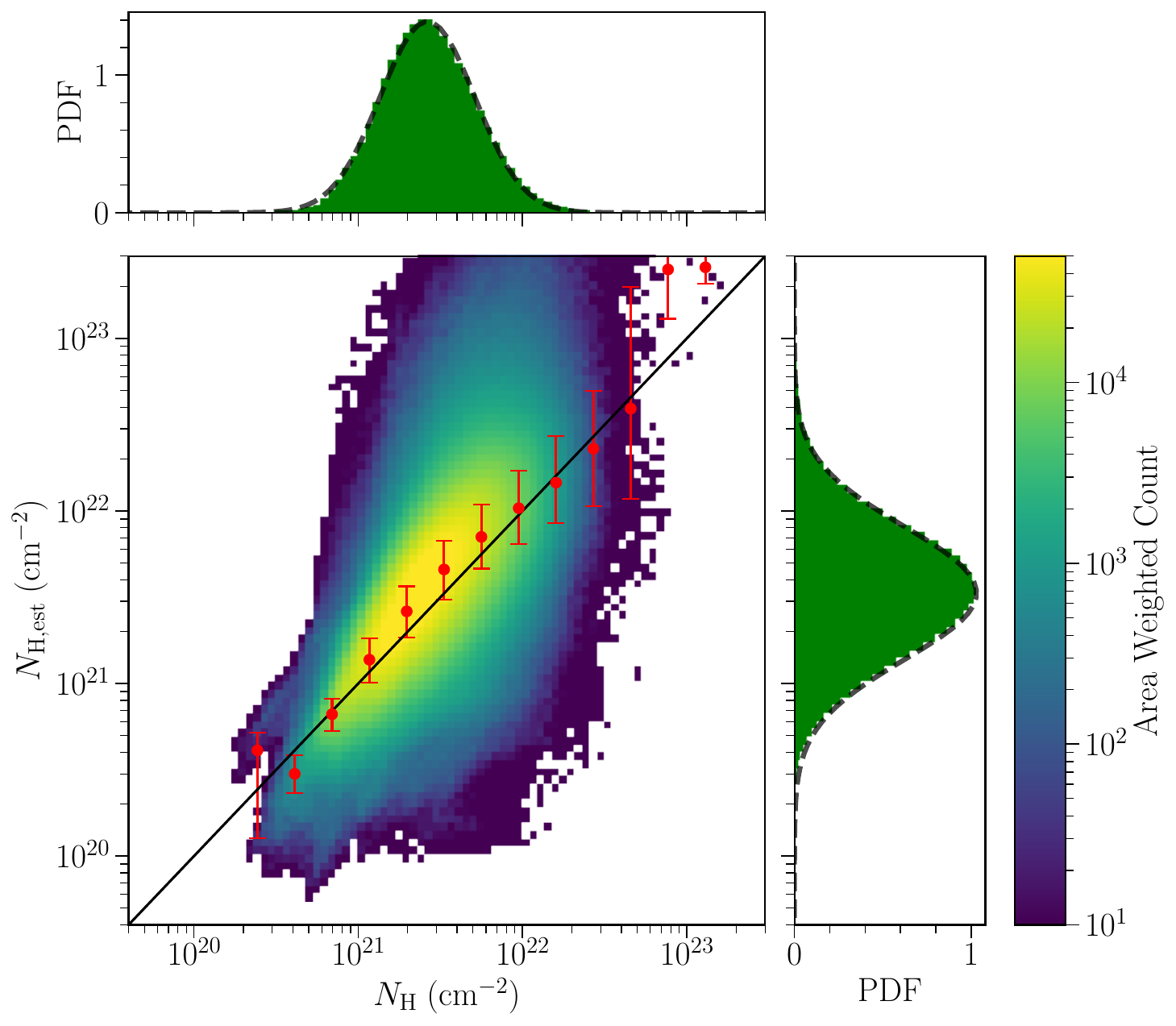}
    \caption{Same as \autoref{fig:heatingrate} but for the {\tt LGR4-2pc} simulation between $250$--$350 \Myr$.}
    \label{fig:heatingrate_LG}. 
\end{figure*}

More generally, in \autoref{fig:heatingrate} and \autoref{fig:heatingrate_LG} we show the distributions of gas column $N_\mathrm{H}$ vs. $\mathcal{N}_d/\phi_s$ and $N_\mathrm{H,\rm{est}}$, considering all times for both models.  We see that overall, both versions of the dust heating parameter follow the total gas column with an approximately linear relationship.  Below a column density of $2\times10^{20}$ cm$^{-2}$, the standard deviation in $\mathcal{N}_d/\phi_s$ at fixed $N_\mathrm{H}$ is only $\sim 10-30\%$ of the mean for {\tt R8-4pc}. This increases to much larger variance in $\mathcal{N}_d$ at high $N_\mathrm{H}$, however. We conclude that dust emission statistically traces gas column well, but it should be borne in mind that there will be significant scatter about the mean at high column. The scatter in $N_\mathrm{H,\rm{est}}$ is similar at low column, and reduced compared to $\mathcal{N}_d$ at high $N_\mathrm{H}$.

We fit a log-normal to the PDFs of $N_\mathrm{H}$, $\mathcal{N}_d/\phi_s$, and $N_\mathrm{H,\rm{est}}$. For {\tt R8-4pc}, the standard deviation in the distribution of $\ln(\mathcal{N}_d/\phi_s)$ is 0.94 for {\tt R8-4pc}, which is larger than that in $\ln(N_\mathrm{H})$ of 0.71. The standard deviation of $\ln(N_\mathrm{H,\rm{est}})$ is smaller at 0.67. For {\tt LGR4-2pc}, the standard deviations are 1.2, 0.66, and 0.88 for $\ln(\mathcal{N}_d/\phi_s)$, $\ln(N_\mathrm{H})$, and $\ln(N_\mathrm{H,\rm{est}})$ respectively. Thus, we expect that the varying radiation field will enhance the appearance of structure within the ISM based on dust maps, compared to what is present in the gas itself (if comparable resolution were available). Some of this excess variation can be reduced if a model radiation field is applied, as correction by our simple sum of sources model demonstrates. We note that since \autoref{eq:J_sum_source} uses the mean density in computing optical depth, this method would need to be applied iteratively, with the density obtained from the mean column density of the previous iteration and an estimate of the disk scale height.

Finally, we note that to determine the total dust emission at a given frequency, one would need to take into account optical (OPT), near-UV (NUV), and LyC radiation that emitting grains absorb, in addition to FUV.  Optical radiation from older stars is important in heating larger grains.  Because the older stellar population is relatively smooth, an approximate estimate of $J_\mathrm{OPT}$ may be able make use of the slab model outlined in \autoref{sec:slab} and \autoref{sec:slab_deriv}.

\section{LyC Radiation Field}\label{sec:LyC}

We now discuss the properties of the simulated LyC radiation field. Unlike the FUV radiation, these photons are strongly attenuated by neutral hydrogen and are responsible for ionization of the gas. Therefore, we will explore the ionization state of the \WIM{} in relation to the LyC radiation field. We will also compare the LyC radiation field in the simulations with approximate models.

\subsection{Ionization Equilibrium and the Ionization Parameter $U$}\label{sec:ion_param} 

We can determine the extent to which gas is in ionization equilibrium using the energy density of the ionizing radiation. The ionization parameter, $U$, is defined as:
\begin{equation}\label{eq:U_def}
U = \frac{n_{\rm ph}}{\nH} = \frac{\mathcal{E}_{\rm LyC}/(h \nu_{\rm LyC})}{ \nH},
\end{equation}
where $n_{\rm ph}$ is the number density of ionizing photons, $\nH$ is the total number density of hydrogen, $\mathcal{E}_{\rm LyC}$ is the energy density of the ionizing radiation field, and $h \nu_{\rm LyC}$ is the average photon energy of a single ionizing photon (a blackbody with $T = 3.5\times 10^4 \Kel$ has $h\nu_{\rm LyC} = 18\eV$ and we adopt this as a typical value). In ionization-recombination equilibrium, we would expect 
\begin{equation}
 \alpha_B n_e n_{\rm H^+} =
 n_\mathrm{HI} \sigma_\mathrm{pi} \frac{c\mathcal{E}_{\rm LyC}}{h \nu_\mathrm{LyC}} = n_\mathrm{HI}\zeta_{\rm pi}
\label{eq:equil}
\end{equation}
where $\alpha_B \approx 3.12 \times 10^{-13}(T/8000\Kel)^{-0.83} \cm^{3}\,\second^{-1}$ is the case B recombination rate coefficient, $\sigma_\mathrm{pi}$ is the photoionization cross section, and $\zeta_{\rm pi}$ is the photoionization rate (the probability per unit time of photoionization of a hydrogen atom) \citep{kim2023photchem}. This ignores collisional ionization and cosmic ray ionization, which are unimportant unless the LyC radiation field is very weak, and grain-assisted recombination, which is insignificant in warm ionized gas \citep{Weingartner2001recomb}. Assuming that the molecular fraction is small so that all hydrogen is either \ion{H}{1} or \ion{H}{2} ($x_n = x_{\rm HI} \approx 1- \xHII$) and that most free electrons come from hydrogen ($x_{e} \approx x_{\rm H^+}$), this can be rewritten in terms of $x_\mathrm{n}$ and $U$:
\begin{equation}
\frac{(1 - x_\mathrm{n})^2}{x_\mathrm{n}} \approx \frac{\xHII x_e}{x_{\rm HI}} = \frac{U}{U_\mathrm{crit, pi}} \,.
\label{eq:ion_eq}
\end{equation}
Here,
\begin{multline}
U_\mathrm{crit, pi} \equiv \frac{\alpha_B}{\sigma_\mathrm{pi}c} = 3.34\times 10^{-6} \\
\times 
\bigg(\frac{\alpha_B}{3\times 10^{-13}\textrm{ cm}^3\textrm{ s}^{-1}}\bigg)\bigg(\frac{\sigma_\mathrm{pi}}{3\times 10^{-18}\textrm{ cm}^2}\bigg)^{-1}
\label{eq:Ucrit}
\end{multline}
is the critical ionization parameter defined such that the warm ionized gas would be fully ionized ($x_\mathrm{n} \ll 1$) as long as $U \gg U_{\rm crit,pi}$. If $U/U_{\rm crit,pi}=1/2$, then $x_\mathrm{n} = 1/2$.

\begin{figure*}
    \centering
	\includegraphics[scale = 0.4]{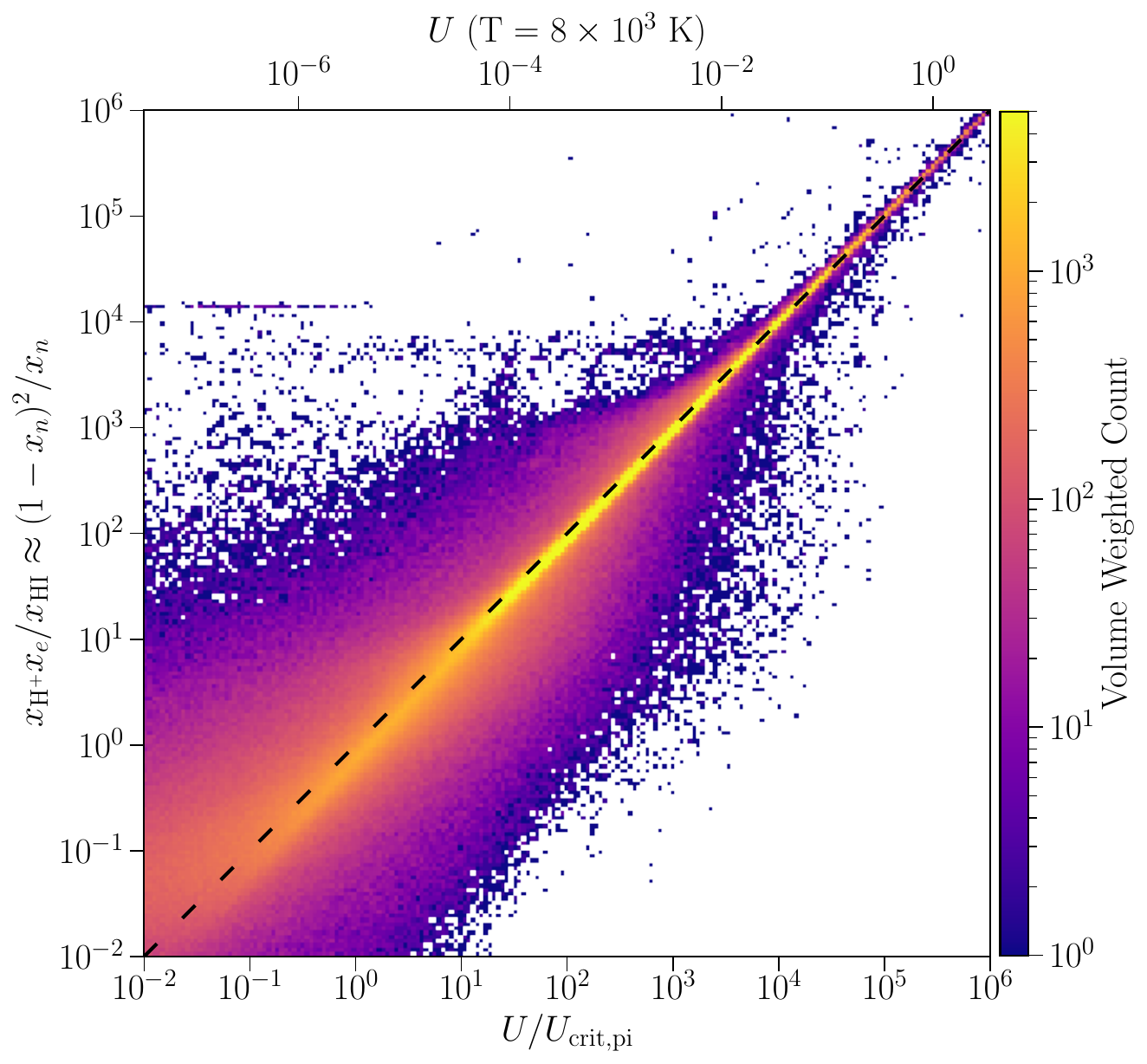}
         \includegraphics[scale = 0.4]{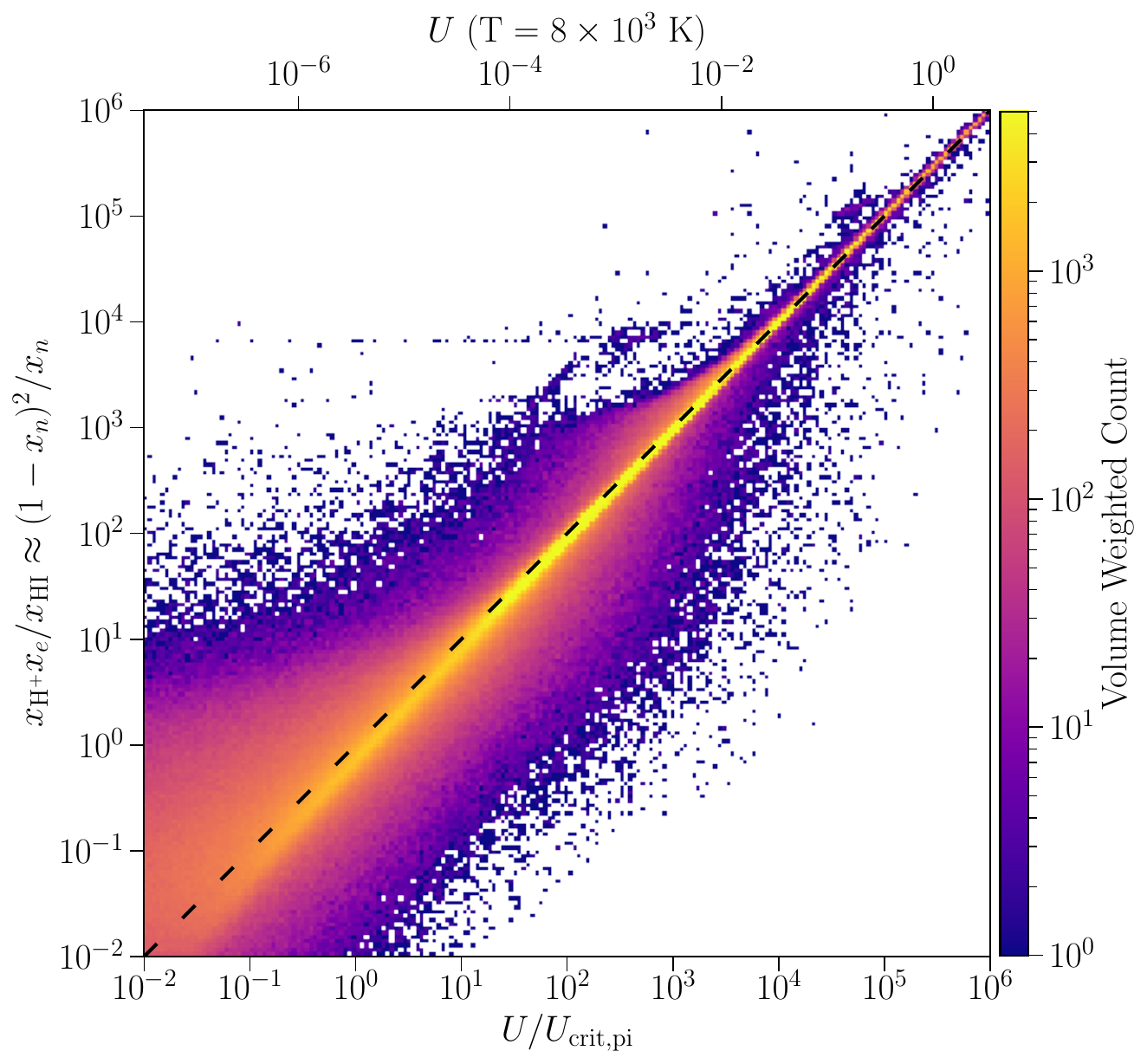}
    \caption{Distribution from {\tt R8-4pc} (left) and {\tt LGR4-2pc} (right) of volume as functions of $U/U_{\rm crit,pi}$ and $\xHII x_e/x_{\rm HI}$ (\autoref{eq:ion_eq}). This includes all warm gas (between $6\times10^3$ and $3.5\times10^4$ K) with non-zero $\mathcal{E}_{\rm LyC}$ and within $|z| < 4$ pc or $|z| < 2$ pc respectively. $U_\mathrm{crit,pi}$ is defined in \autoref{eq:Ucrit}. The black dashed line represents photoionization equilibrium (see  \autoref{eq:ion_eq}); most locations with $\xHII x_e/x_{\rm HI}> 10$ ($x_n \lesssim 0.1$) are consistent with this. Top axes indicate the ionization parameter $U$ for $T=8000 \Kel$.}
    \label{fig:ion_eq}
\end{figure*}

In \autoref{fig:ion_eq}, we present a two-dimensional histogram of the ionization parameter, $U/U_\mathrm{crit,pi}$, in relation to $\xHII x_e/x_{\rm HI}$. We show the total distribution for both models across time, restricting to gas at the midplane. We include only warm gas with temperature between $T=6 \times 10^3$ and $3.5 \times 10^4 \Kel$, and further limit to gas with $\mathcal{E}_{\rm LyC} > 0$. With this distribution, we include a black dashed line defined by \autoref{eq:ion_eq}. For $\xHII x_e/x_{\rm HI}>10$ (corresponding to $x_n \lesssim 0.1$), the density peak in the histogram closely follows this line and we conclude that the gas near the midplane is close to ionization-recombination equilibrium.

The absorption rate of LyC photons by dust is $\sigma_d \nH c\mathcal{E}_{\rm LyC}/(h \nu_\mathrm{LyC})$, so we can define a second critical ionization parameter as the maximum value such that the photoionization rate exceeds the dust absorption rate.  For well-ionized gas, this is 
\begin{multline}
U_\mathrm{crit,d} \equiv  \frac{\alpha_B}{\sigma_{\rm d} c} = 0.01 \\
\times 
\bigg(\frac{\alpha_B}{3\times 10^{-13}\textrm{ cm}^3\textrm{ s}^{-1}}\bigg)\bigg(\frac{\sigma_\mathrm{d}}{10^{-21}\textrm{ cm}^2}\bigg)^{-1}.
\label{eq:Ucrit_d}
\end{multline}
For $U_\mathrm{crit,pi} < U < U_\mathrm{crit,d}$, the gas will be well ionized and most of the ionizing photons will be absorbed by hydrogen rather than dust.

\begin{figure*}
    \centering
	\includegraphics[scale = 0.32]{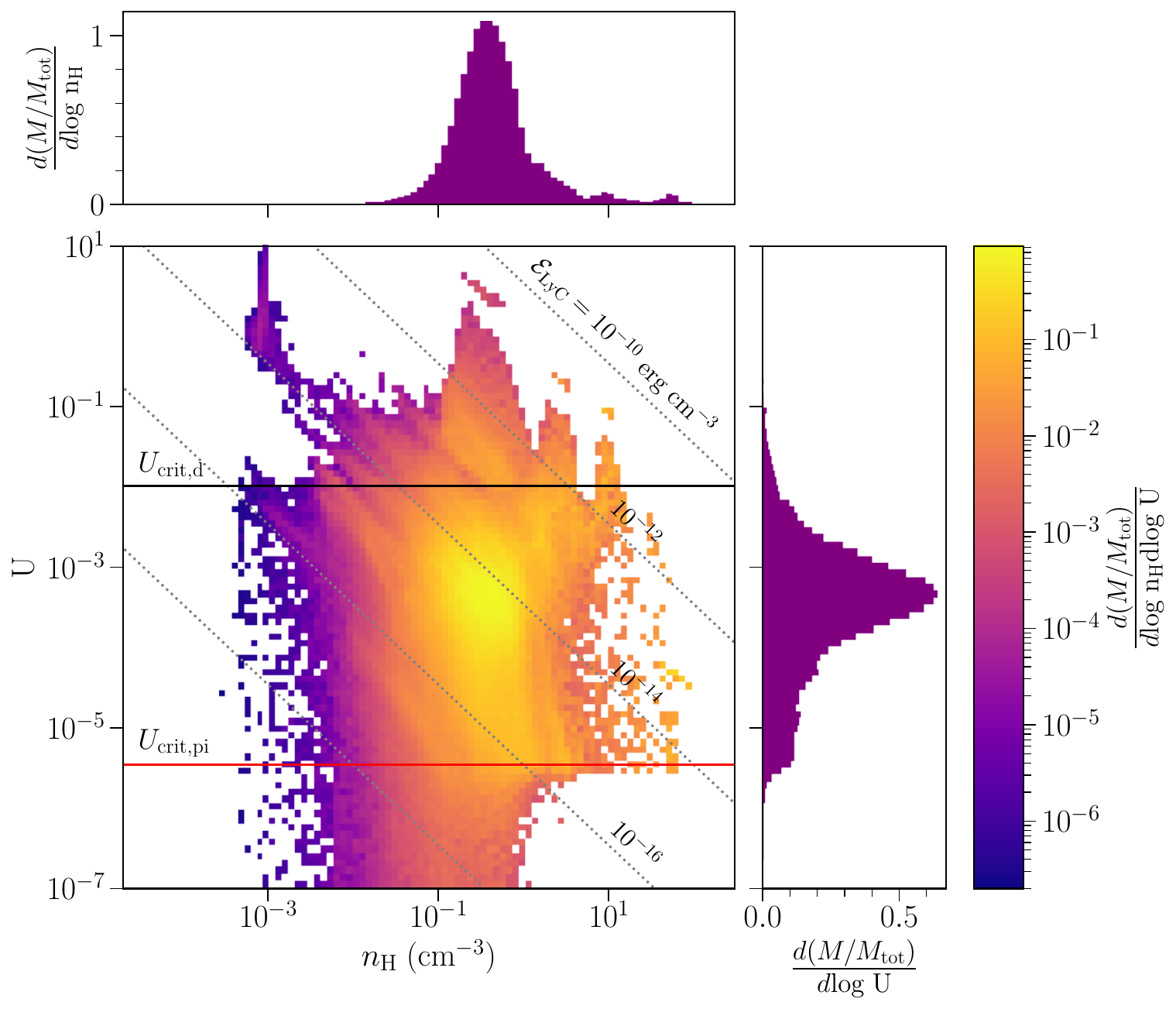
}
\includegraphics[scale = 0.32]{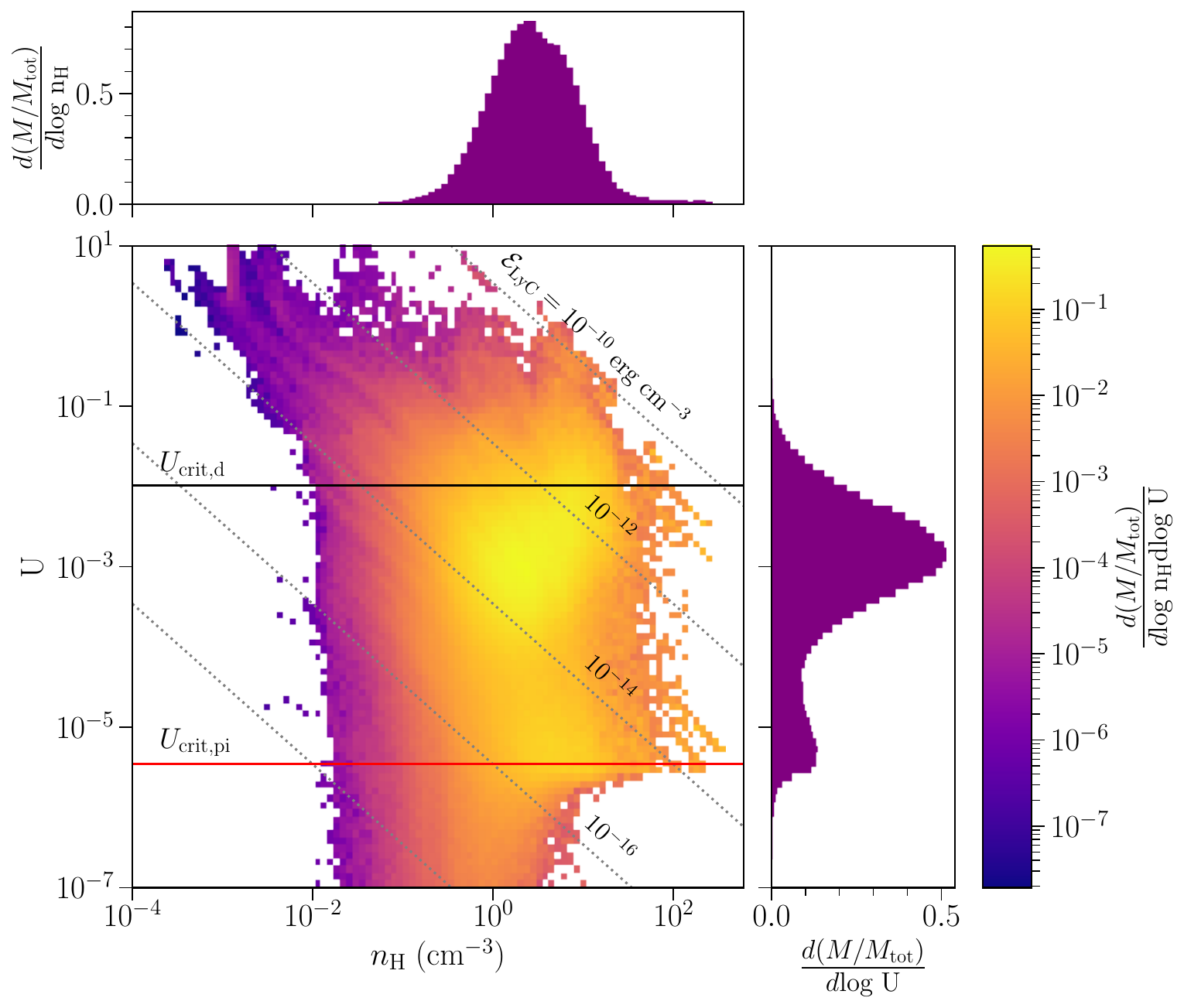}
   \caption{The mass weighted distribution of ionization parameter $U$ and gas density $\nH$ for all \WIM\ at $|z|<300 \pc$. The left panel shows the {\tt R8-4pc} simulation at $t = 430 \Myr$. The right panel shows the {\tt LGR4-2pc} at $t = 298 \Myr$. The top and side panels show the projection of this distribution into distributions of $U$ and $\nH$ alone. Dotted diagonal lines show the corresponding value of the LyC energy density at each gas density under the assumption of ionization equilibrium (\autoref{eq:equil}). Below $U_\mathrm{crit,d}$,  absorption by neutral hydrogen exceeds dust absorption (see \autoref{eq:Ucrit_d}). The black and red horizontal lines show the critical ionization parameters for $\alpha_{\rm B}=3 \times 10^{-13}\cm^{3} \second^{-1}$ (\autoref{eq:Ucrit} and \autoref{eq:Ucrit_d}).}
    \label{fig:U_nH}
\end{figure*}

\begin{figure*}
    \centering
	\includegraphics[scale = 0.32]{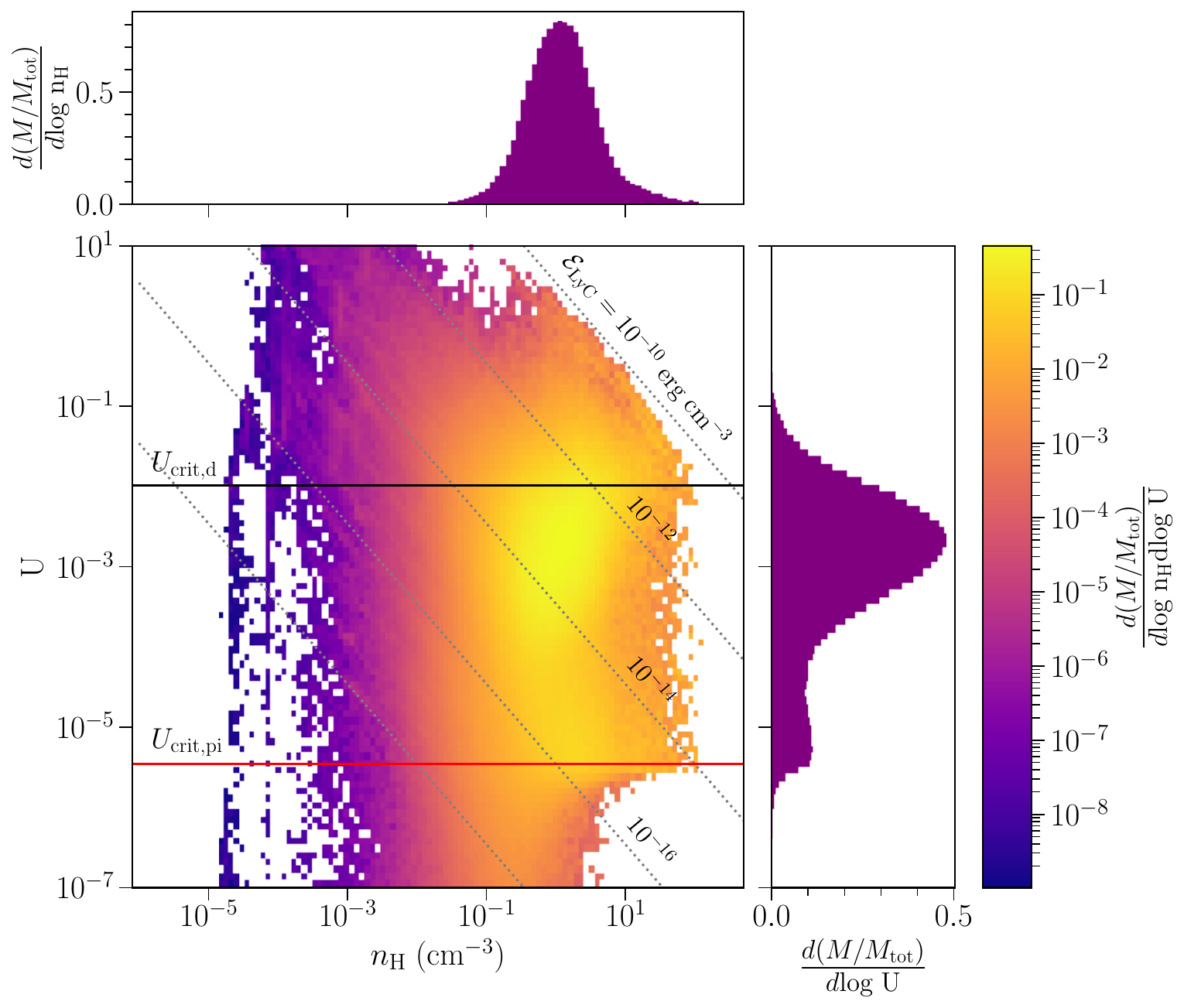}
 \includegraphics[scale = 0.32]{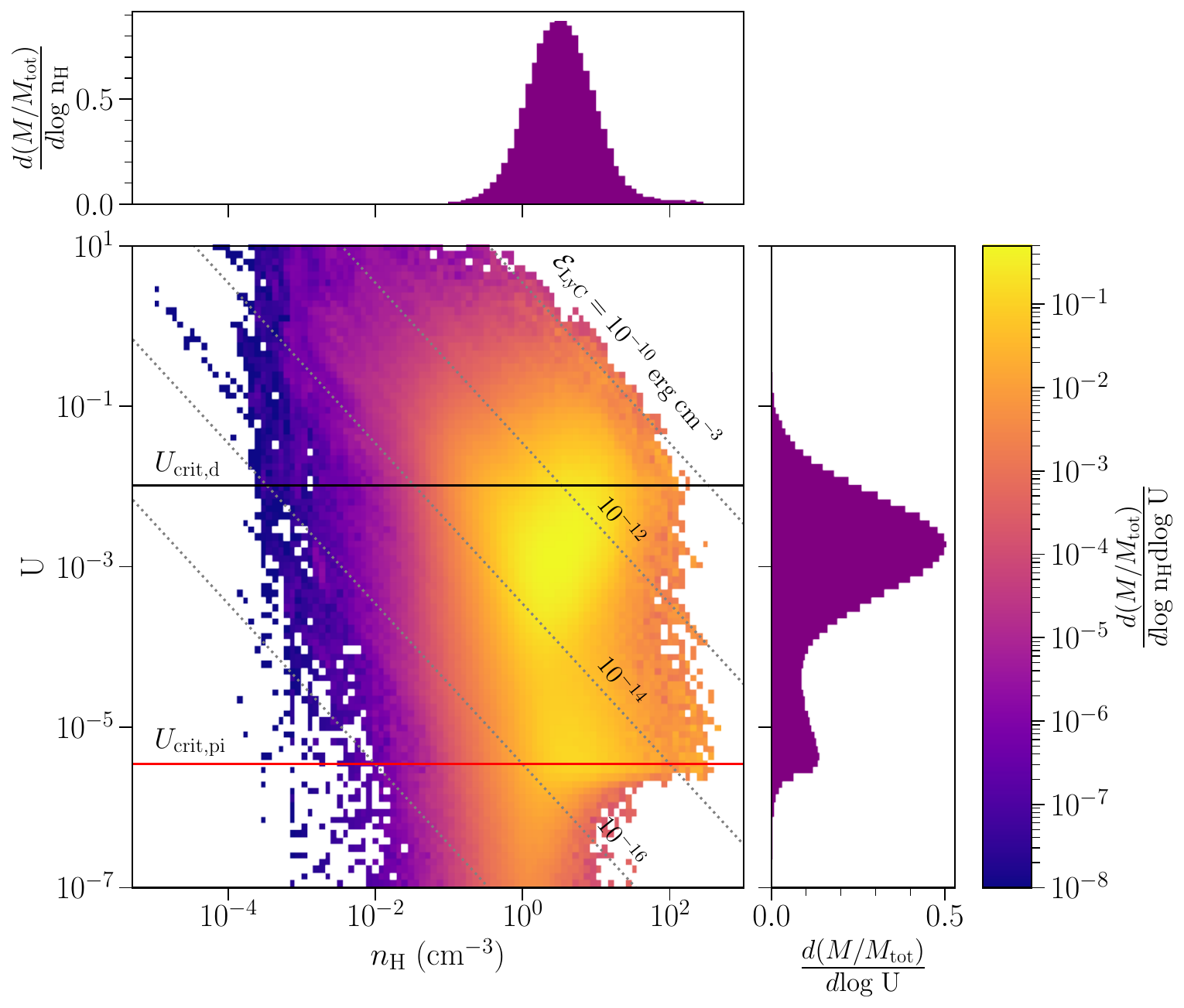}
   \caption{The same as \autoref{fig:U_nH} but for midplane \WIM\ across all snapshots. The left panel shows the {\tt R8-4pc} simulation for $t=250$--$450\Myr$. The right panel shows the {\tt LGR4-2pc} simulation for $t= 250$--$350\Myr$. The gas is limited to that within $|z|<4 \pc$ or $|z|<2 \pc$ respectively.}
    \label{fig:all_U_nH}
\end{figure*}

In \autoref{fig:U_nH} we show the two-dimensional distributions of $U$ and $\nH$ at $t = 430 \Myr$ for the {\tt R8-4pc} simulation and $t = 298 \Myr$ for {\tt LGR4-2pc}. We include only the \WIM\ within $|z|<300\pc$. Each cell is weighted by mass. \autoref{fig:all_U_nH} shows the same values for each model but includes all snapshots, while being limited to gas at the midplane ($|z|<4\pc$ or $|z|<2\pc$ respectively). All of the distributions have a single primary peak, as well as an extension to lower value of $U$.  Physically, this extension represents the edges of ionized regions, where the radiation field drops off sharply and gas becomes neutral.

The red and black horizontal lines in these figures represent $U_\mathrm{crit,pi}$ (\autoref{eq:Ucrit}) and $U_\mathrm{crit,d}$ (\autoref{eq:Ucrit_d}), respectively. Most of the distribution falls between the two critical ionization parameters, meaning that it is well ionized and that the photoionization rate exceeds the dust absorption rate.

The diagonal lines represent constant values of the LyC energy density $\mathcal{E}_{\rm LyC}$ for a given photon energy (18 eV) (see \autoref{eq:equil}). Note that for $\mathcal{E}_{\rm LyC}=10^{-14}\,{\rm erg}\,\cm^{-3}$ and $\sigma_\mathrm{pi} = 3\times 10^{-18}\cm^{2}$, the corresponding photoionization rate is $\zeta_{\rm pi} = 3.1 \times 10^{-11}\second^{-1}$. Along each axis, we show projections of the two-dimensional distribution into distributions of $\nH$ and $U$ alone. In \autoref{fig:U_nH}, the overall peaks of the distributions occur at $U\sim 4\times 10^{-4}$ and $\nH\sim 0.4$ cm$^{-3}$ in the {\tt R8-4pc} snapshot and $U\sim 1\times 10^{-3}$ and $\nH\sim 2.2$ cm$^{-3}$ in the {\tt LGR4-2pc} snapshot. 

Based on the theory of pressure-regulated, feedback-modulated star formation \citep[see][and references therein]{OK22}, we expect thermal pressure and therefore the density of the warm (WNM and WIM) gas, $\nH$, to vary approximately linearly with $\Sigma_{\rm SFR}$ when comparing different galactic environments. Additionally, we would expect a roughly linear dependence of $\mathcal{E}_{\rm LyC}$ on $\Sigma_{\rm SFR}$. This argues that the mean value of $U \propto \mathcal{E}_{\rm LyC}/\nH$  (\autoref{eq:U_def}) would be independent of galactic environment. \autoref{fig:all_U_nH} confirms that the peak of the distributions, based on all snapshots, is at $U \approx 2\times10^{-3}$ in both the {\tt R8-4pc} and {\tt LGR4-2pc} models. We can understand the typical value of $U$ in {\tt R8-4pc} based on the photon rate per unit area, $\Phi_{\rm LyC} \sim 4 \times 10^{50} \second^{-1} \kpc^{-2}$ from \autoref{tab:quantities}, and the typical density in the warm gas at the midplane, $\bar{n}_H\sim 0.4$. With $l_*^{-2}$ the number of sources per unit area, we can estimate the mean photon rate per source to be $L_*/(h\nu) = \Phi_{\rm LyC} l_*^{2}$. If the nearest source is at a distance of $\sim l_*/2$, the photon flux is roughly $F = L_*/(4\pi(l_*/2)^2)=\Phi_{\rm LyC}/\pi$. Therefore, the photon number density is $n_{\rm ph} = F/c \approx \Phi_{\rm LyC} /(\pi c)$. We can then estimate $U=n_{\rm ph}/\nH \sim \Phi_{\rm LyC} /(\pi c \nH) \sim 10^{-3}$. A similar calculation in {\tt LGR4-2pc} would lead to the prediction $U \sim \Phi_{\rm LyC} /(\pi c \nH) \sim 3\times 10^{-3}$.

Along any given line of sight, the total optical depth in the LyC band can be written as
\begin{eqnarray}
    \tau_{\rm LyC} &=& \int \nH \left(\sigma_{\rm d}   + x_n \sigma_\mathrm{pi}\right) ds \nonumber \\
    &=& \sigma_\mathrm{pi} \int \nH \frac{U_\mathrm{crit,pi}}{U}\left(\frac{U}{U_\mathrm{crit,d}}   +  1\right) ds. 
    \label{eq:tauLyC}
\end{eqnarray}
In the second line we assume that ionization equilibrium holds with $U/U_\mathrm{crit,pi} > 1$ so $x_e\approx 1$.  \autoref{eq:tauLyC} shows that provided that $U/U_\mathrm{crit,d}<1$, the dust absorption term may be neglected, i.e. $ \tau_{\rm LyC} \approx \sigma_\mathrm{pi} \int \nH x_n ds$. \autoref{eq:tauLyC}  also allows us to write the mean free path for LyC photons within photoionized gas as
\begin{eqnarray}
\ell_\mathrm{LyC} &=& \frac{1}{\sigma_\mathrm{pi}\langle\nH  U_\mathrm{crit,pi}/U\rangle}=\frac{c/\alpha_B}{ \langle \nH/{U}\rangle }\nonumber\\
&\sim & 100\pc \left\langle \left(\dfrac{\nH}{0.5\,{\rm cm}^{-3}}\right) \left( \dfrac{2\times 10^{-3}}{U}\right) \right\rangle^{-1},
\end{eqnarray}
where the angle brackets indicate a volume-weighted average, and in the second line we have normalized $\nH$ and $U$ to typical diffuse ISM values for solar neighborhood conditions.  This shows that within diffuse ionized gas (DIG), LyC photons are expected to be able to traverse distances approaching the disk scale height -- while still not photoionizing the whole ISM, which would require larger $U$. At large $|z|$, the density drops and the ionization parameter remains roughly constant which would lead to an increase in the mean free path.  Given that a significant volume fraction in the ISM is occupied by hot gas with which the LyC photons do not interact, we can understand how LyC radiation escapes from near the disk midplane and is able to maintain an extended  photoionized layer at high altitude \citep[see also][]{kado2020diffuse}.

\subsection{LyC Sum of Sources Model}
\label{sec:atten_ion}

In this section, we compare a simple model to the solution of the ray-tracing for the LyC radiation. The model we explore is similar to that described in \autoref{sec:FUV_source} for FUV radiation, and also to the model presented in \cite{belfiore2022tale} with some modifications. As in these models, we estimate the LyC radiation field at each point by summing the contribution from each of the source particles. For the LyC radiation, we must include the absorption by both neutral hydrogen and dust. The angle-averaged estimated radiation field at each point is then described by
\begin{equation}
J_\mathrm{LyC,src} = \frac{1}{4\pi}\sum_\mathrm{sources\ i} \frac{L_i}{4 \pi r_i^2}e^{- r_i \langle \nH \rangle (\sigma_\mathrm{pi}x_{\rm n, eff} + \sigma_\mathrm{d})}.
\label{eq:belf_J}
\end{equation}
In this expression, $L_i$ is the LyC luminosity of source $i$ and $r_i$ is the distance from the source to a given point. As in \autoref{sec:FUV_source}, we sum over sources in a horizontally extended domain with positions determined by shearing-periodic boundary conditions. In the exponent, the average density is set equal to total mean hydrogen number density within one scale height of the midplane for each snapshot. Attenuation comes from both dust, represented with the cross section $\sigma_\mathrm{d}$, and photoionization, represented by the product of the cross section  $\sigma_\mathrm{pi}$ and the effective neutral fraction, $x_{\rm n, eff}$, since only neutral hydrogen absorbs photons. All values in this expression are taken directly from the simulation outputs except for $x_{\rm n, eff}$, which is left as a free parameter. This can be used to fit the model to the simulation.

After calculating $J_\mathrm{LyC,src}$ at each point with \autoref{eq:belf_J}, we want to use this to create an observable; since LyC produces ionized gas, the most natural observable is the emission measure, most commonly obtained from H$\alpha$. In terms  of the electron and proton number density, the emission measure is calculated as
\begin{equation}
{\rm EM} = \int_{-H}^H n_e n_{\rm H^+} dz.
\label{eq:EM1}
\end{equation}
This expression can be used to calculate EM directly from the simulation, in which both $n_e$ and $n_{\rm H^+}$ are known at each point based on the ray-tracing and (time-dependent) photochemistry. For the model, we treat $n_e n_{\rm H^+}$ as unknown. We rewrite \autoref{eq:EM1}  under the assumption that the gas is in ionization equilibrium, which from \autoref{eq:equil} yields 
\begin{equation}
{\rm EM} = \frac{4\pi\sigma_\mathrm{pi}}{\alpha_B {h\nu}_\mathrm{LyC}} \int_{-H}^H \nH   x_{\rm n,eff}  J_\mathrm{LyC} dz. 
\label{eq:EM2}
\end{equation}
In \autoref{eq:EM2}, we use the constant value $\langle \nH \rangle$ for $\nH$;  $x_{\rm n, eff}$  is also treated as a constant (this is the fitting parameter of the model), and we use  \autoref{eq:belf_J} for $J_\mathrm{LyC}$. The value of $\alpha_B$ is for $T = 8000$ K. We restrict our model to within one scale height of the midplane.

For any given simulation snapshot, for both the simulation and the model, we find EM at each $(x,y)$ coordinate using  \autoref{eq:EM1} and \autoref{eq:EM2} respectively. We find the distributions of EM across all cells, then compare the model and simulation distributions $f({\rm EM})$ for different values of the fitting parameter $x_\mathrm{n,eff}$. To compare the model and data, we define an error function

\begin{equation}
\textrm{error} = \sqrt{\frac{\sum_i (f_{\textrm{model}}(\textrm{EM}_i)-f_{\textrm{sim}}(\textrm{EM}_i))^2\textrm{EM}_i}{\sum_i \textrm{EM}_i}}
\label{eq:belf_err}
\end{equation}
where the sum is over bins of EM in log space. 

We omit points within a projected distance of 50 pc from each source, as an approximate approach to masking out classical \ion{H}{2} regions. We compare the distributions of EM rather than comparing cell by cell so that differences in regions with very low EM do not dominate. Our method weights the error towards high EM so that the model fits the brightest regions best. By varying $x_\mathrm{n,eff}$ we can then find the model which best fits the simulation.

The model in \cite{belfiore2022tale} is similar to ours, except that rather than $\langle \nH \rangle (\sigma_\mathrm{pi} x_\mathrm{n,eff} + \sigma_\mathrm{d})$ in the exponent of \autoref{eq:belf_J}, they assume a single galaxy-wide constant $k_0$ (which has units 1/distance), and in computing the EM they adopt an overall scaling factor for the product of the disk thickness and the ratio $f_\mathrm{esc}/(1-f_\mathrm{esc})$, which is normalized from the data; this $f_\mathrm{esc}$ represents the escape fraction from \ion{H}{2} regions.  They use their model to compute a predicted H$\alpha$ emission per unit area, which is proportional to EM (they note that their model does not   correct for extinction in the diffuse gas).  The overall best-fit value of $k_0$ is obtained by taking a ratio of model H$\alpha$ to data H$\alpha$ (which does not correct for extinction), after masking the \ion{H}{2} regions. While they present the ratio of model to data as a function of H$\alpha$ emission, they do not weight by emission in selecting the best-fit value of $k_0$.  

\begin{figure*}
    \centering
	\includegraphics[scale = 0.5]{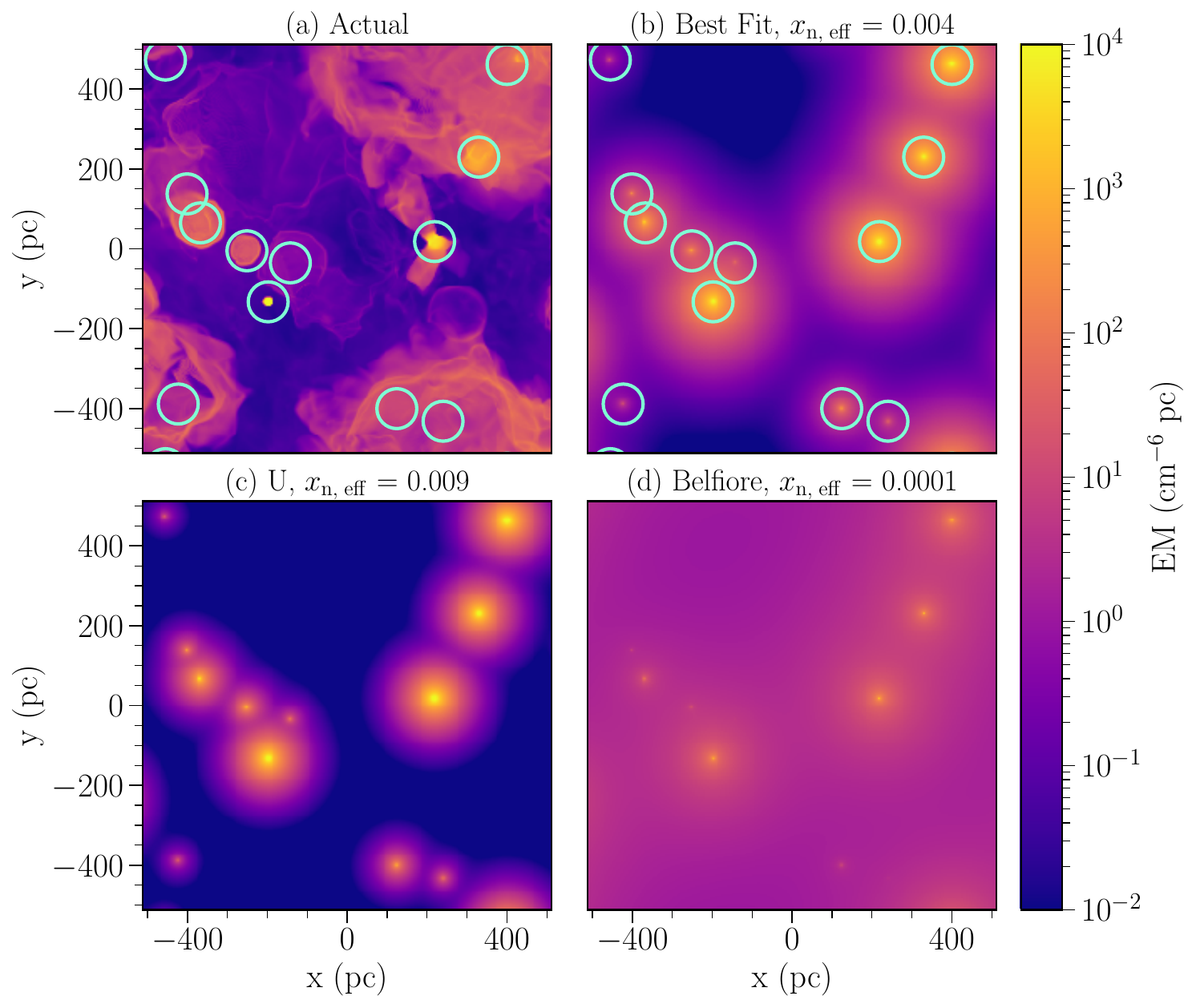}
    \caption{Projection on the disk midplane of the EM from the simulation at t = 430 Myr, panel (a), in comparison to model EM determined from the attenuated ionizing multi-source model, using different values of $x_\mathrm{n,eff}$ in \autoref{eq:belf_J} and \autoref{eq:EM2}. In panel (b), we show the best fit value based on minimizing the difference in EM as illustrated in \autoref{fig:belf_err} ($x_\mathrm{n,eff}=0.004$). In the first two panels, we include blue circles representing the regions around each source which are masked for the fit. The value based on the peak of the emission pdf in $U$ and $\nH$ in \autoref{fig:U_nH} ($x_\mathrm{n,eff}=0.009$) is shown in panel (c). A value consistent with a similar modeling approach reported in \cite{belfiore2022tale} ($x_\mathrm{n,eff}=0.0001$) is shown in panel (d).}
    \label{fig:belf}
\end{figure*}

\begin{figure}
    \centering
	\includegraphics[scale = 0.4]{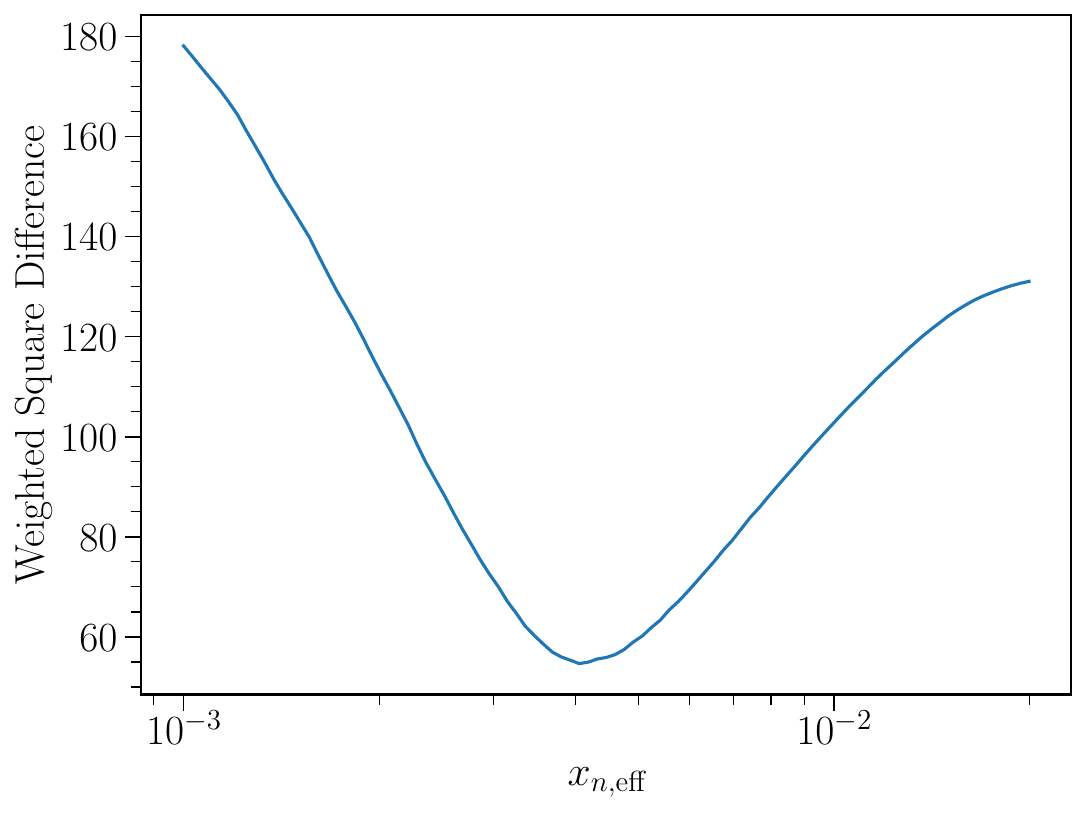}
    \caption{Weighted square difference (\autoref{eq:belf_err}), summed over the map, between the simulation value of EM and the EM calculated with the attenuated ionizing multi-source model, for different values of the parameter $x_n$ in \autoref{eq:belf_J} and \autoref{eq:EM2}. The snapshot is at $t = 430 \Myr$. This allows us to identify the best-fit value as $x_\mathrm{n,eff}=0.004$.  See text in \autoref{sec:atten_ion} for model details.
    }
    \label{fig:belf_err}
\end{figure}

The comparison of the simulated EM to the best fit model for one snapshot from {\tt R8-4pc} is shown in \autoref{fig:belf}. The first panel (a) represents the map of EM determined from the simulated values of $n_e$ and $n_{\rm H^+}$. The second panel (b) represents the modeled EM with the best fit value of $x_\mathrm{n,eff}=0.004$,
which minimizes the error for this snapshot (see 
\autoref{fig:belf_err}). In the modeled EM, contributions from individual sources
are spherically symmetric, unlike in the simulated EM where complex structures are visible. For example, the large-scale moderate-brightness regions that appear in panel (a) are generally associated with photoionized gas on the inner surfaces of hot bubbles; these hot bubbles can be seen in the midplane slices of \autoref{fig:snap_h}.

In \autoref{fig:belf}, we also include comparisons using values of $x_\mathrm{n,eff}$ obtained in alternate ways. One method to find $x_\mathrm{n,eff}$ is by using the ionization parameter, $U$, through \autoref{eq:ion_eq}. As $x_\mathrm{n,eff}$ is small, we find $x_\mathrm{n,eff} \approx U_\mathrm{crit, pi}/U$. The motivation for this approach is that $U$ can be obtained in observations from line ratios.  For a given snapshot, we find the distribution of ionization parameter including only the \WIM\ (as in \autoref{fig:U_nH}). We take the mass weighted median value of $U$ and use this to calculate $x_\mathrm{n,eff}$. This value can be applied in \autoref{eq:belf_J} as before to obtain $J_\mathrm{LyC,src}$, and then the distribution of $J_\mathrm{LyC,src}$ is used  to compute the EM map from \autoref{eq:EM2}. For the snapshot shown, the median $U=4\times10^{-4}$ (from the distribution shown in \autoref{fig:all_U_nH})  corresponds to $x_\mathrm{n,eff}=0.009$, similar to the value we obtained through fitting the model to the data. The model using the value of $x_{\rm n,eff}$ derived from the ionization parameter is shown in the third panel (c) of \autoref{fig:belf}.  

Finally, we include an EM map using the results from \cite{belfiore2022tale}. They found a typical $k_0=0.52\ \mathrm{kpc}^{-1}$ (i.e. mean free path of 1.9 kpc), which for our mean density of $\langle \nH \rangle = 0.6\ {\rm cm}^{-3}$ in the snapshot  shown translates to $x_\mathrm{n,eff} = 0.52\, {\rm kpc}^{-1} / (\langle \nH \rangle \sigma_\mathrm{pi})\approx0.0001$.  The EM map using this value of $x_{n,\rm{eff}}$ is shown in panel (d) of \autoref{fig:belf}.  Because the constant in the attenuation law is nearly two orders of magnitude lower than our best fit or the $U$-derived map, the appearance is very different. There is significantly more diffuse emission because attenuation by neutral hydrogen is underestimated.  

\begin{figure*}
    \centering
	\includegraphics[scale = 0.53]{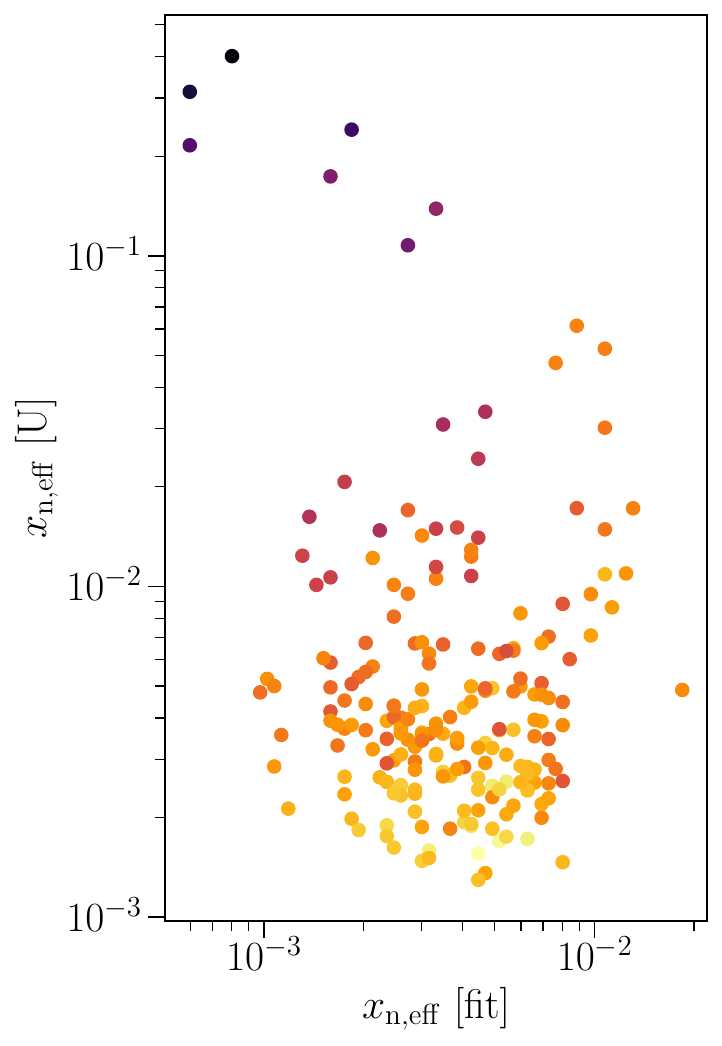}
	\includegraphics[scale = 0.53]{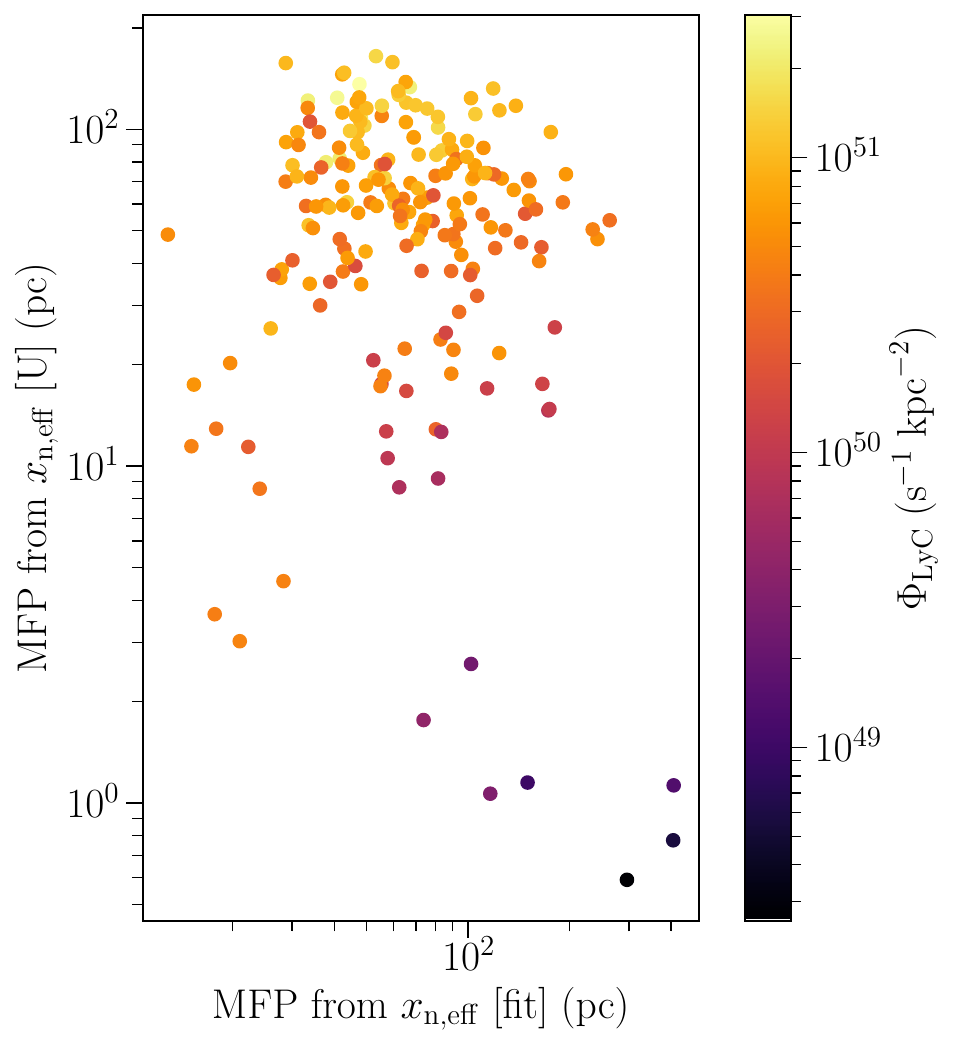}
    \caption{
    For 200 different snapshots of model {\tt R8-8pc}, the {\it left panel} shows a comparison of $x_\mathrm{n,eff}$ from minimizing the difference between EM from the simulation and EM calculated via the attenuated ionizing multi-source model ($x_\mathrm{n,eff}\,[{\rm fit}]$; see \autoref{sec:atten_ion}), and the most likely value of $x_\mathrm{n,eff}$ based on the ionization parameter and density distribution ($x_\mathrm{n,eff}\,[U]$; see \autoref{sec:ion_param}). The {\it right panel} shows the corresponding mean free path of ionizing photons $\ell_\mathrm{LyC} = (\langle \nH \rangle \sigma_\mathrm{pi} x_\mathrm{n})^{-1}$ for each value of $x_\mathrm{n,eff}\,[{\rm fit}]$ or $x_\mathrm{n,eff}\,[U]$.}
    \label{fig:xn_mfp}
\end{figure*}

Although the values of $x_\mathrm{n,eff}$ from both the best fit of the model and as determined with $U$ are similar, they do not completely agree. We repeat this process for many simulation snapshots of the {\tt R8-8pc} model to find the distribution of $x_\mathrm{n,eff}$. We find that the two methods do not give values of $x_\mathrm{n,eff}$ that are well correlated, but both fall in a limited range. The fitted value is in the range $x_\mathrm{n,eff}=10^{-3}$--$10^{-2}$ for almost all snapshots, as shown in the left panel of \autoref{fig:xn_mfp}. Most values derived from $U$ are in the same range, although some are as large as $\sim10^{-1}$. These snapshots tend to have low values of $\Phi_{\rm LyC}$ and very little \WIM\ gas. The corresponding mean free path for ionizing radiation is $\ell_\mathrm{LyC}= (\sigma_\mathrm{pi} \langle \nH \rangle x_\mathrm{n,eff})^{-1}$, which we show in the lower panel of \autoref{fig:xn_mfp}. The mean free path from our fits is $\ell_\mathrm{LyC}\sim 20-300$ pc.

As a comparison, we apply this model to the {\tt LGR4-2pc} model at t = 298 Myr. We cannot mask 50 pc around each source in this case because the majority of the area is removed. Without the mask, we find $x_\mathrm{n,eff} = 0.002$. This is of the same magnitude as the value of $x_\mathrm{n,eff}$ from the {\tt R8-4pc} model. The estimated value of $x_\mathrm{n,eff}$ from the ionization parameter, $U$, is also similar between the two models. For the {\tt LGR4-2pc} snapshot we find $x_\mathrm{n,eff} = 0.003$. We expect the ionization parameter to be relatively insensitive to galactic environmental conditions, as discussed in \autoref{sec:ion_param}, and therefore the value of $x_\mathrm{n,eff}$ estimated from U would be similar between the two models.

We note that the smaller attenuation constant (or larger mean free path) obtained by \cite{belfiore2022tale} would correspond to an order of magnitude higher ionization parameter than we typically find in the {\tt R8-4pc} model.  One possibility is that the parameter regime for their sample of galaxies is simply somewhat different from what we are exploring in our simulations. However, as noted above, we expect similar values of $U$ and $x_\mathrm{n,eff}$ in different environments. For a given $x_\mathrm{n,eff}$, we would expect shorter, rather than longer mean free path for LyC photons in environments at higher density, which may be more similar to those explored by \cite{belfiore2022tale}.   
Another alternative is that the details of our respective fitting methods are responsible for the differences in our conclusions.  In particular, it is interesting that when \cite{belfiore2022tale} set $k_0=10\ \mathrm{kpc}^{-1}$ (similar to our best-fit value for $\langle \nH \rangle \sigma_\mathrm{pi} x_\mathrm{n,eff}$), their Fig. 6 shows the model is quite close to the data relatively near spiral arms where \ion{H}{2} regions are concentrated, with a departure only in very low density regions.  For their preferred value, $k_0\sim 1 \mathrm{kpc}^{-1}$, the model exceeds the data relatively near to spiral arms, indicating that radiation is insufficiently attenuated. It is possible that if their fit had given more weight to the regions with higher EM, as we do, they would have obtained more similar results to ours.  

\section{Summary}\label{sec:summary}

In this work, we analyze  characteristics of the FUV and LyC radiation fields in TIGRESS-NCR simulations with ray tracing and explore applications of our results to observations and other theoretical modeling.  We focus on one model with conditions similar to the solar neighborhood and another similar to the inner galaxy.

Key results are as follows:

\begin{itemize}
    \item The time-averaged midplane FUV radiation intensity for our solar neighborhood model is very similar to observational estimates for the local Milky Way.  For both the solar neighborhood and inner galaxy models, the FUV intensity decreases exponentially with distance from the midplane.   
    \item The statistical distribution of the FUV intensity is well characterized by a model similar to that presented in \citet{draine2007a}. This combines a strong power law component near sources with a diffuse component, which we find is well fit by a lognormal.  We fit expressions for the local FUV attenuation that depend  just on density.  
    \item We introduce a normalized FUV radiation intensity, ${\cal J}$, by factoring out the total FUV luminosity per unit area per unit solid angle from all radiation sources in the disk and compare simulation values to three approximate models for ${\cal J}$.  We find the plane parallel approximation follows the simulated values closely, and therefore represents an attractive approach in simulations when full ray tracing is infeasible but an accurate radiation field is required to compute the photoelectric heating rate.  The other two approximations are within 50\% of the simulation value, and are useful for simple estimates of the radiation field.  
    \item We convolve the FUV radiation with density to make synthetic maps of dust heating in external galaxies. To the extent that the dust abundance and emissivity are spatially uniform, these heating maps are a proxy for IR dust emission.  We demonstrate that given the spatial distribution and luminosities of star clusters, it is straightforward to correct for nonuniform heating to obtain an estimator of the gas column, $N_\mathrm{H,est}$. The estimator is  linear in $N_{\rm H}$ up to columns exceeding $10^{22}\cm^{-2}$, and has small scatter.  This approach has potential applications in calibrating PAH emission observations to obtain high resolution maps of gas surface density. 
    \item For both of our simulations, the majority of warm ionized gas is diffuse, with densities $<1 \pcc$. We show that ionization-recombination equilibrium is satisfied for gas in the midplane that is exposed to LyC radiation and has  $x_n <0.1$.  The ionization parameter has a typical value  $U\sim 10^{-3}$ in both simulations. This result is consistent with the theoretical expectation that $U$ in diffuse gas will be independent of galactic environment.
    \item We model the LyC radiation field using a sum of sources model accounting for attenuation by neutral gas and dust similar to that in \citet{belfiore2022tale}, using this model to create  EM maps. While these maps do not reproduce the full structure of the EM from our ray-tracing simulations, the fitted value of the neutral fraction, $x_\mathrm{n,eff}$, is comparable to the true value within the diffuse ionized gas.  The corresponding value of the 
    effective mean free path for ionizing photons is of order 100 pc.
\end{itemize}

Beyond the many interesting individual results from the two simulations we have analyzed here, our presentation bears witness to the value of implementing accurate and efficient radiative transfer methods in numerical simulations of the multiphase ISM. The TIGRESS-NCR framework with ART has now been applied to a much wider range of galactic conditions, including environments with far lower metallicity \citep{2024arXiv240519227K}.  These new simulations will provide a diverse testbed in which to investigate connections between the radiation field and the physical state and observational diagnostics of interstellar material.  

\acknowledgements
This work was supported in part by grant 510940 from the Simons Foundation to E.~C.\ Ostriker. The work of C.-G.K.\ was supported in part by NASA ATP grant No.\ 80NSSC22K0717. J.-G.K. acknowledges support from the EACOA Fellowship awarded by the East Asia Core Observatories Association. Computational resources were provided by the Princeton Institute for Computational Science and Engineering (PICSciE) and the Office of Information Technology's High Performance Computing Center at Princeton University.

\software{Athena \citep{stone2008, stone_gardiner2009}, Matplotlib \citep{Hunter:2007}, NumPy \citep{harris2020array}, SciPy \citep{2020SciPy-NMeth}, Astropy \citep{astropy:2013, astropy:2018, astropy:2022}, xarray \citep{hoyer2017xarray}, pandas \citep{mckinney-proc-scipy-2010}}

\bibliography{references}

\appendix

\section{Approximate FUV Radiation Models}\label{sec:slab_deriv}

\citet{OML10} provided an expression for the angle-averaged radiation field at the midplane of a uniformly emitting slab with uniform FUV dust opacity $\kappa$, given by 
\begin{equation}
\label{eq:J_OML}
J=\frac{\Sigma_\mathrm{FUV}}{4 \pi \tau_\mathrm{\perp} }\left[1- E_2(\tau_\mathrm{\perp}/2)   \right],
\end{equation}
where $\Sigma_\mathrm{FUV}$ is the FUV luminosity per unit area from all sources, $\tau_\mathrm{\perp}=\Sigma_\mathrm{slab}\kappa$ is the FUV optical depth through the slab for $\Sigma_\mathrm{slab}$ its gas surface density, and $E_2$ is the second exponential integral.   Note that in this appendix we streamline the notation from the main text, dropping ``FUV'' subscripts for all variables where there is no ambiguity.  The expressions given here would also apply for any other continuum band (e.g. OPT, NUV) at frequency $\nu$, with appropriate substitutions $\Sigma_\mathrm{FUV} \rightarrow \Sigma_\mathrm{\nu}$, $\kappa \rightarrow \kappa_\mathrm{\nu}$, etc. \citep[e.g., Appendix C of][]{kim2023photchem}.

To derive \autoref{eq:J_OML}, first note that the emissivity per unit gas mass is 
$j=\Sigma_\mathrm{FUV}/\Sigma_\mathrm{slab}$, so that the source function is $S\equiv j/(4 \pi \kappa) = \Sigma_\mathrm{FUV}/(4 \pi \tau_\perp)$.  In the direction $\theta = \arccos(\mu)$ with respect to vertical, the midplane intensity is 
\begin{equation}
\label{eq:Imid}
    I(z=0;\mu) = \int_0^{\tau_\perp/(2\mu)}
    S e^{-\tau^{\prime}} d \tau^{\prime}= S[1 - e^{-\tau_\perp/(2\mu)}].
\end{equation}  
The angle-averaged midplane intensity is then, considering the symmetry above and below the midplane, 
\begin{eqnarray}
    J &=& S\int_0^1  [1 - e^{-\tau_\perp/(2\mu)}]d\mu=S\left[1 -\int_1^\infty 
    e^{-y \tau_\perp/2 } y^{-2} dy \right]\nonumber \\ 
    &=& S \left[1- E_2(\tau_\perp/2) \right];
\end{eqnarray}
here we have changed variables to $y=1/\mu$ and used the definition of $E_2$.
Substituting in for $S$, we obtain \autoref{eq:J_OML}. If we further divide by $\Sigma_\mathrm{FUV}/(4 \pi)$, we obtain the normalized midplane intensity ${\cal J}_{\rm slab,mid}$ given by \autoref{eq:J_norm_OML} of the text.  

\begin{figure*}[!t] 
    \centering
    \includegraphics[scale=0.4]{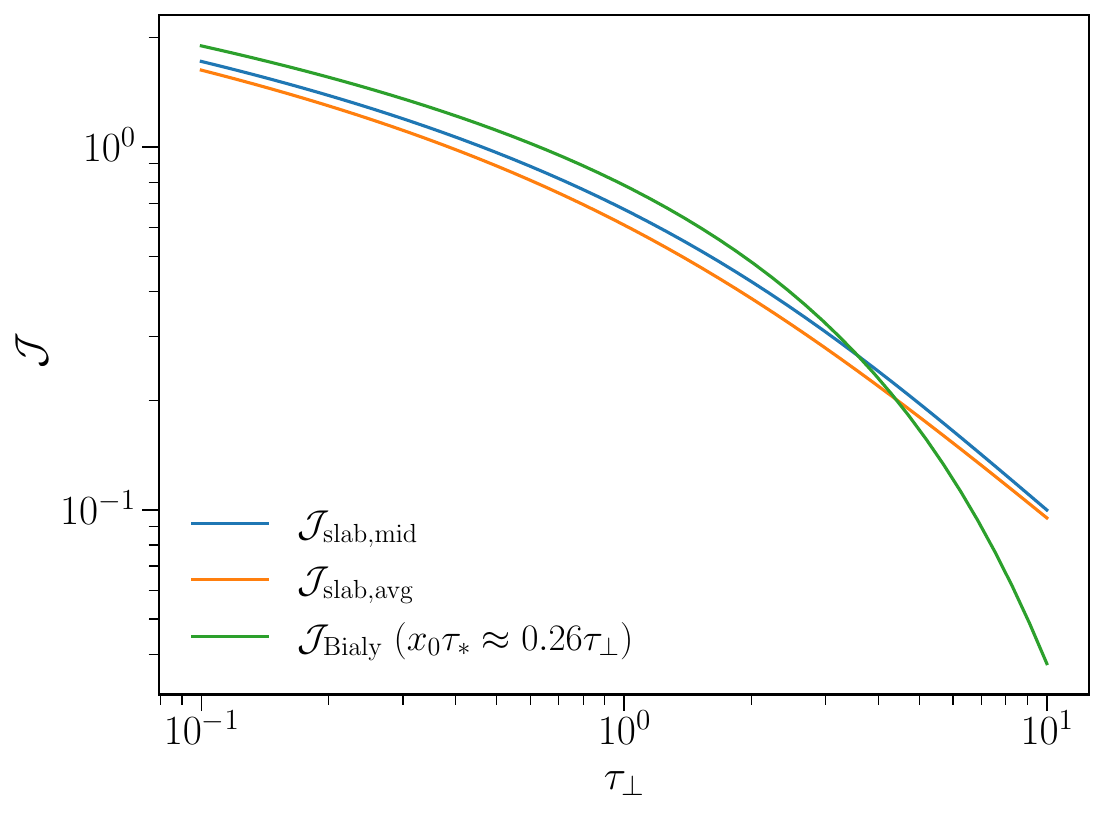}
    \caption{Comparison of the normalized FUV radiation field using different approximations as a function of $\tau_\perp$. For the ${\cal J}_{\rm Bialy}$ model, $\tau_\perp$ is defined using average values from the {\tt R8-4pc} simulation such that $x_0 \tau_* = 0.26 \tau_\perp$.}
    \label{fig:rad_model_comp}
\end{figure*}

For the vertically-averaged mean intensity, a similar calculation to \autoref{eq:Imid} gives
\begin{equation}
I(z; \mu)=  S\begin{cases}
   1- e^{- \tau_\perp(1-z/H)/(2\mu)}   &  \mu >0 \\
  1- e^{\tau_\perp(1+z/H)/(2\mu)}  &  \mu <0,
\end{cases}    
\end{equation}
where $2H$ is the thickness of the slab.
Averaging over $z$ and over $\mu$ then leads to the expression for ${\cal J}_{\rm slab, avg}$ given in \autoref{eq:J_norm_slab_ave} of the text.  
In \autoref{fig:rad_model_comp} we show the comparison between ${\cal J}_{\rm slab,mid}$ (\autoref{eq:J_norm_OML}) and ${\cal J}_{\rm slab,avg}$ (\autoref{eq:J_norm_slab_ave}) as a function of $\tau_\perp$.   

The \cite{bialy2020far} model depends on the product of the separation of sources $l_\star$, the midplane density $\nH$,  and the FUV crossection $\sigma_\mathrm{FUV}$.  The characteristic optical depth in  \autoref{eq:taustar} can also be 
written as $\tau_* \equiv l_\star \nH \sigma_\mathrm{FUV} = \tau_\mathrm{\perp} l_*/(2 H)$. Thus, given the parameter $\tau_\perp$ for the slab model, an additional parameter entering the \cite{bialy2020far} model is the ratio $l_*/(2 H)$.

For the {\tt R8-4pc} model, the mean $l_{\star} = 290\pc$ while $H = \Sigma_\mathrm{slab}/(2 \nH \mu_H m_p) \rightarrow 270\pc $ if we use the mean total gas surface density $10.5\;M_\odot\;\mathrm{pc}^{-2}$ for $\Sigma_\mathrm{slab}$ and $\nH = 0.57 \pcc$, which yields $x_0 \tau_\star = 0.24 \tau_\perp l_\star/H \sim 0.26\tau_\perp$ as the argument of ${\cal J}_\mathrm{Bialy}$ in \autoref{eq:bialy_F}. In \autoref{fig:rad_model_comp}, we compare the slab model scaled mean intensity to ${\cal J}_\mathrm{Bialy}$, assuming this factor (see upper axis).  The value of ${\cal J}_\mathrm{Bialy}$ is above that of ${\cal J}_{\rm slab,mid}$ and  ${\cal J}_{\rm slab,avg}$.  As noted in the text, a larger value of $l_\star \sim 2 H$ would be required for the estimates to yield similar values.

\section{Mass of diffuse ionized gas compared to classical HII Regions}\label{sec:ionmass}

For photoionized gas, ionization-recombination equilibrium in a given component of the ISM requires
\begin{equation}
Q_\mathrm{i,eff}\approx \alpha_{\rm B} n_{\rm i,rms}^2 V_{\rm i}\,,
\end{equation}
where for that component $Q_\mathrm{i,eff}$ is the photon rate that goes into ionization, $V_{\rm i}$ is the volume, and $n_{\rm i,rms}$ is the rms density of ionized gas. The mass of ionized gas in that component is then 
\begin{equation}
M_{\rm i} = \mu_{\rm H}m_{\rm H} n_{\rm i,mean} V_{\rm i} = \mu_{\rm H}m_{\rm H} f_{\rm ion}Q_\mathrm{i,eff} t_{\rm rec} \mathcal{C}_{n_e}
\end{equation}
where $\mu_{\rm H}m_{\rm H}=2.34 \times 10^{-24}\,{\rm g}$, $n_{\rm i,mean}$ is the mean density, $t_{\rm rec} = 1/(\alpha_{\rm B} n_{\rm i,rms})$ is the recombination time scale, and $\mathcal{C}_{n_e} \equiv n_{\rm i,mean}/n_{\rm i,rms} \le 1$ is the clumping correction factor \citep[e.g.][]{kado2020diffuse}.

Under the assumption that all diffuse ionized gas (DIG) is photoionized, the mass ratio between DIG and classical \HII\ regions would then be
\begin{equation}
    \frac{M_{\rm DIG}}{M_{\rm II}} = \dfrac{
    (Q_\mathrm{i,eff}
    )_{\rm DIG} t_{\rm rec,DIG} \mathcal{C}_{n_e,{\rm DIG}}}{(
    Q_\mathrm{i,eff}
    )_{\rm II} t_{\rm rec,II} \mathcal{C}_{n_e,{\rm II}}}
\end{equation}
For DIG, we assume that the source of ionizing photons is those escaping from classical \HII\ regions which are not absorbed by dust,  $Q_{\rm i,eff,DIG} 
= (1 - f_{\rm dust,DIG})f_{\rm esc,II}Q_{\rm i}$, for $Q_{\rm i}$ the total injected photons. For \HII\ regions,
$Q_\mathrm{eff,i,II} = (1 - f_{\rm dust,II} - f_{\rm esc,II})Q_{\rm i}$, so that  
\begin{equation}
    \frac{M_{\rm DIG}}{M_{\rm II}} \approx \dfrac{(1 - f_{\rm dust,DIG})f_{\rm esc,II}}{(1 - f_{\rm dust,II} - f_{\rm esc,II})} \times \dfrac{n_{\rm i,rms,II}}{n_{\rm i,rms,DIG}} \times \dfrac{\mathcal{C}_{n_e,{\rm DIG}}}{\mathcal{C}_{n_e,{\rm II}}}
\end{equation}
The escape fraction, dust absorption fraction, and clumping factors for DIG and classical \HII\ regions can be quantified from TIGRESS and cloud-scale simulations, respectively. The first and third terms are expected to be of order unity 
\citep{kimjg2019,kado2020diffuse}, but $n_{\rm i,rms,II} \gg n_{\rm i,rms,DIG}$. We therefore conclude that $M_{\rm DIG} \gg M_{\rm II}$ is expected.

\end{document}